%% file: main.tex
\documentclass[acmtog]{acmart}

\usepackage{multirow}
\usepackage{makecell}
\usepackage{tabularx}
\usepackage{wrapfig}
\usepackage{tabularx}
\usepackage{diagbox}
\usepackage{booktabs}
\usepackage{adjustbox}
\usepackage{array}
\usepackage[usestackEOL]{stackengine}
\usepackage{cuted}
\usepackage{capt-of}
\usepackage{listings}
\usepackage{enumitem,kantlipsum}
\usepackage[ruled]{algorithm2e}

\DeclareMathOperator*{\argmin}{arg\,min}

\AtBeginDocument{%
  \providecommand\BibTeX{{%
    \normalfont B\kern-0.5em{\scshape i\kern-0.25em b}\kern-0.8em\TeX}}}

\setcopyright{rightsretained}
\acmJournal{TOG}
\acmYear{2024} \acmVolume{43} \acmNumber{6} \acmArticle{217} \acmMonth{12}\acmDOI{10.1145/3687964}

\acmSubmissionID{823}

\setcitestyle{square}

\begin{document}

%%
%% The "title" command has an optional parameter,
%% allowing the author to define a "short title" to be used in page headers.
% \title{Medial Skeletal Diagram: a Compact and Complete Representation for 3D Shapes}
\title[Medial Skeletal Diagram]{Medial Skeletal Diagram: A Generalized Medial Axis Approach for 3D Shape Representation}

%%
%% The "author" command and its associated commands are used to define
%% the authors and their affiliations.
%% Of note is the shared affiliation of the first two authors, and the
%% "authornote" and "authornotemark" commands
%% used to denote shared contribution to the research.
\author{Minghao Guo}
\authornote{Both authors contributed equally to this research.}
\email{guomh2014@gmail.com}
\orcid{0000-0003-3408-4997}
\author{Bohan Wang}
\authornotemark[1]
\email{wangbh11@gmail.com}
\orcid{0000-0003-1439-1455}
% \affiliation{%
%   \institution{Institute for Clarity in Documentation}
%   \streetaddress{P.O. Box 1212}
%   \city{Dublin}
%   \state{Ohio}
%   \country{USA}
%   \postcode{43017-6221}
% }

\author{Wojciech Matusik}
\email{wojciech@csail.mit.edu}
\orcid{0000-0003-0212-5643}
\affiliation{%
  \institution{CSAIL, MIT}  
  \city{Cambridge, MA}
  \country{USA}}

%%
%% By default, the full list of authors will be used in the page
%% headers. Often, this list is too long, and will overlap
%% other information printed in the page headers. This command allows
%% the author to define a more concise list
%% of authors' names for this purpose.
\renewcommand{\shortauthors}{Guo and Wang, et al.}

\newcommand{\mh}[1]{#1}
\newcommand{\bh}[1]{#1}
\newcommand{\hl}[1]{#1}
\newcommand{\hln}[1]{#1}
% test
%\newcommand{\hlca}[1]{\textcolor{orange}{#1}}
\newcommand{\hlca}[1]{#1}

\begin{strip}\centering
\includegraphics[width=0.96\textwidth]{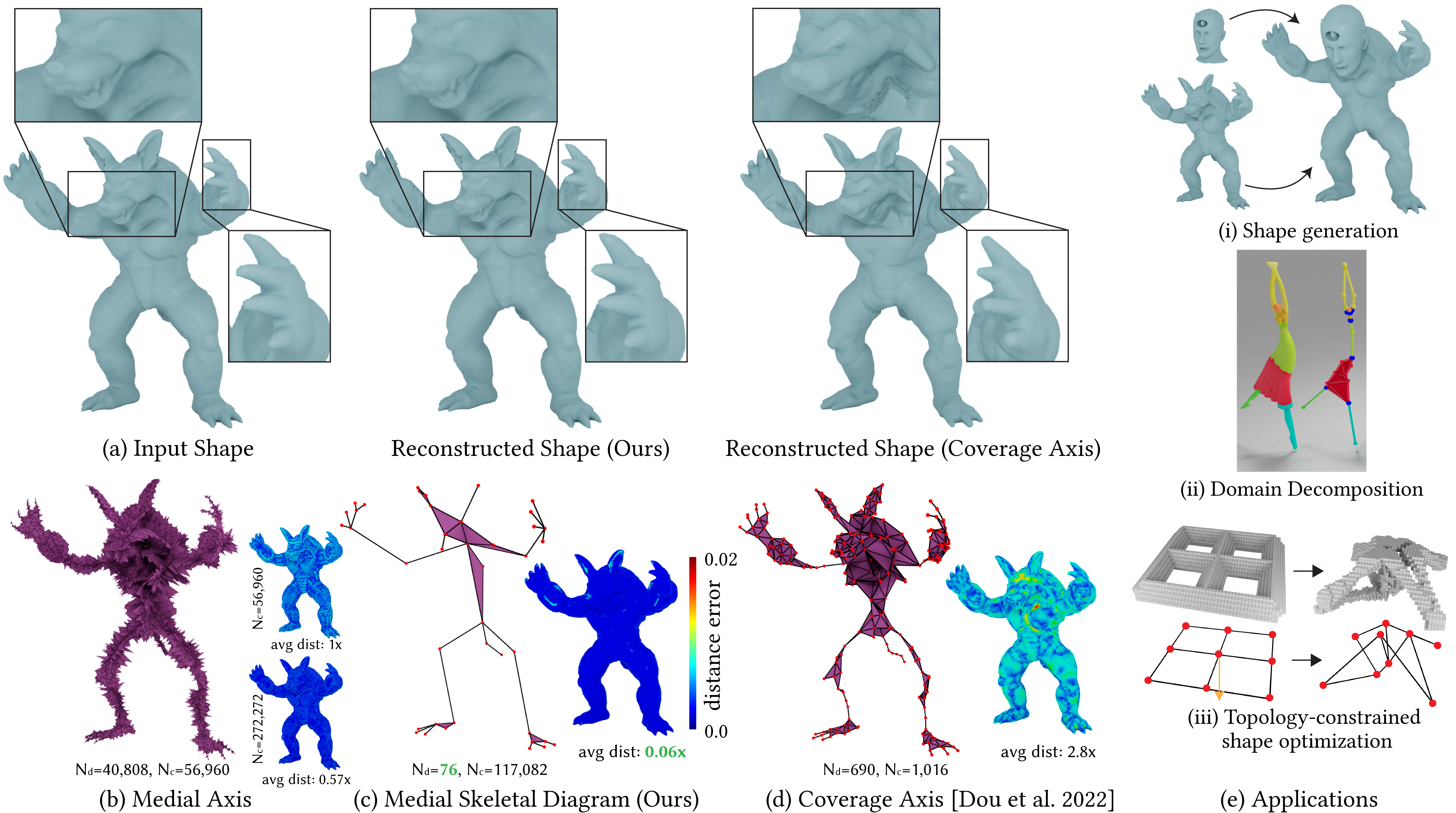}
\vspace{-0.3cm}
\captionof{figure}{
\textbf{We introduce the \textbf{Medial Skeletal Diagram} -- \hlca{a representation with a sparse skeleton} that generalizes the medial axis.}
\hln{
Our representation reduces the number of discrete elements ($N_d$) required compared to both standard and simplified medial axes ((b) and (d)), using only 76 elements compared to their 40,808 and 690, respectively.
This reduction is achieved
while delivering the highest reconstruction accuracy (c).
% Whereas the medial axis and the coverage axis require 40,808 and 690 discrete elements, respectively, 
% our representation needs only 99 elements.
The \hlca{sparsity} of discrete elements and the completeness of our method facilitate a broad range of applications, 
such as shape generation, mesh decomposition, and shape optimization (e).
In terms of continuous parameters ($N_c$), our method uses 117,082 scalars, whereas 
the medial axis and coverage axis use 56,960 and 1,016 scalars. 
Experimentally, 
even a $4.8\times$ increase in the number of scalars for the medial axis ($2.3\times$ our method) -- 
undesirably resulting in an approximate proportional increase in the number of discrete elements to 201,323 -- 
still yields an average reconstruction error that is 9$\times$ higher than that of our method.}
}
\vspace{-0.3cm}
\label{fig:teaser}
\end{strip}

%%
%% The abstract is a short summary of the work to be presented in the
%% article.
\begin{abstract}
\input{sections/abstract}
\end{abstract}

%%
%% The code below is generated by the tool at http://dl.acm.org/ccs.cfm.
%% Please copy and paste the code instead of the example below.
%%

%% Keywords. The author(s) should pick words that accurately describe
%% the work being presented. Separate the keywords with commas.

% \received{20 February 2007}
% \received[revised]{12 March 2009}
% \received[accepted]{5 June 2009}

%%
%% This command processes the author and affiliation and title
%% information and builds the first part of the formatted document.
\maketitle

\input{sections/introduction}

\input{sections/related_work}

\input{sections/approach_representation}

\input{sections/approach_computation}

\input{sections/experiments}

\input{sections/limitation}

\input{sections/conclusion}

\begin{acks}
The authors would like to express their gratitude for the constructive feedback provided by the reviewers. 
The authors also thank the authors of MATFP~\citep{wang2022computing}, LS~\citep{baerentzen2021skeletonization}, Coverage Axis~\citep{dou2022coverage}, Point2Skeleton~\citep{lin2021point2skeleton}, and CoACD~\citep{wei2022approximate} for making their code publicly available for research purposes.
\end{acks}

\bibliographystyle{ACM-Reference-Format}
\bibliography{sample-base}

%%
%% If your work has an appendix, this is the place to put it.
\appendix
\input{sections/appendix}
\end{document}

%%
%% The "title" command has an optional parameter,
%% allowing the author to define a "short title" to be used in page headers.
\title[Supplementary Technical Details for ``Medial Skeletal Diagram'']{Supplementary Technical Details for ``Medial Skeletal Diagram: A Generalized Medial Axis Approach for 3D Shape Representation''}

%%
%% The "author" command and its associated commands are used to define
%% the authors and their affiliations.
%% Of note is the shared affiliation of the first two authors, and the
%% "authornote" and "authornotemark" commands
%% used to denote shared contribution to the research.
\author{Minghao Guo}
\authornote{Both authors contributed equally to this research.}
\email{guomh2014@gmail.com}
\orcid{0000-0003-3408-4997}
\author{Bohan Wang}
\authornotemark[1]
\email{wangbh11@gmail.com}
\orcid{0000-0003-1439-1455}
% \affiliation{%
%   \institution{Institute for Clarity in Documentation}
%   \streetaddress{P.O. Box 1212}
%   \city{Dublin}
%   \state{Ohio}
%   \country{USA}
%   \postcode{43017-6221}
% }

\author{Wojciech Matusik}
\email{wojciech@csail.mit.edu}
\orcid{0000-0003-0212-5643}
\affiliation{%
  \institution{CSAIL, MIT}  
  \city{Cambridge, MA}
  \country{USA}}

%%
%% By default, the full list of authors will be used in the page
%% headers. Often, this list is too long, and will overlap
%% other information printed in the page headers. This command allows
%% the author to define a more concise list
%% of authors' names for this purpose.
\renewcommand{\shortauthors}{Guo and Wang, et al.}
  
%%
%% This command processes the author and affiliation and title
%% information and builds the first part of the formatted document.
\maketitle

%% forcing some vspace => need \phantom{} else the layout ignores it
%\phantom{.}
%\smallskip

%% 
%% Main content
\input{sections/supp-refinement.tex}
%%
%% The next two lines define the bibliography style to be used, and
%% the bibliography file.
\bibliographystyle{ACM-Reference-Format}
\bibliography{sample-base}

%% file: sections/abstract.tex
We propose the Medial Skeletal Diagram, a novel skeletal representation that tackles the prevailing issues around \hlca{skeleton sparsity} and reconstruction accuracy in existing skeletal representations.
Our approach augments the continuous elements in the medial axis representation to effectively shift the complexity away from the discrete elements.
To that end, we introduce generalized enveloping primitives, 
an enhancement over the standard primitives in the medial axis,
which ensure efficient coverage of intricate local features of the input shape and substantially reduce the number of discrete elements required.
Moreover, we present a computational framework for constructing a medial skeletal diagram from an arbitrary closed manifold mesh.
\hl{Our optimization pipeline ensures that the resulting medial skeletal diagram comprehensively covers the input shape with the fewest primitives.
}
Additionally, each optimized primitive undergoes a post-refinement process to guarantee an accurate match with the source mesh in both geometry and tessellation.
We validate our approach on a comprehensive benchmark of 100 shapes,
demonstrating the \hlca{sparsity} of the discrete elements and superior reconstruction accuracy across a variety of cases.
\hl{
Finally, we exemplify the versatility of our representation in downstream applications such as shape generation, mesh decomposition, shape optimization, mesh alignment, mesh compression, and user-interactive design.
}

%% file: sections/introduction.tex
\section{Introduction}
Geometric shape representations are fundamental to diverse areas of computer graphics and geometric modeling. 
Among them, the shape skeleton stands out as a powerful and intuitive tool for understanding and manipulating shapes, with a broad range of applications including 3D reconstruction~\citep{tang2019skeleton}, shape segmentation~\citep{lin2020seg}, character animation~\citep{baran2007automatic}, and shape correspondence~\citep{wong2012skeleton}.
As a prominent example among skeletal representations, the \emph{medial axis}~\citep{blum1967transformation} is ubiquitous in shape approximation, simplification, and analysis tasks. 
Its approximative counterpart, \emph{medial axis transform} (MAT), comprises a \emph{discrete} simplicial complex (termed \emph{medial mesh}) and a \emph{continuous} radius function defined on each vertex of the complex. 
Consequently, MAT enjoys many advantages including 
% lower dimensionality than the original shapes, 
capture of components and protrusions, and preservation of homotopy~\citep{lieutier2003any}.

Nonetheless, like most existing skeletal representations, the medial axis representation encounters an inevitable trade-off between \hlca{\emph{sparsity}} in the skeletal structure and \emph{completeness} in the reconstruction.
\hl{The skeletal structure} should ideally be as simple and \hlca{sparse} as possible to allow for easy interpretation and modification.
However, it is equally important to retain sufficient complexity to faithfully reconstruct all geometric details and features.
Balancing such conflicting requirements poses a significant challenge. 
The excessive complexity hinders the application of medial axis to shape generation and interactive shape editing, which require a \hlca{sparser} skeleton but a complete representation.
Numerous research efforts have been dedicated to simplifying the discrete part of the skeletal representation to enhance its manageability and visual cleanliness~\citep{li2015q, dou2022coverage, lin2021point2skeleton, tagliasacchi20163d}.
However, despite these advancements, the simplified skeleton is still bounded by excessively many elements with compromised reconstruction accuracy, resulting in degraded quality of the skeleton and a loss of geometric detail.

To conquer these limitations, we present a novel skeletal representation that significantly reduces the number of discrete elements while preserving the completeness characteristic of the medial axis.
% remarkable \mh{reconstruction accuracy -- almost identical to the original shape in both geometry and tessellation., complete}
Our method stems from the insight that the complexity of the medial axis skeleton can be decimated by an augmented representation of the \emph{continuous} elements.
\hln{Our representation shifts the burden from the number of discrete elements to the number of continuous parameters, 
achieving a significant reduction in discrete elements at the cost of increasing the number of continuous parameters.}
Discrete elements are often challenging to optimize due to their combinatorial nature, which results in severe computational bottlenecks.
In contrast, continuous elements can be formulated and solved efficiently using off-the-shelf optimization frameworks. 

Our versatile skeletal representation, named the \emph{medial skeletal diagram} (MSD), extends the medial axis using \emph{generalized enveloping primitives}.
Compared with the uniform, linearly interpolated primitives used in the medial axis, our generalized enveloping primitives are \emph{non-uniform} and \emph{nonlinearly interpolated}. 
For example, the medial axis assumes consistent and uniform radii,
whereas in our medial skeletal diagram, each medial sphere has varying radii along different directions.
Such generalized primitives can effectively cover complex local regions, which in turn reduces the number of primitives required to represent a given shape.
As a result, the medial skeletal diagram can represent arbitrary closed manifold surfaces, with the raw medial axis being a special case. 

\hl{
To construct a medial skeletal diagram for an arbitrary input, 
we start by building a forward procedure. 
This procedure takes a subset of points from the raw medial axis as its input and produces an MSD as its output. 
Utilizing \emph{restricted Voronoi diagrams} (RVD) and \emph{restricted Delaunay triangulation} (RDT), 
we determine the discrete skeletal structure based on the input set of points, which effectively ``remesh'' the medial axis.
Following this, generalized enveloping primitives are generated at each element of the discrete skeletal structure to align with the input mesh. 
With this established procedure, transforming a subset of medial axis points into a medial skeletal diagram, we then introduce an optimization pipeline.
This pipeline is designed to identify the optimal subset of points, ensuring that the resulting medial skeletal diagram covers the input shape with the fewest necessary primitives. In the final stage, the optimized primitives undergo a refinement phase, aimed at achieving precise alignment with the target mesh in both its geometric form and tessellation.
}

The contributions of this paper are summarized as follows:
\begin{itemize}
    \item We propose a novel 3D shape representation (in Sec.~\ref{sec:msdRepresentation}) that extends the medial axis approach using generalized enveloping primitives, enabling both \hlca{sparse} skeletal structures and accurate representation for arbitrary closed manifold surfaces. 
    \item We introduce a computational pipeline to construct the medial skeletal diagram from any given shape (in Sec.~\ref{sec:msdComputation}).
    \hl{ Our pipeline finds the best MSD which maximizes the coverage of the input shape with a minimal number of primitives.
    }
    A subsequent refinement process precisely aligns optimized primitives with the target mesh, both in terms of geometry and tessellation.
    % Our implementation utilizes exact arithmetic and predicates to achieve superior accuracy and robustness.
    \item We validate the medial skeletal diagram on a benchmark of $100$ shapes to demonstrate the \hlca{sparsity} of the skeleton and the completeness of the representation (in Sec.~\ref{sec:exp}).
    \hl{Furthermore, we showcase the versatility and advantages of our representation in the context of shape optimization, shape generation, mesh decomposition, mesh alignment, mesh compression, and user-interactive design.}
\end{itemize}

%% file: sections/related_work.tex
\section{Related Work}

\paragraph{Medial axis transformation.}
MAT represents a 3D shape using thin-centered structures and is able to jointly describe shape topology and geometry.
The rich history of MAT starts from early algorithms that employ seam tracing techniques to calculate the exact MAT for polyhedra~\cite{milenkovic1993robust, sherbrooke1996algorithm, culver2004exact}.
These exact algorithms primarily concentrate on simple shapes with up to~$20$ faces.
A significant amount of subsequent research has shifted towards approximated MAT, which is more suitable for practical applications~\cite{amenta2001powera, pizer2003multiscale, dey2004approximating, chazal2005lambda, miklos2010discrete, sobiecki2014comparison, saha2016survey}.
For a more comprehensive discussion, we refer the reader to survey papers by~\citet{pizer2003multiscale} and~\citet{tagliasacchi20163d}.

\paragraph{Curve skeletonization.}
% Curve skeletonization has been widely researched in the realms of computer graphics and computer vision. 
% \mh{mention rigging}
% The most relevant shape representation to our work is curve skeletons, which have been widely studied in rigging and interactive shape modeling~\cite{borosan2012rigmesh}.
One type of shape representation relevant to our work is curve skeletons, which have been widely studied in rigging and interactive shape modeling~\cite{borosan2012rigmesh}.
Traditional approaches compute curve skeletons following hand-crafted rules that encode geometric information. 
For instance, \citet{ma2003skeleton} extract skeletons using radial basis functions. 
\citet{sharf2007fly} apply deformable model evolution to capture the volumetric shape of an object for curve skeleton approximation. 
\citet{au2008skeleton} compute the curve skeleton via mesh extraction. 
\citet{livesu2012reconstructing} use visual hulls to reconstruct curve skeletons.
Mean curvature skeletonization~\cite{tagliasacchi2012mean} creates curve skeletons by deploying mean curvature flow on the shape. 
Reeb graphs~\cite{tierny2008enhancing} are also considered curve skeletons, which are constructed by tracing the isocontours of a given height function and contracting connected components into a single point.
However, they are often suboptimal as skeleton junctions can be very close to the surface.
Recently, \citet{baerentzen2021skeletonization} and \citet{baerentzen2023multilevel} treat 3D shapes as spatially embedded graphs and compute curve skeletons using local operators.
In shape modeling, ~\citet{baerentzen2014interactive} use polar-annular mesh to obtain a co-representation of both the mesh and the curve skeleton. 
\citet{pandey2022face} obtain a curve skeleton from a sequence of face-loop modeling operations on a quad mesh.
% Meanwhile, learning-based techniques for predicting curve skeletons have emerged~\cite{xu2019predicting}. 
Although curve skeletons are well-suited for representing organic, articulated shapes, they often suffer from capturing
% an excessive number of elements to capture 
large thin flat components, which are prevalent in CAD models. 
Our representation incorporates triangles and effectively captures large, flat regions with fewer elements.
% \mh{TODO Reeb graph, copy from Skeletonization via LS}
% \mh{add necessary discussion for \url{http://www.dgp.toronto.edu/~karan/papers/pam_siggraph14.pdf}}
% Nevertheless, while curve skeletons excel at representing tubular shapes, they struggle to depict arbitrary shapes, particularly those with flat components.
\vspace{-0.2cm}
\paragraph{Medial axis Simplification.}
The medial axis representation is complete but sensitive to boundary noise. This has led to a large body of research on its simplification while preserving significant and stable parts. 
Angle-based filtering methods~\cite{amenta2001powerb, foskey2003efficient, dey2002approximate, sud2005homotopy} consider the angle in MAT during simplification but struggle to maintain the original topology. 
$\lambda$-medial axis methods~\cite{chazal2005lambda, chaussard2011robust} use \hlca{the radius of the closest medial sphere} as a pruning criterion but have limitations in feature preservation at different scales~\cite{attali2009stability}.
The Scale Axis Transform (SAT)~\cite{miklos2010discrete} effectively prunes spikes by identifying unstable medial axis points.
However, SAT has high computational costs and risks topology disruption by introducing new topological structures at large scales. 
Delta Medial Axis~\cite{marie2016delta} and Bending Potential Ratio pruning~\cite{shen2011skeleton} primarily focus on MAT simplification but cannot precisely reconstruct the original shape. 
Q-MAT~\cite{li2015q} collapses edges of MAT based solely on local information, which may lead to a suboptimal spatial distribution of medial vertices.
Voxel Cores~\cite{yan2018voxel} approximate the medial axis of any smooth shape while guaranteeing correct topology but require fine voxel resolutions and high computational cost to achieve high geometric accuracy.
Erosion Thickness~\cite{yan2016erosion} computes the simplified medial axis by introducing a burning process over the medial axis but is unable to reconstruct the original shape exactly.
Coverage Axis~\cite{dou2022coverage} aims to encompass all mesh surface points while minimizing the number of internal medial spheres.
MATFP~\cite{wang2022computing} computes MAT using a restricted power diagram 
while ensuring the preservation of both external mesh surface features and 
internal medial axis features.
Both Coverage Axis and MATFP serve as competitive baselines for comparison in our experiments.
Another line of research on point cloud data skeletonization includes methods like $l1$-medial skeleton~\cite{huang2013l1}, LSMAT~\cite{rebain2019lsmat}, and meso-skeleton based approaches~\cite{wu2015deep, tagliasacchi2012mean}. 
However, these methods often lack topological constraints or exhibit large errors.
Recently, deep learning has been used for skeleton construction including P2MAT-NET~\cite{yang2020p2mat}, Deep medial fields~\cite{rebain2021deep}, and Point2Skeleton~\cite{lin2021point2skeleton}. They do not guarantee accurate geometric features and suffer from generalization issues due to training data dependency.
Compared to these medial axis simplification methods, our approach outputs a \hlca{sparser} skeleton and employs a global optimization process to obtain the skeleton as opposed to a suboptimal heuristic search.
% Our approach offers several key advantages compared to previous medial axis simplification methods. 
% First, we propose generalized medial primitives that result in a significantly more compact skeleton. 
% it does not rely on the medial primitives in the standard medial axis; instead, we propose  that results in a significantly simpler skeleton. 
% Second, our method employs a global optimization process to obtain the skeleton as opposed to a suboptimal heuristic search.
These improvements allow our approach to efficiently generate higher-quality skeletons for various applications.
% , including shape generation, shape optimization, and interactive editing.

\paragraph{Shape Decomposition.}
The generalized enveloping primitives in our method share similarities with several variants proposed in the field of shape decomposition~\citep{simari2005extraction, zhou2015generalized, wei2022approximate}.
Examples of such methods include the use of generalized cylinders~\citep{zhou2015generalized} and ellipsoidal surface regions~\citep{simari2005extraction}, both involving the use of predefined primitives.
Our method has a distinct primary objective from shape decomposition as we aim to extract a \hlca{sparse} and meaningful skeleton.
By contrast, most shape decomposition methods do not yield a skeleton directly based on the decomposed domains.
Although some of these methods, including generalized cylinder decomposition~\citep{zhou2015generalized}, can produce a curve skeleton, they share the typical limitations of other curve skeleton techniques and do not attain the level of \hlca{sparsity} that our approach offers.
% \paragraph{Medial axis transformation.}
% MAT represents a 3D shape using thin-centered structures and jointly describes the topology and the geometry of the shape.
% Computing MAT has a rich history, dating back to early algorithms that employ seam tracing techniques for the exact calculation of MAT for polyhedra~\cite{milenkovic1993robust, sherbrooke1996algorithm, culver2004exact}.
% The exact computation algorithms primarily concentrate on calculating the MAT for simple shapes with up to approximately $20$ faces. 
% A significant amount of subsequent research has shifted towards approximated MAT, which is more suitable for practical real-world applications~\cite{amenta2001powera, pizer2003multiscale, dey2004approximating, chazal2005lambda, miklos2010discrete, sobiecki2014comparison, saha2016survey}.
% For a more comprehensive discussion, we refer readers to the survey papers~\cite{pizer2003multiscale, tagliasacchi20163d}.

%% file: sections/approach_representation.tex
\section{Preliminaries}\label{sec:preliminaries}
\subsection{Medial Axis and Medial Mesh}

Consider a closed, oriented shape $\mathbf{\mathcal{S}}$ in $\mathbb{R}^3$. The medial axis $\mathbf{\mathcal{M}}$ is defined as 
the set of centers of maximal spheres inscribed within $\mathbf{\mathcal{S}}$. 
\begin{wrapfigure}[9]{r}[0pt]{0.19\textwidth}
\begin{center}
\hspace{-1cm}
\includegraphics[width=0.22\textwidth]{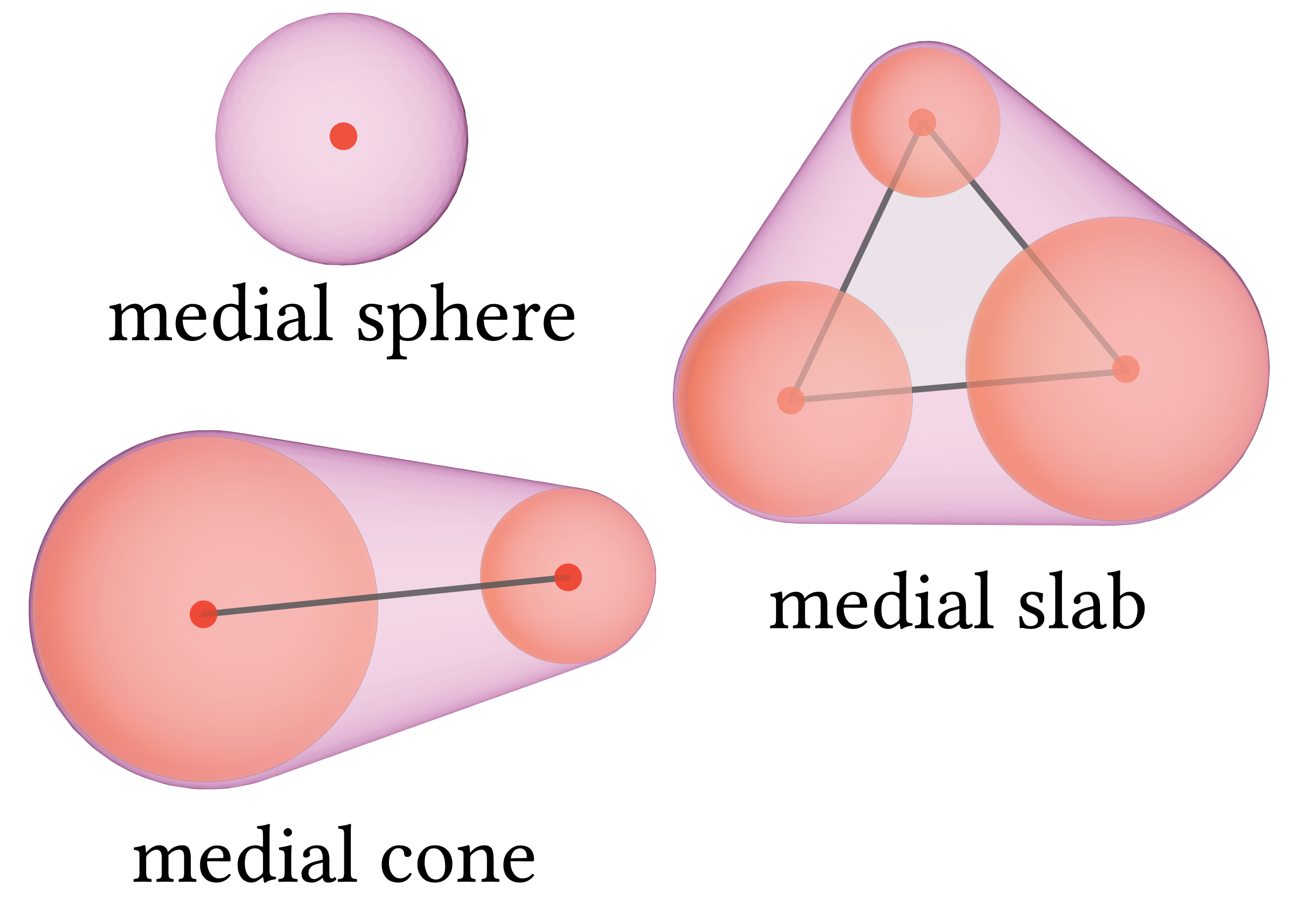}
\end{center}
% \caption{Birds}
\end{wrapfigure}
Each sphere is tangent to at least two points on the boundary of $\mathbf{\mathcal{S}}$ and does not contain any other boundary points in its interior. 
% The medial axis transformation (
The MAT comprises both $\mathbf{\mathcal{M}}$ and the radius function associated with each sphere center~\cite{li2015q}. 
% \bh{probably need a citation. check what other people cite}

Typically, MAT is approximated using a 2D simplicial complex known as the \emph{medial mesh}~\citep{li2015q, sun2015medial}.
The medial mesh is a triangle mesh that includes three types of basic elements: 
\emph{medial vertex}, denoted by $\mathbf{v}$;
\emph{medial edge}, a line segment connecting two medial vertices, denoted by $\mathbf{e}$; and
\emph{medial face}, the convex combination of three medial vertices, defined by $\mathbf{f}$.
Each of these three elements corresponds to an \emph{enveloping primitive} in $\mathbb{R}^3$:
1) \emph{medial sphere} which is a uniform sphere centered at the corresponding medial vertex;
% , $\mathbf{m} = ({P}, r)$,  where ${P}\in\mathbb{R}^3$ represents the position of the corresponding medial vertex, and $r$ denotes its associated radius value; 
2) \emph{medial cone},
% $\mathbf{e} = (1-t)\mathbf{m}_1 + t \mathbf{m}_2, t\in[0,1]$, 
which is the linear interpolation of two spheres connected by a medial edge; and
3) \emph{medial slab}, the convex hull of three spheres connected by a medial triangle face.
See the inset figure for an illustration.
% \mh{mention topology information, need to change the notation to avoid overlapping}

\subsection{Restricted Voronoi Diagram and Delaunay Triangulation}
Given a finite set of points $\mathbf{V} = \{\mathbf{v}_k\} \subset \mathbb{R}^n$, the \emph{Voronoi cell} of a point $\mathbf{v}_k$ comprises all points in $\mathbb{R}^n$ whose distance to $\mathbf{v}_k$ is no greater than their distance to any other point in $\mathbf{V}$. It can be defined as
\begin{align*}
    \mathrm{Vor}(\mathbf{v}_k, \mathbf{V}) = \big\{ \mathbf{x}\in\mathbb{R}^n, ||\mathbf{v}_k - \mathbf{x}|| \leq ||\mathbf{v}_l - \mathbf{x}||, \forall l\neq k\big\}.
\end{align*}
The \emph{Voronoi diagram} is the collection of all Voronoi cells.
The \emph{Delaunay triangulation} is the complex with a dual structure to the Voronoi diagram.
Each $d$-dimensional Voronoi element is dual to an $(n{-}d)$-dimensional Delaunay element, which is defined as the convex hull of points in $\mathbf{V}$ whose Voronoi cells have the considered Voronoi element on their boundary.
For instance, when $n=3$, a Delaunay vertex corresponds to a Voronoi volumetric cell, a Delaunay edge corresponds to a Voronoi surface, and vice versa.

\begin{wrapfigure}[8]{r}{0.21\textwidth}
  \begin{center}
  \vspace{-0.5cm}
  \hspace{-0.5cm}
    \includegraphics[width=0.21\textwidth]{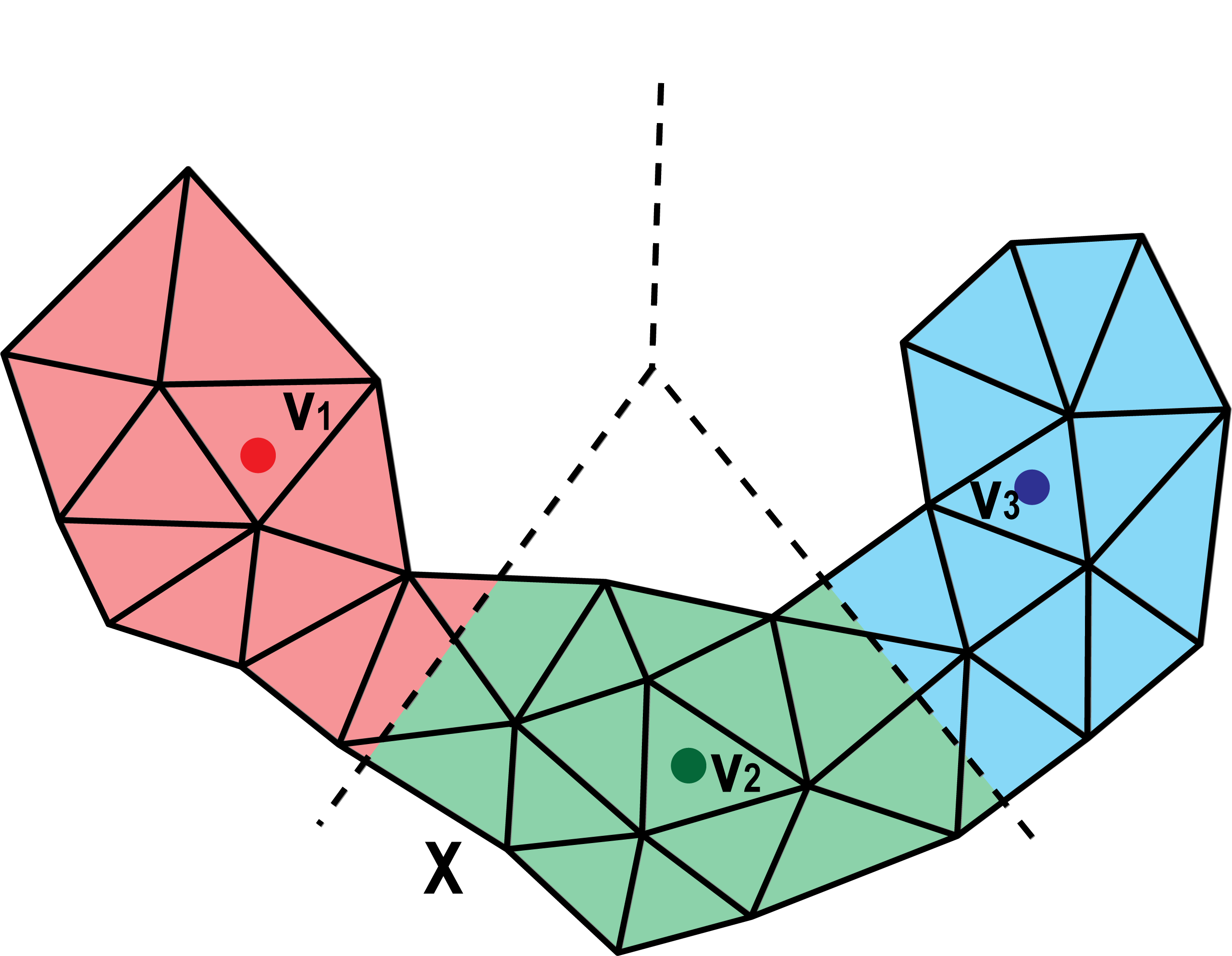}
  \end{center}
  % \caption{Birds}
\end{wrapfigure}
The Voronoi diagram restricted within a given domain $\mathbf{X}\subseteq\mathbb{R}^n$ is called \emph{restricted Voronoi diagram} (RVD). 
RVD is defined as the collection of \emph{restricted Voronoi cells} (RVC), each of which is the intersection between the Voronoi cell and the domain:
\begin{align}\label{eq:RVD}
\begin{aligned}
    \mathrm{RVC}(\mathbf{v}_k, \mathbf{V}) &= \mathrm{Vor}(\mathbf{v}_k, \mathbf{V}) \cap \mathbf{X} \\
    &= \big\{\mathbf{x}\in\mathbf{X}, ||\mathbf{v}_k - \mathbf{x}|| \leq ||\mathbf{v}_l - \mathbf{x}||, \forall l\neq k\big\},
\\
    \mathrm{RVD}(\mathbf{V}, \mathbf{X}) &= \big\{ \mathrm{RVC}(\mathbf{v}, \mathbf{V}), \ \ \forall \mathbf{v}\in \mathbf{V}, \mathrm{Vor}(\mathbf{v}, \mathbf{V}) \neq \varnothing \big\}.
\end{aligned}
\end{align}
Likewise, the \emph{restricted Delaunay triangulation} (RVT) is the dual of the restricted Voronoi diagram, with the constraint that all its elements lie within the domain $\mathbf{X}$.

\section{Medial Skeletal Diagram Representation}\label{sec:msdRepresentation}
\begin{figure}[t]
\centerline{\includegraphics[width=0.75\linewidth]{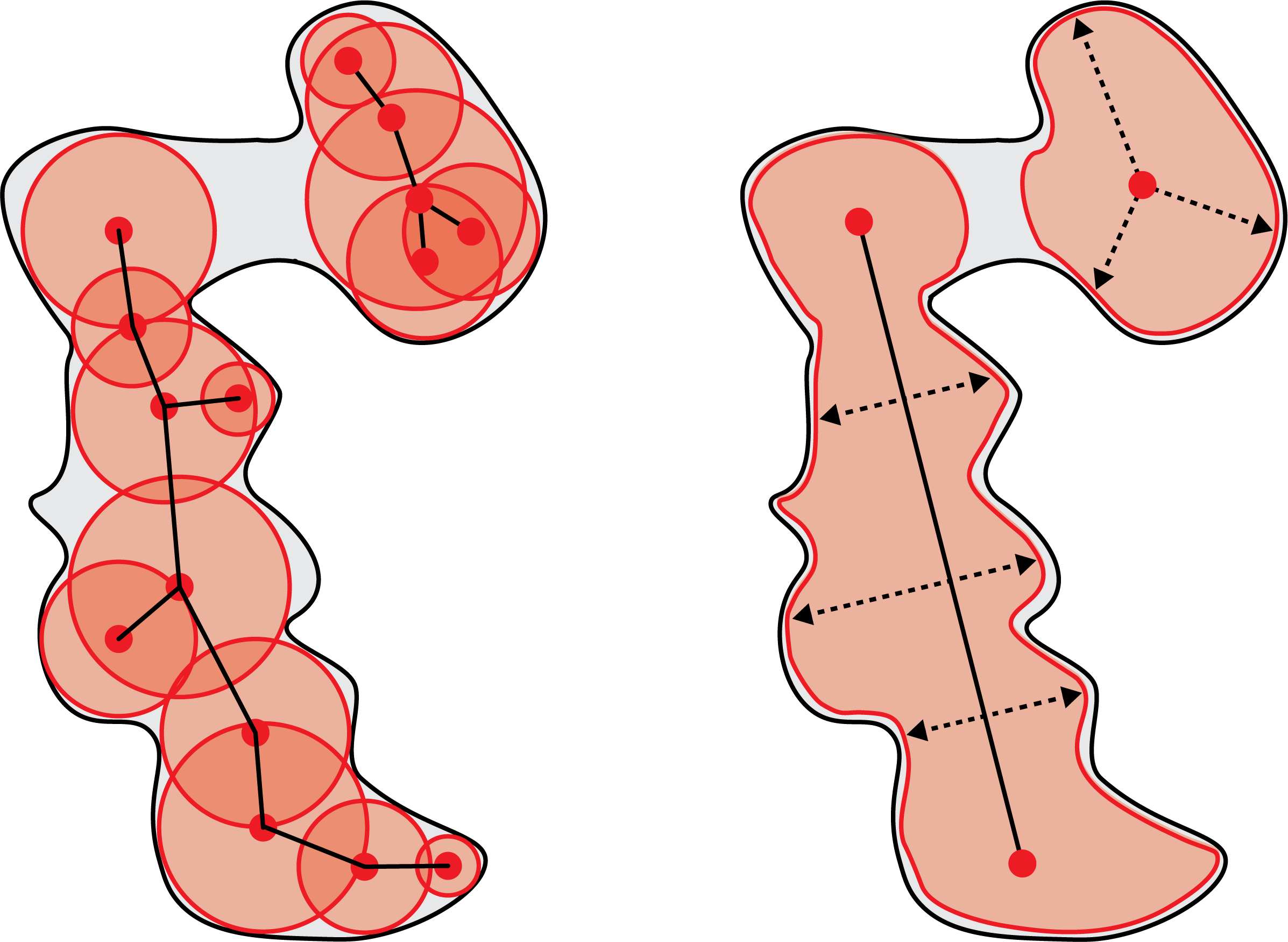}}
\caption{\textbf{Generalized enveloping primitives in 2D.} An illustration of employing our generalized enveloping primitives (right) for representing a 2D region. Compared to the standard medial primitives (left), our representation exhibits enhanced \hlca{the sparsity of the skeleton} when dealing with intricate and high-curvature geometries.}
\vspace{-0.25cm}
\label{fig:generalizedEP2D}
\end{figure}

In this section, we present how our proposed medial skeletal diagram represents 3D shapes (closed, oriented 2D manifolds).
We leverage the medial axis to capture the topological and geometric information of 3D shapes but greatly improve the \hlca{sparsity} of the skeleton while maintaining its accuracy.
Our fundamental insight is that while the convex nature of medial enveloping primitives offers advantages such as simplicity and rapid proximity query, it \hlca{restricts} the ability to accurately represent intricate local regions, especially those with high-curvature sharp features.
This issue is depicted in Fig.~\ref{fig:generalizedEP2D}, where it is clear that a substantial number of medial spheres are needed to encompass all vertices in sharp local regions, resulting in a complex and unwieldy medial axis.
Furthermore, a majority of these spheres intersect with only a few vertices of the target mesh, often merely covering two vertices, which leads to considerable inefficiency. This problem is exacerbated by the fact that the medial axis becomes overly complicated and noisy even for shapes with simple topologies.
\hln{
An intuitive solution to these issues is to increase the complexity of the primitives associated with each skeleton element by making them more general, rather than restricting them to simple spheres.
However, using arbitrarily general primitives can be problematic, as it may lead to a poorly defined skeleton. 
% one could equip each skeleton element with a more general primitive by using an arbitrary shape homotopy to the skeleton element as an alternative.
% However, such a primitive would result in a poorly defined skeleton. 
For example, as shown in Fig.~\ref{fig:general-primitive}, by simply dividing the original shape into generic subregions, there is a risk that the skeleton could potentially lie outside of the input shape. 
Therefore, a careful design of such general primitives is needed.}
\hl{To this end, 
we introduce \emph{generalized enveloping primitives} for medial spheres, cones, and slabs.
These new primitives have been crafted to cover complex local regions effectively with fewer units, thus facilitating the creation of a simpler skeleton.}

\subsection{Generalized Enveloping Primitives}\label{sec:continuousEP}
We begin by outlining the formulation of generalized enveloping primitives. 
The basic idea is to replace the uniform, linearly interpolated primitives in the original medial axis with \emph{non-uniform}, \emph{nonlinearly interpolated} ones. 

The generalized enveloping primitive is defined for all three types of medial mesh elements: $\mathbf{v}$, $\mathbf{e}$, and $\mathbf{f}$.
For convenience, we define ${P}:=\mathrm{cvx}(\{v_k\}_{k=1}^K)\in\{\mathbf{v}, \mathbf{e}, \mathbf{f}\}$, where $\mathrm{cvx}(\cdot)$ denotes the convex hull, $K\in\{1,2,3\}$ corresponds to the three types of medial mesh elements, respectively.
\hln{We leverage the concept of \emph{$\epsilon$-neighborhood}~\cite{munkres2018analysis, lee2012smooth} to define our generalized enveloping primitive. % $\mathcal{E}_{{P}}$. 
The {$\epsilon$-neighborhood} of ${P}$, denoted by ${P}^{\epsilon}$, is defined as ${P}^{\epsilon}:=\big\{y\in\mathbb{R}^3 \big| ||y - x||_2 < \epsilon(x) \ \text{for some} \ x\in {P}\big\}$.
We consider the unit normal vector for each point $y$ on the boundary $\partial{P}^{\epsilon}$ of ${P}^{\epsilon}$.
% TODO add something related to normal bundle, and the generalized enveloping primitive is defined based on the directional vectors.
We associate a set of directional vectors for each point $p\in{P}$, denoted as
\begin{align}
    d_p({P}) := \big\{||n||_2 = 1\big|g(y, n)=p, \forall (y, n)\in N{\partial P}^{\epsilon}\big\},
\end{align}
where $N{\partial P}^{\epsilon}= \{(y, n)|y\in\partial{P}^{\epsilon}, n\in N_p\partial{P}^{\epsilon}\}$ denotes the unit \emph{normal bundle} of ${\partial P}^{\epsilon}$ and $g: N{\partial P}^{\epsilon}\mapsto{P}$ is a surjective mapping between the normal bundle and the points in $P$.
Appendix~\ref{app:welldefine} provides a detailed analysis of the well-definedness of this surjection.}
We denote the set of all directional vectors of ${P}$ as
\begin{align*}
    d({P}) = \big\{d_p({P})| \forall p\in{P}\big\}.
\end{align*}

\begin{figure}[t]
\centerline{\includegraphics[width=0.75\linewidth]{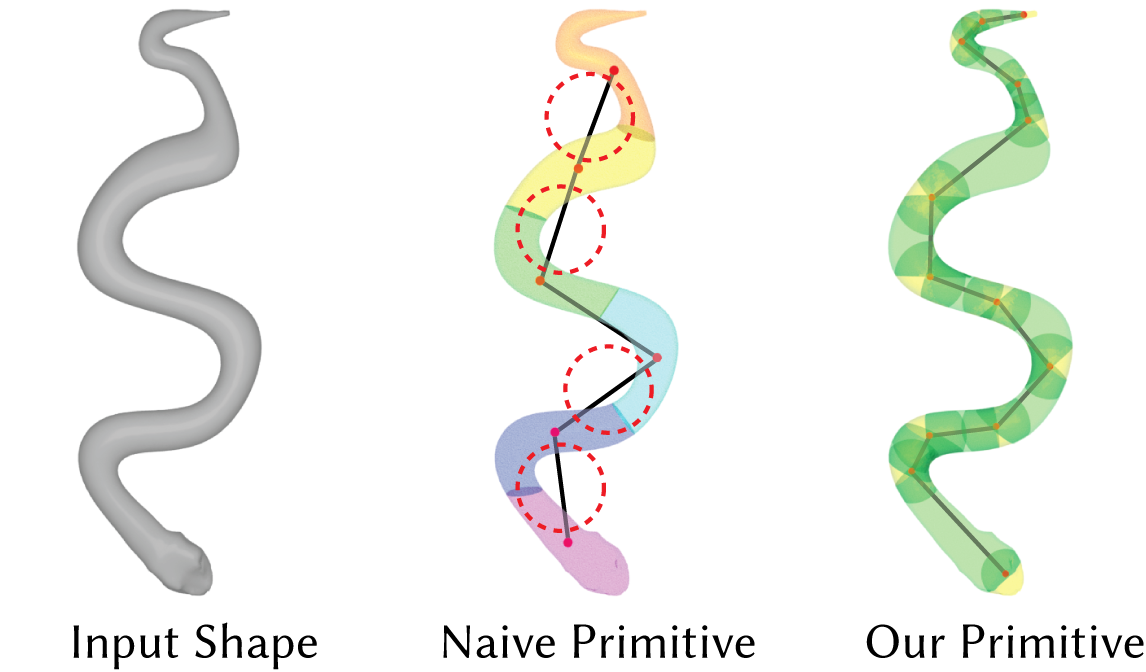}}
\vspace{-1ex}
\caption{\textbf{Naive primitives -- created by simply dividing the original shape into generic subregions -- can result in an invalid skeleton.} 
\hlca{For illustration purpose, we selected a sparse set of points for naive primitives using the Coverage Axis~\cite{dou2022coverage}.
The corresponding skeletons are not always contained within the primitive, 
and thus may not reside within the shape, as indicated by the red circles.
While increasing the number of sampled points could create more subregions and enhance the likelihood of the skeleton being inside the mesh, 
these naive primitives do not guarantee that the skeleton is within the mesh.
In contrast, our method ensures this containment. 
%Additionally, the definitions of primitives associated with edges and triangles in the naive approach remain unclear.
}}
\vspace{-0.25cm}
\label{fig:general-primitive}
\end{figure}

\begin{figure}[t]
\centerline{\includegraphics[width=0.95\linewidth]{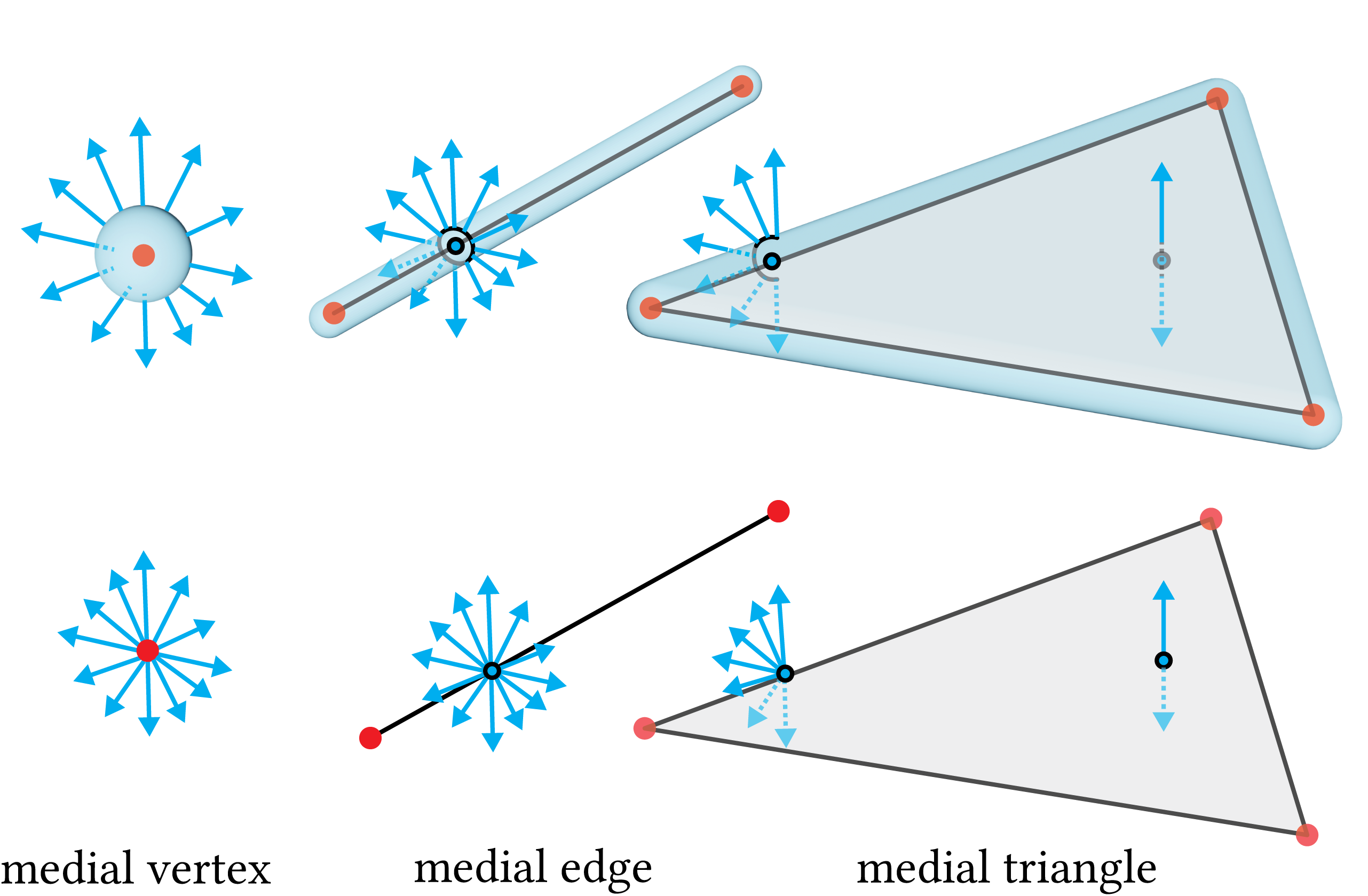}}
\vspace{-1ex}
    \caption{
    \textbf{$\epsilon$-neighborhoods and corresponding normal bundles.} Visualization of the boundary of ${P}^{\epsilon}$ for three medial mesh element types and the directional vectors $d({P})$ (only at selected points for clarity).}
\label{fig:epsilon}
\end{figure}

Fig.~\ref{fig:epsilon} shows $\partial{P}^{\epsilon}$ and $d({P})$ for all three types of medial mesh elements.
In the case of a medial vertex, all the vectors in the set $d_p({P})$ belong to 2-sphere ${S}^2$.
For a medial edge, the set of associated vectors $d_p({P})$ for any interior point $p$ forms a 1-sphere ${S}^1$. 
Meanwhile, at either endpoint of the edge, $d_p({P})$ forms a 2-hemisphere.
For a medial triangle, the set $d_p({P})$ exhibits different characteristics for interior and boundary points. 
For each point located in the interior of the medial triangle, 
$d_p({P})$ consists of the normal vector of the triangle plane and its negative. 
For points in the interior of the three edges of the medial triangle, 
$d_p({P})$ comprises 1-hemisphere.
For each of the three vertices at the corners of the medial triangle corners, 
$d_p({P})$ consists of a sphere wedge, 
whose angle is equal to the external angle of the triangle at the respective corner.

By leveraging the directional vectors $d_p({P})$, we have definition of a \emph{generalized enveloping primitive} as follows:
\begin{definition}[Generalized enveloping primitive]\label{def:primitive}
Given a medial mesh element ${P}=\mathrm{cvx}(\{v_k\}_{k=1}^K)$ and a smooth radius function $r(\cdot): d({P})\mapsto\mathbb{R}_{+}$, the \emph{generalized enveloping primitive} is defined by an implicit function as
\begin{align}
   \mathcal{E}_{{P}}(x) = ||x - p_x||_2 - r(d_{p_x}), d_{p_x} = (x-p_x) / ||x-p_x||_2,
\end{align}
where $p_x = \mathrm{cp}(x;{P})$ is the point in ${P}$ which is closest to $x\in\mathbb{R}^3$.
\end{definition}

\noindent 
% We make $r(\cdot)$ to be smooth to comply with Theorem~\ref{thm:eNeighbor}. 
One can observe that if $r(\cdot)\equiv\epsilon$, then the zero set of $\mathcal{E}_{{P}}(x)$ is $\partial{P}^{\epsilon}$.
Fig.~\ref{fig:generalizedEP2D} and~\ref{fig:MSDRepHand} illustrate
the generalized enveloping primitives representing local regions of a 2D and a 3D shape, respectively. 
Compared to the ones used in the standard medial mesh, our generalized enveloping primitives can cover a diverse range of local regions using fewer primitives.
% \mh{$P^{e}, e_P$, difference}

It is crucial to highlight that our generalized enveloping primitive formulation encompasses the standard enveloping primitives used in the medial axis as a special case, where the radius function $r(\cdot)$ is uniform and linearly interpolated (Fig.~\ref{fig:primitive-generalization}). 
More specifically, for a medial sphere,
\begin{align*}
r(d) \equiv \bar{r},\ \forall d\in d({P}).
\end{align*}
For a medial cone ($K=2$) or a medial slab ($K=3$) with radii $\{\bar{r}_k\}_{k=1}^K$ at its corners, for each point $ p\in{P}$, 
\begin{align*}
r(d) \equiv \sum_{k=1}^{K}w_k(p)\bar{r}_k,\ \forall d\in d_p({P}),
\end{align*}
where $w_k(p)$ is the barycentric coordinates of $p$ on the medial element (edge or triangle). 

\hln{
In contrast to arbitrarily general primitives (as shown in Fig.~\ref{fig:general-primitive}), 
by utilizing the normal bundle of the $\epsilon$-neighborhood to define the generalized enveloping primitive, 
our method ensures that the skeleton elements are contained within the primitives.
Therefore, the skeleton does not extend outside the input shape or intersect inappropriately with it.}
% Another potential idea is to define a generalized enveloping primitive 

% \bh{Need more detailed discussion about the primitives and the advantages for each of them.}
% general: arbitrary shape, special case, local region
% implicit function: collision detection, shape optimization
% compress, subspace, harmonics, wavelets

\subsection{Medial Skeletal Diagram}
Our medial skeletal diagram (MSD) is constructed as a non-manifold triangular mesh with a much-reduced number of elements compared to the medial axis. 
Each element of this mesh is equipped with a generalized enveloping primitive. 
Formally,

\begin{definition}[Medial skeletal diagram]\label{def:MSD}

\begin{figure}[t]
\centerline{\includegraphics[width=1.0\linewidth]{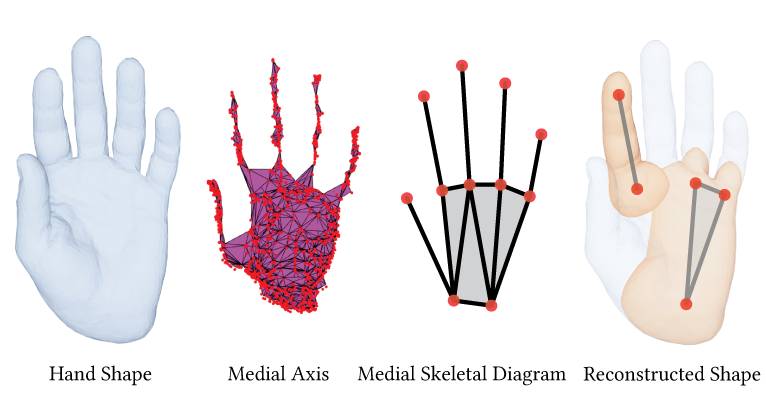}}
\vspace{-3ex}
\caption{\textbf{Medial skeletal diagram on Hand example.} An illustration of employing our representation for a 3D hand shape. Compared to the medial axis, our medial skeletal diagram \hlca{has a sparser skeleton} while maintaining the ability to accurately capture the detailed geometry. Our medial skeletal diagram offers a \emph{complete} representation for 3D shapes.}
\vspace{-1ex}
\label{fig:MSDRepHand}
\end{figure}

A \emph{medial skeletal diagram} \hl{$\mathcal{D}$} is a non-manifold triangulated mesh $\mathcal{M}_S = (\mathsf{V}, \mathsf{E}, \mathsf{F})$ equipping with a generalized enveloping primitive $\mathcal{E}_{{P}}$ for each element ${P}\in \mathsf{V}\cup \mathsf{E}\cup \mathsf{F}$.
\hl{$\mathcal{D}$ represents a 3D shape $\hat{\mathcal{S}}$ by $\hat{\mathcal{S}} = \bigcup_{{P}}\mathcal{E}_{{P}}$.}
\end{definition}

\begin{figure}[t]
\centerline{\includegraphics[width=1.0\linewidth]{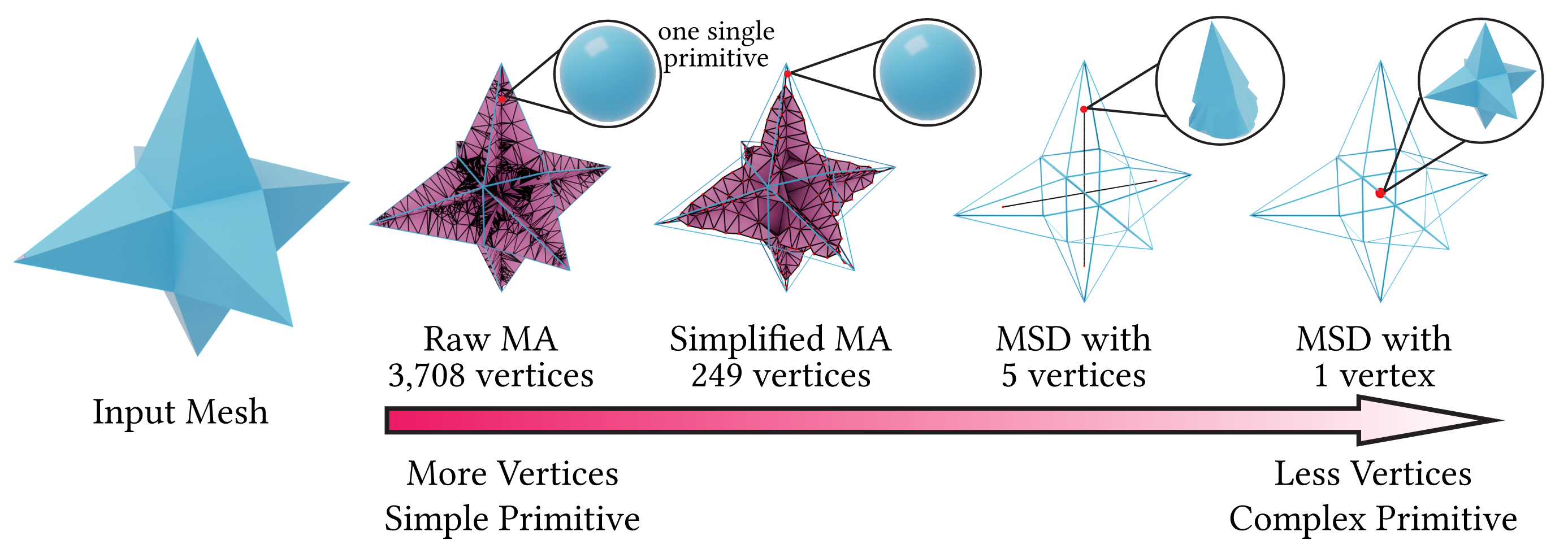}}
\vspace{-1ex}
    \caption{\textbf{Generalization of our MSD.} 
    Our MSD spans a broad range of the skeleton representation spectrum.
    At one extreme lies the medial axis representation, 
    characterized by simple primitives but intricate skeletons.
    \hln{At the opposite end, we have star-shaped regions that encompass just one complex primitive.}
    The complexity of the MSD skeleton can be modulated by adjusting the fitting capability of each primitive.
}
\vspace{-1ex}
\label{fig:primitive-generalization}
\end{figure}

Medial skeletal diagram provides a \hlca{sparse} skeleton of 3D shapes while maintaining the ability to accurately capture complex local regions and sharp features.
As demonstrated in Fig.~\ref{fig:MSDRepHand}, under the same accuracy, the traditional medial axis requires 1,056 medial spheres and 2,449 medial slabs to represent the whole shape, resulting in a complex skeleton.
By contrast, our method can efficiently and accurately cover the whole hand shape with 11 generalized medial spheres, 5 generalized medial cones, and 4 generalized medial slabs.
% Our medial skeletal diagram offers a \emph{complete} representation for 3D shapes and is a \emph{generalized} formulation of the medial axis. 
% Since it inherently encompasses the standard medial axis as a special case and any 3D shape can be converted into a medial axis, our medial skeletal diagram is capable of representing arbitrary 3D shapes (closed, oriented 2D manifolds, as similarly required by the medial axis).
Furthermore, our medial skeletal diagram provides a \emph{complete} representation for 3D shapes, incorporating the standard medial axis as a special case. It is capable of representing arbitrary 3D shapes (closed, oriented 2D manifolds, akin to the medial axis requirements).
Additionally, it pledges a more versatile representation compared to the standard approach. 
Under a worst-case scenario concerning the number of primitives needed, our method is not worse than the standard medial axis representation.
\hl{This is because the upper limit of primitives necessary to fit the mesh in our model equals that in the medial axis, a correlation depicted in Fig.~\ref{fig:primitive-generalization}.
In such instances, while the primitive remains simple, there is a trade-off in the form of an amplified skeletal complexity.
Conversely, our diagram can encapsulate the entity with a single vertex for shapes representable through a solitary spherical function, such as the star domain demonstrated in Fig.~\ref{fig:primitive-generalization}.
}
\hln{
With the aforementioned advantages, 
our generalized enveloping primitives offer several benefits:
\begin{itemize}
\item Provide a substantially simpler skeleton compared to conventional medial primitives, as demonstrated in Sec.~\ref{sec:exp-cmp}.
This \hlca{sparse} skeletal structure is advantageous in various applications, 
as highlighted in Sec.~\ref{sec:app}.
\item Incorporate a well-researched implicit function formulation, 
making them suitable for a range of tasks such as shape optimization~\cite{chen2007shape},
as further detailed in Sec.~\ref{sec:exp-shapeopt}.
\item Compress easily, thereby reducing the storage. 
In Sec.~\ref{sec:mesh-compression},
we showcase a simple strategy to compress our primitives.
\end{itemize}
}

%% file: sections/approach_computation.tex
\section{Medial Skeletal Diagram Computation}
\label{sec:msdComputation}

% By incorporating the ability to use fewer generalized primitives to cover diverse local regions, our representation is more efficient and flexible.
% \begin{figure*}[t]
% \centerline{\includegraphics[width=0.95\linewidth]{./figures/pipeline.png}}
%     \caption{\textbf{Pipeline of medial skeletal diagram computation.} Our pipeline, built on a local-global optimization paradigm, takes a target shape and its medial axis as input. The discrete structure of medial skeletal diagram is constructed based on a small set of points from the medial axis. At the local step, we fit a discretized enveloping primitive to each element of the current medial skeletal diagram. At the global stage, we optimize the medial skeletal diagram. 
%     After the local-global optimization converges, the final fitted primitives undergo 
%     feature-preserving refinement to match the target mesh.}% in both geometry and tessellation.}
% \label{fig:pipeline}
% \end{figure*}
\hl{
To produce our medial skeletal diagram \(\mathcal{D}\) for a given closed manifold triangular mesh \(\mathcal{S} = (V_{\mathcal{S}}, E_{\mathcal{S}}, F_{\mathcal{S}})\),
we begin with a target shape and its medial axis.
Using an arbitrary set of points from the raw medial axis,
we deterministically create the medial skeleton as detailed in Sec.~\ref{sec:MSDConstruct}.
Subsequently, we fit our generalized enveloping primitives to each vertex, edge, and triangle of the skeleton to match the target shape,
as discussed in Sec.~\ref{sec:primitiveFitting}.
The collective union of these primitives defines the final shape.
This process forms an automated workflow to compute an MSD from any selection of vertices on the input medial axis.
}

Additionally, we introduce an efficient optimization pipeline to compute an optimal MSD.
\hln{It aims to minimize the count of primitives while capturing the input shape in detail, including both geometry and tessellation.
Utilizing the aforementioned procedure, 
we employ a gradient-free optimizer to determine a vertex set ensuring that 
the skeleton is simple and the resultant 3D shape aligns with the input shape, as explained in Sec.~\ref{sec:globalOpt}.}
After optimization, the fitted primitives are refined through a feature-preserving refinement process considering both geometry and tessellation to achieve a precise match with the target mesh, as described in Sec.~\ref{sec:refinement}.

\subsection{Medial Skeleton Construction}\label{sec:MSDConstruct}
\newcommand{\MD}{{$\mathcal{M}_S$}} 
The idea behind constructing the medial skeleton $\mathcal{M}_S$ of a medial skeletal diagram $\mathcal{D}$ is to perform a ``remeshing'' of the medial axis by selecting a subset of vertices on the medial axis as constraints.
%Given that the medial axis is homotopy equivalent to the shape and the remeshing operation generally preserves topology, the resulting \MD{} can embody most of the topological information of the shape.
\hl{We presume the input medial axis is high-quality and densely sampled.}
Meanwhile, by limiting the number of constrained vertices, we can maintain a \hlca{sparse} number of discrete elements in \MD{}.

There are primarily two ways for conducting vertex-constrained remeshing on the non-manifold raw medial mesh: 
1) using mesh simplification operations, including vertex removal, edge collapse, and triangle collapse, while maintaining topology preservation through link conditions~\cite{dey1999topology}; 
or 2) constructing a Voronoi diagram restricted to the raw medial mesh and using its dual, restricted Delaunay triangulation (RDT), for remeshing. 
\hln{Although mesh simplification operations can provide a strict guarantee of topology preservation, 
the remeshing result is affected by the order in which these operations are conducted~\cite{liang20203d}. Consequently, the resulting simplified skeleton from mesh simplification cannot be reduced to a significantly small required number. 
% This is particularly true when the user-required count is significantly smaller than the \mh{original, as noted in.}
}

Conversely, the RDT method is order-independent but requires additional conditions to satisfy homotopy~\cite{yan2009isotropic, edelsbrunner1994triangulating, amenta1998surface}.
In our framework, we opt for the RDT approach, because the number of constrained vertices required to construct \MD{} is remarkably smaller than that in the raw medial axis. 
This means that a significant number of order-dependent operations would be required if the mesh simplification methods were used.
% We refer the readers to Sec.~\ref{sec:limitations} for more discussions regarding topology preservation.
% \mh{Fig. ?} illustrates the pipeline of medial skeletal diagram construction.

\begin{figure}[t]
\centerline{\includegraphics[width=1.0\linewidth]{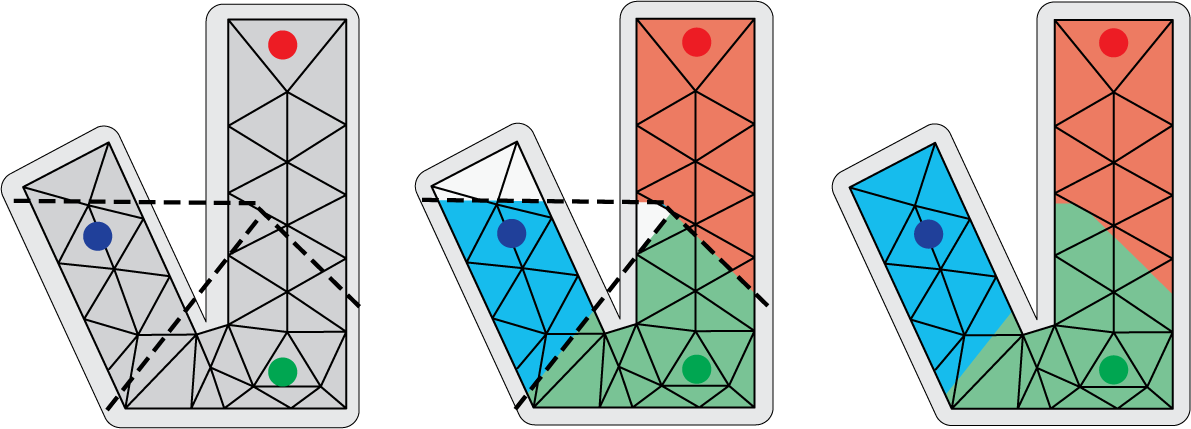}}
\vspace{-1ex}
    \caption{\textbf{Illustration of RVC and its modification.} The initial RVC (bounded by dash lines) may contain more than one connected component (CC) of the medial mesh $\mathcal{M}$. We assign each separate unassigned CC (white regions in the middle figure) to the nearest neighboring CC. }
\label{fig:RVDDisconnectedCC}
\vspace{-1ex}
\end{figure}

More specifically, given the target mesh $\mathcal{S}$ and its raw medial mesh $\mathcal{M}$, 
\hln{the algorithm of \MD{} construction takes a set of points on the medial mesh as inputs,
namely, $\mathsf{V} \subset \mathcal{M}$.}
These vertices can either be manually selected or obtained from the global optimization described in Sec.~\ref{sec:globalOpt}.
The restricted Voronoi diagram $\mathrm{RVD}(\mathsf{V}, \mathcal{M})$ is then constructed following Eq.~\eqref{eq:RVD}.
We split any element in $\mathcal{M}$ that intersects with a Voronoi face into two parts at the intersection and insert necessary vertices or edges to maintain it as a triangle mesh.
Each vertex $\mathbf{v}$ corresponds to a restricted Voronoi cell $\mathrm{RVC}(\mathbf{v}, \mathsf{V})$.
Note that the $\mathrm{RVC}(\mathbf{v}, \mathsf{V})$ might have more than one connected component (CC) of $\mathcal{M}$.
% \mh{references, there is a work dealing with disconnected components}
Similar to~\cite{xin2022surfacevoronoi}, we modify each RVC to contain only one CC for the use of the following Delaunay triangulation.
Specifically, we first perform a breadth-first search (BFS) rooted at $\mathbf{v}$ on $\mathcal{M}$ within each $\mathrm{RVC}(\mathbf{v}, \mathsf{V})$. 
We then assign the CC expanded by BFS to $\mathbf{v}$. 
For those CCs not assigned to any vertices, i.e., they do not contain any vertex from the input set,
we assign them to their nearest neighboring CC based on the distance between the RVC site and the center of the CC. 
Fig.~\ref{fig:RVDDisconnectedCC} shows an example of the RVC modification.
Since the raw medial mesh has only one CC for a closed shape, this assignment ensures that every element of $\mathcal{M}$ has an associated $\mathbf{v}$ and that each modified RVC is a single CC. 
% We denote the refined RVCs as $\hat{\mathrm{RVC}}(\mathbf{v}, \mathbf{V})$.

Then, the medial skeleton is obtained by RDT, which is dual to the RVD.
Each $\mathbf{v}{\in}\mathsf{V}$ corresponds to a volumetric cell, i.e., the modified RVC. 
Each $\mathbf{e}{\in}\mathsf{E}$ connects two $\mathbf{v}$ if their corresponding modified RVCs share a common face of $\mathcal{M}$. 
Each $\mathbf{f}{\in}\mathsf{F}$ joins the vertices whose corresponding RVCs share a common edge of $\mathcal{M}$.
We triangulate the polygonal surfaces of $\mathbf{f}$ to ensure $\mathsf{F}$ only consists of triangle faces.
Note that RDT can yield volumetric objects when several RVC cells share a common vertex of $\mathcal{M}$.
To address this, we adopt the thinning process described in~\cite{wang2022computing} to convert all tetrahedra in the RDT into two-dimensional sheets, thereby ensuring \MD{} does not contain any solid components.

\begin{figure}[t]
\centerline{\includegraphics[width=0.78\linewidth]{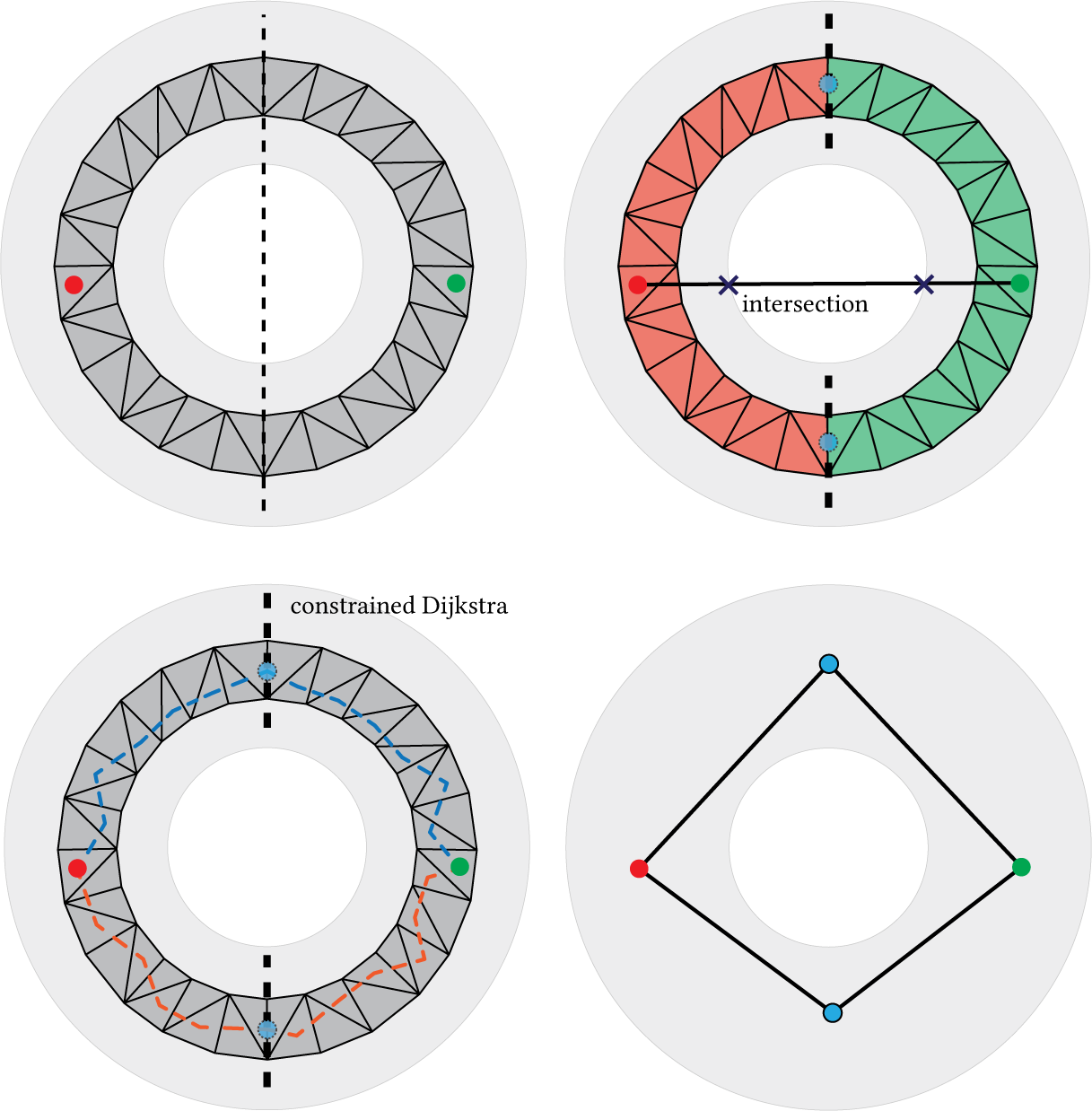}}
\vspace{-1.5ex}
    \caption{\textbf{Revising RDT for a torus.} 
    The initial RVCs for two vertices of the torus exhibit two interfaces (bold dash lines), and the edge connecting these vertices intersects with the mesh (top right).
    We add additional vertices at each interface and compute the constrained shortest path (bottom left).
    The revised RDT (bottom right) remains entirely within the mesh and effectively captures the topology of the shape.
    }
\vspace{-1ex}
\label{fig:MSDConstructTorus}
\end{figure}

While the standard RDT approach successfully constructs the \MD{} for most shapes, 
there are two additional considerations needed, 
particularly for shapes of non-zero genus (example shown in Fig.~\ref{fig:MSDConstructTorus}): 1) neighboring RVCs can share more than one isolated interface (bold dash lines in Fig.~\ref{fig:MSDConstructTorus}); and
2) elements of the \MD{} might intersect with the target mesh $\mathcal{S}$, resulting in parts of the element being located outside the mesh.

To handle both situations, we first revise the edges of \MD{} and then update the associated RDT triangles.
To address 1), we perform a BFS to identify all shared elements of \MD{}
between every pair of neighboring RVCs, allowing us to locate their isolated interfaces.
For each interface, we add a piecewise linear curve to \MD{} in $\mathcal{S}$ that originates from and terminates at the two vertices corresponding to the RVCs and passes through this interface.
\hl{For situation 2), if an edge intersects with $\mathcal{S}$, we iteratively subdivide the edge until the tessellated edge is in $\mathcal{S}$. For each newly added point, we project it to the shortest path between these two endpoints on $\mathcal{M}$.}
Upon revising all the edges, we proceed to update the triangles incorporating any edges that have been revised in the previous two steps.
Due to the addition of the edges, the triangles change to polygons.
Therefore, we then perform triangulation of them.
This approach effectively avoids triangle intersections with the target mesh for all the samples we tested in the paper.
As an alternative, subdividing intersecting triangles and deforming them to fit inside the mesh could be explored in future work.
\hln{Note that performing this revision results in a final medial element set for the medial skeletal diagram that may contain more vertices than the initial input set $\mathsf{V}$.}
% This is due to the additional vertices added to meet these conditions. 
We denote the final constructed medial skeleton as $\mathcal{M}_S = (\hat{\mathsf{V}}, \hat{\mathsf{E}}, \hat{\mathsf{F}})$, where $\mathsf{V}\subseteq \hat{\mathsf{V}}$.

\subsection{Local Primitive Fitting}\label{sec:primitiveFitting}
Primitive fitting involves discretizing the generalized enveloping primitive $\mathcal{E}_{{P}}$ in the space of the directions $d({P})$ for a given medial element $P$.
This is achieved by the discretization of the $\epsilon$-neighborhood boundary $\partial P^{\epsilon}$, which provides a spatial tessellation for $\mathcal{E}_{{P}}$.
A piecewise linear triangulated surface, denoted as $\mathcal{T}_{\partial P^{\epsilon}} = (V_{\epsilon}, E_{{\epsilon}}, F_{{\epsilon}})$, is used to discretize $\partial P^{\epsilon}$.
For meshes scaled to the unit cube, we set $\epsilon = 0.1$. 
This leads to the spatial discretization of directional vectors $d({P})$ aligning with the vertex normal vectors of $\mathcal{T}_{\partial P^{\epsilon}}$. 
We denote these unit discretized directional vectors as $\mathbf{d}\in\mathbb{R}^{|V_{\epsilon}|\times3}$.

The discretized generalized enveloping primitive, denoted as $\mathcal{T}_{\mathcal{E}_{P}}=(V_{\mathcal{T}}, E_{\mathcal{T}}, F_{\mathcal{T}})$, is a triangle mesh derived by equipping $\mathcal{T}_{\partial P^{\epsilon}}$ with a discretized radius function $\mathbf{r}\in\mathbb{R}^{|V_{\epsilon}|}_{+}$. 
Each vertex of $\mathcal{T}_{\mathcal{E}_{P}}$ can only move along the direction $\mathbf{d}_i$, with distance indicated by $r_i$. 
Here, $E_{\mathcal{T}} = E_{{\epsilon}}$, $F_{\mathcal{T}} = F_{{\epsilon}}$, and $V_{\mathcal{T}} = \{v_i+{r}_i\mathbf{d}_i, \forall v_i\in V_{\epsilon}\}$.
Corresponding to the continuous case as in Def.~\ref{def:primitive}, 
we use $\mathcal{E}(x;\mathcal{T}_{\mathcal{E}_{P}})$ to denote the implicit function of $\mathcal{T}_{\mathcal{E}_{P}}$ for any point $x\in\mathbb{R}^3$.

\begin{figure}[t]
\centerline{\includegraphics[width=0.9\linewidth]{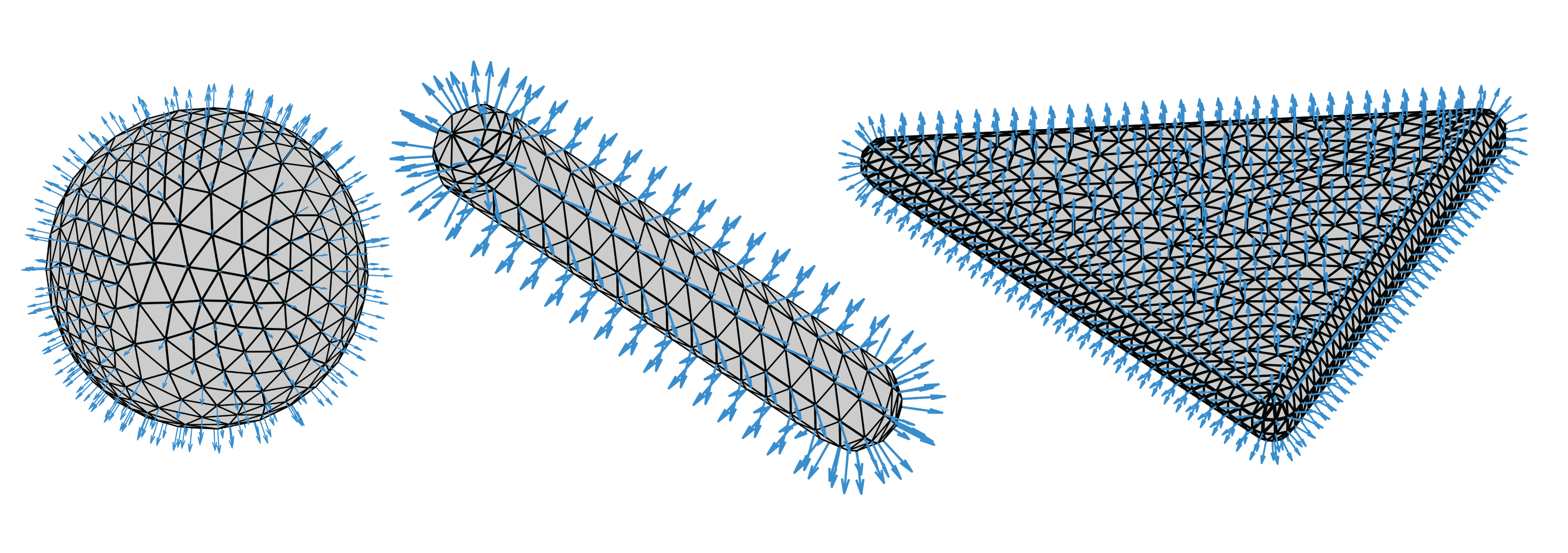}}
\vspace{-3ex}
    \caption{\textbf{Discretized generalized enveloping primitives.} We use triangle meshes to represent the discretized primitives. The directional vectors $d(P)$ become vertex normals (the blue vectors) in the discrete cases.}
\label{fig:discreteEPs}
\vspace{-0.1cm}
\end{figure}

\hl{To be more specific, for the three types of medial elements, when we have a medial vertex $\mathbf{v}$, $\mathcal{T}_{\partial P^{\epsilon}}$ is represented by an isotropically meshed unit sphere centered at $\mathbf{v}$ with radius $\epsilon$.
By modifying $\mathbf{r}$, each vertex of $\mathcal{T}_{\mathcal{E}_{P}}$ moves along the direction extending from the sphere center to the vertex itself.
For a medial edge $\mathbf{e}$, we represent $\mathcal{T}_{\partial P^{\epsilon}}$ using two components: a uniformly meshed cylindrical wall and two end cap hemispheres.
For a medial triangle $\mathbf{f}$, 
$\mathcal{T}_{\partial P^{\epsilon}}$ consists of three parts: two triangle planes, three half-cylindrical walls, and three spherical wedges.
}
Fig.~\ref{fig:discreteEPs} illustrates the discretized mesh $\mathcal{T}_{\mathcal{E}_{P}}$ and the discretized directional vectors $\mathbf{d}$ for three cases.
\hl{In general, 
the resolution of the primitive meshes is set 
such that it matches the input mesh resolution.}
For the generalized enveloping primitives of the medial vertices, 
we use a mesh with 1,588 vertices.
For medial edges and triangles, 
the mesh resolution is determined by the size of the element.
On average, the primitive mesh for an edge has 4,037 vertices 
and the mesh for a triangle has 1,360 vertices.

The goal of fitting primitive $\mathcal{T}_{\mathcal{E}_{P}}$ to the target mesh $\mathcal{S}$ is to find an optimal $\mathbf{r}^{*}$ such that $\mathbf{r}^{*}$ is smooth across $\mathcal{T}_{\mathcal{E}_{P}}$ and matches the local region of $\mathcal{S}$.
% The smoothness of the radius function is 
% a prerequisite to upholding submersion in Thm.~\ref{thm:eNeighbor} and the requirement according to the definition of generalized enveloping primitive in Def.~\ref{def:primitive}.\bh{if this is needed}
\hl{
% Moreover, 
The smoothness prevents the occurrence of excessively large radii along certain directions, guaranteeing the locality of the primitives.}
We define the following energy optimization problem:
\begin{equation}\label{eq:energies}
	\begin{aligned}		
    \mathbf{r}^{*} = \argmin_\mathbf{r\in\mathbb{R}^{|V|}_{+}}\ & E_{\mathrm{smooth}}(\mathbf{r})+ wE_{\mathrm{expansion}}(\mathbf{r})  \\
    \mathrm{s.t.} \ & \mathbf{r} - \mathbf{r}^{\mathrm{max}} \leq \mathbf{0},
    \end{aligned}
\end{equation}
where the energies are defined as
\begin{equation}
    E_{\mathrm{smooth}}(\mathbf{r}) = \mathbf{r}^\intercal \mathbf{L}\mathbf{r}, \ \ 
    E_{\mathrm{expansion}}(\mathbf{r}) = ||\mathbf{r} - \mathbf{r}^{\mathrm{tgt}}||_2^2.
\end{equation}

\begin{figure}[t]
\centerline{\includegraphics[width=1.0\linewidth]{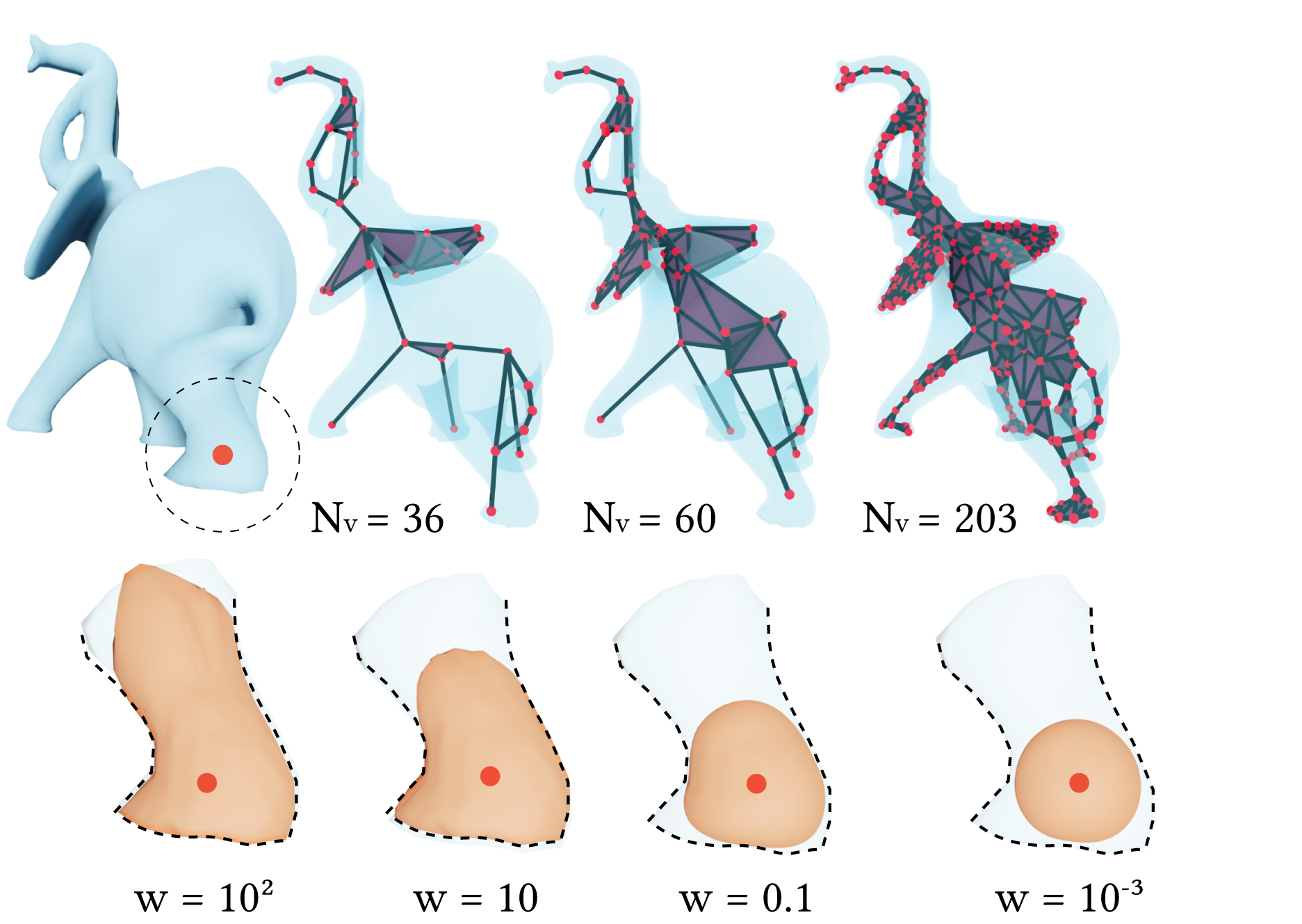}}
\vspace{-2ex}
\caption{\textbf{Effect of $w$ in local primitive fitting.} 
The weight $w$ balances the trade-off between expansion and smoothness.
An overly large $w$ can induce excessively long radii, 
especially in tubular shapes, 
while a notably small $w$ may over-smooth the primitive, 
failing to capture necessary geometry details \hl{(bottom row). 
As a result, a more smoothed primitive requires more skeleton elements,
and vice versa (top row).}
We choose $w=10$ in our implementation.}
\vspace{-1ex}
\label{fig:smoothness}
\end{figure}

In the optimization problem,
$E_{\mathrm{smooth}}$ penalizes non-smoothness of the radii across $\mathcal{T}_{\mathcal{E}_{P}}$.
We choose $\mathbf{L}$ to be the graph Laplacian matrix of the initial $\mathcal{T}_{\mathcal{E}_{P}}$.
$\textbf{r}^{\mathrm{tgt}}$ determines how much the primitives grow.
$E_{\mathrm{expansion}}$ aims to expand $\mathcal{T}_{\mathcal{E}_{P}}$ and increase the radii $\mathbf{r}$ to a larger target value $\mathbf{r}^{\mathrm{tgt}}$,
while the constraint prevents the penetration of $\mathcal{T}_{\mathcal{E}_{P}}$ into the target mesh $\mathcal{S}$.
$\mathbf{r}^{\mathrm{max}}\in\mathbb{R}^{|V|}_{+}$ is the maximal expansion radii of all directions.
This is precomputed by measuring the distance from $v_i$ to the first intersection point between the target mesh $\mathcal{S}$ and the ray that originates at $v_i$ with directional vector $\mathbf{d}_i$.
In Fig.~\ref{fig:smoothness}, 
we show how weight $w$ balances the trade-off between the expansion and the mitigation of exceedingly long radii.
In our implementation, we choose $w = 10$ consistently for all the examples.

% Although all the energy terms in ~\eqref{eq:energies} are quadratic, the objective cannot be minimized by a single-step linear solver since the contact constraint should be dynamically updated.
Eq.~\eqref{eq:energies} is a quadratic programming problem that can be solved using standard techniques, such as the interior point method. 
\hl{To improve performance, we employ an alternating optimization approach inspired by~\cite{bouaziz2014projective} instead of a Newton-type solver.
For each iteration, 
we first identify the index set $\mathcal{Z}$ which includes all directions having a radius larger than $\mathbf{r}_i^{\mathrm{max}}$.}
Then, we optimize the following modified objective:
\begin{align}
\label{eq:pdEng}
&\tilde{\mathbf{r}}^{(t)} = 
\argmin_\mathbf{r} \ E_{\mathrm{smooth}}(\mathbf{r})+ 
wE_{\mathrm{expansion}}(\mathbf{r}) + \sum_{i \in \mathcal{Z}}\Delta{^{(t)}}(r_i), \\
&\Delta^{(t)}(r_i) = \hat{w}||r_i - r_i^{\mathrm{max}}||_2^2,
\end{align}
\hl{where $\hat{w}$ is a constant ($\hat{w}=10^4$) for the contact constraints.
We execute up to 15 iterations; however, our solver typically converges in fewer than 5 iterations, with an average fitting time of 0.08 seconds per primitive.}
Compared to Newton-type optimization methods, 
the objective is quadratic and can be minimized with a single linear solver, 
thereby speeding up the optimization process.
\hln{This fast optimization is achievable thanks to the definition of our primitives, where the deformation is restricted to the 
% aforementioned simplification is possible
% because we restricted our deformation to the 
direction $\mathbf{d}$. 
Compared to allowing arbitrary deformations, 
our approach reduces the DOFs for $\mathbf{r}$ by a factor of three, 
thereby simplifying penetration constraints.
Moreover, the positive $\mathbf{r}$ keeps the skeleton inside the primitive.
}

In our implementation, 
the initial radii of enveloping elements are computed based on the closest distance from the shape to their centers. 
To increase the robustness of primitive expansion, we progressively increase $\textbf{r}^{\mathrm{tgt}}$ by $10\%$ of the initial radius for each ray direction and solve Eq.~\eqref{eq:energies} at each progression.
$\mathbf{r}^{\mathrm{tgt}}$ is increased $25$ times in our experiments.

\subsection{Global Optimization}\label{sec:globalOpt}

\begin{figure}[t]
\centerline{\includegraphics[width=1.0\linewidth]{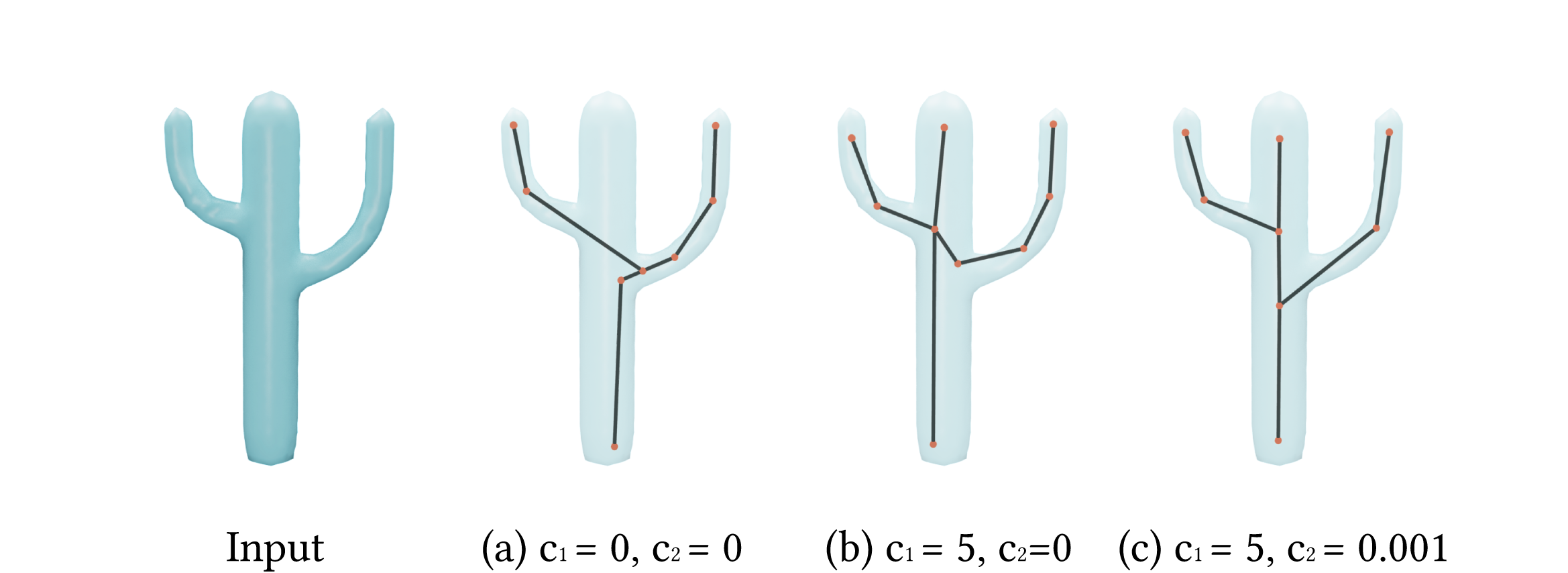}}
% \vspace{-2ex}
\caption{\hl{\textbf{Effect of each energy term.}
(a) Without any regularization, the skeleton vertices are unevenly distributed along the input medial axis.
(b) Incorporating \(E_{\mathrm{centrality}}\) addresses this issue.
(c) Furthermore, introducing \(E_{\mathrm{count}}\) reduces the vertex count (from 9 vertices to 8 vertices), 
leading to a simpler skeleton.
}}
\vspace{-1ex}
\label{fig:abl-opt-term}
\end{figure}

\begin{figure*}[t]
\centerline{\includegraphics[width=1.0\linewidth]{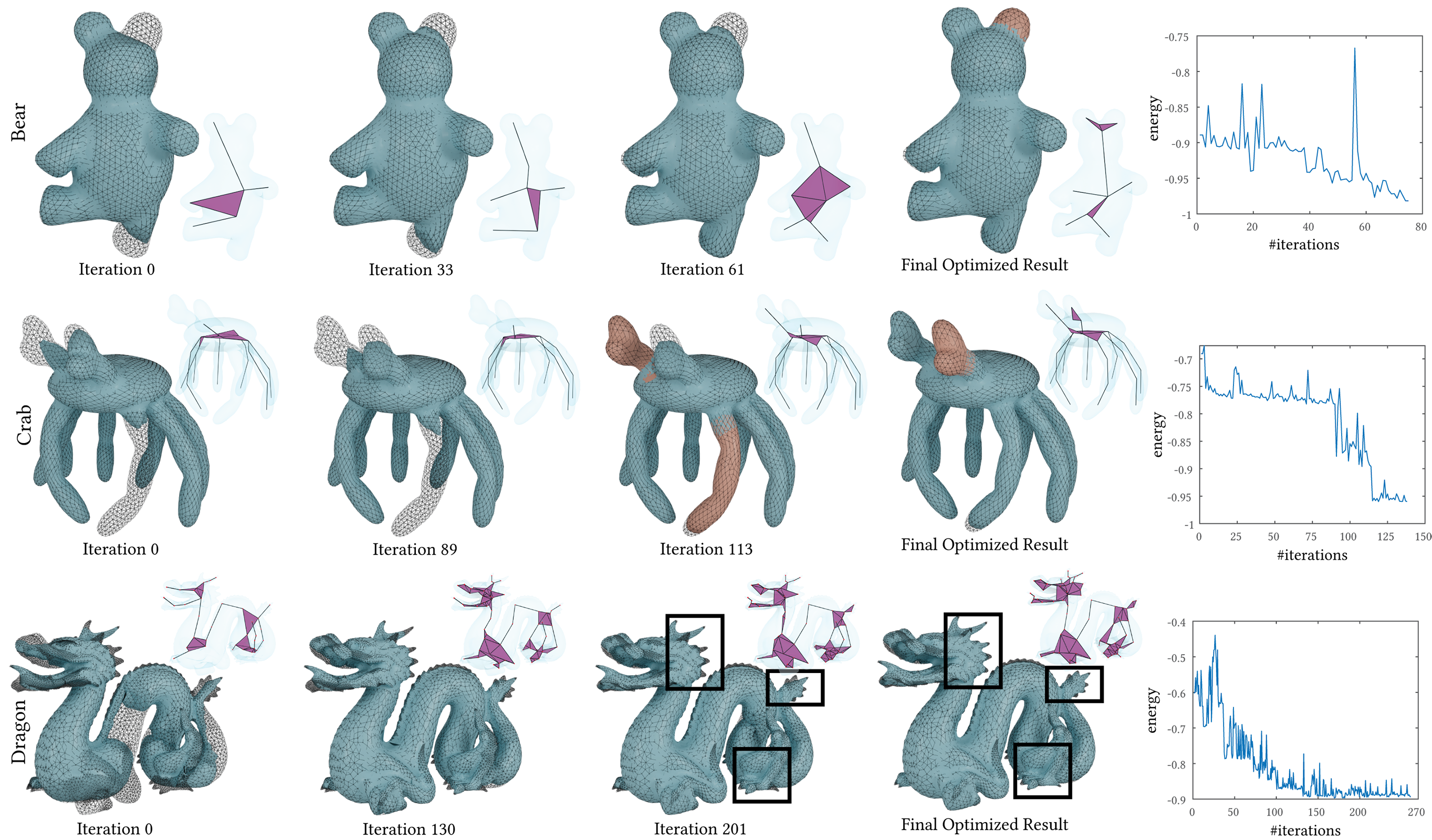}}
% \vspace{-0.08cm}
\caption{\textbf{Intermediate results and energy convergence curves of the global optimization.} 
The global optimization is framed to obtain an optimal medial skeletal diagram that strives to maximize target mesh coverage while minimizing the number of primitives used.
A gradient-free optimization method is carried out to efficiently optimize the energies. 
During the optimization, we incrementally add new points to each isolated uncovered volume and fit primitives (shown in the orange color).
The geometric details such as dragon horns, tails, and claws are
successfully captured, as highlighted by the rectangles.
\hl{Note that energy spikes are encountered due to the introduction of new trial points by the NM optimizer, as it does not ensure a consistent decrease in energy.}}
% \vspace{-0.3cm}
\vspace{-0.02cm}
\label{fig:global-opt}
\end{figure*}
% a sliding bar
% optimization for simplification
% exact
In the previous sections, 
we discussed how the construction of the medial skeletal diagram $\mathcal{D}$ for a given target mesh is determined by the input set of candidate points \hln{on the given medial mesh $\mathcal{M}$.
We denote these candidate points by $\mathsf{V}$ and the resulting medial skeletal diagram by $\mathcal{D}(\mathsf{V}) = (\mathcal{M}_S(\mathsf{V}), \{\mathcal{T}_{\mathcal{E}_{P}}\}_{P\in\mathcal{M}_S(\mathsf{V})})$.}
\hln{
The problem remains to find an optimal $\mathsf{V}^*$.
While existing methods such as \citet{dou2022coverage}, which simplifies a medial axis, might intuitively be adapted to our medial skeletal diagram representation for obtaining optimal candidate points, they face several limitations: 
% Note that simply applying the coverage axis method to our problem would not work~\cite{dou2022coverage}. 
~\citet{dou2022coverage} only considers sphere primitives when formulating the linear programming optimization,
which is inadequate for 
% accurately approximate
cylinders and slabs. 
As the authors note~\cite{dou2022coverage}, there is an intrinsic trade-off between the number of sphere primitives used and the reconstruction error, which is governed by the dilated radius of each sphere.
% This either results in a large number of primitives (using a small dilation radius) 
% or reconstruction errors (using a large dilation radius), .
In our representation, however, every primitive type 
% plays an important role
contributes to the reconstruction of the original shape.
Therefore, there is a need for an optimization algorithm that not only considers all primitives but also strives to find a minimal set of primitives necessary for an accurate reconstruction.}
We present a global optimization scheme that aims to obtain such optimal $\mathcal{D}(\mathsf{V}^*)$
% that strives to maximize target mesh coverage while minimizing the number of primitives used. 
The global optimization is formulated as follows:
\begin{equation}
\mathsf{V}^* =\argmin_{\mathsf{V}\subset \mathcal{M}}\  
    E_{\mathrm{coverage}}(\mathcal{D}(\mathsf{V})) + 
    c_1 E_{\mathrm{centrality}}(\mathsf{V}) + 
    c_2 E_{\mathrm{count}}(\mathsf{V}),
    \label{eq:globalopt}
\end{equation}
where each energy term is defined as,
\begin{align*}
E_{\mathrm{coverage}}(\mathcal{D}(\mathsf{V})) 
%&= \sum_{v\in V_\mathcal{S}} a(v,\{\mathcal{T}_{ \mathcal{E}_{P}}\})\\
&= \sum_{v\in V_\mathcal{S}} 
\begin{cases}
   -A(v) & \text{if } \exists P\in\mathcal{M}_S(\mathsf{V}), \ \text{s.t. }\mathcal{E}(v;\mathcal{T}_{\mathcal{E}_{P}}) < \delta \\
    0 & \text{o.w.}
\end{cases}, \\
E_{\mathrm{centrality}}(\mathsf{V}) 
&= \sum_{\mathbf{v}\in\mathsf{V}}||\mathbf{v} - c(\mathrm{RVC}(\mathbf{v}, \mathsf{V}))||_2^2,\\
E_{\mathrm{count}}(\mathsf{V}) &= ||\mathsf{V}| - |\hat{\mathsf{V}}||.
\end{align*}
Here, $A(v)$ represents the normalized vertex area of $v$ with respect to the total surface area of $\mathcal{S}$,
$P\in\hat{\mathsf{V}}\cup\hat{\mathsf{E}}\cup\hat{\mathsf{F}}$ is the primitive, and
$c(\cdot)$ denotes the centroid of the restricted Voronoi cell.
% ,and $\hat{\mathbf{V}}\in\mathbb{R}^{|\mathsf{V}|}$ represents all the vertices in the constructed medial skeletal diagram using $\mathsf{V}$.
% \bh{this has been mentioned in the previous section.}

The objective in Eq.~\eqref{eq:globalopt} consists of three energy terms. 
$E_{\mathrm{coverage}}(\cdot)$ measures the target mesh coverage of $\mathcal{D}$.
This energy is calculated by enumerating all the vertices of the target mesh $\mathcal{S}$;
if a vertex $v$ is located within a small distance $\delta$ \hlca{($\delta=10^{-4}$)} from any fitted primitive, 
it is then considered to be covered by $\mathcal{D}$, 
and its vertex area contributes negatively to the energy value.
$E_{\mathrm{centrality}}(\cdot)$ enforces the uniformity in the distribution of RVCs within the space.
% We draw inspiration from 
This is inspired by Lloyd's algorithm for Voronoi relaxation~\cite{lloyd1982least}, which aims to achieve uniformly sized Voronoi cells by iteratively moving each input point toward the center of its corresponding Voronoi cell.
% to achieve uniformly sized Voronoi cells.
In our context, we strive for a uniformly distributed medial skeletal diagram by penalizing the distance between the input selected point and the corresponding RVC centroid.
$E_{\mathrm{count}}(\cdot)$ imposes a penalty for the additional vertices introduced 
during the construction of the medial skeletal diagram to minimize the complexity of the resulting structure.
In Fig.~\ref{fig:abl-opt-term}, we show that each term plays an important role in regularization.
Given the non-differentiability of the objective, we use the Nelder-Mead (NM)  algorithm~\cite{richardson1973algorithm}, 
a gradient-free optimization method, to optimize Eq.~\eqref{eq:globalopt}
\hln{and find the optimal set $\mathsf{V}^*$.}

\begin{algorithm}[h]
\caption{\hln{Compute Energy}}\label{alg:overview}
\DontPrintSemicolon
\KwIn{medial axis $\mathcal{M}$, target shape $\mathcal{S}$, and set of points $\bar{\mathsf{V}}$}
\KwOut {energy value $E$}
\;
$\mathsf{V} \gets $ $\bar{\mathsf{V}}$ closest-distance projection onto $\mathcal{M}$\;
Construct the medial skeleton $\mathcal{M}_S(\mathsf{V})$ (Sec.~\ref{sec:MSDConstruct})\;
Fit primitives $\{\mathcal{T}_{\mathcal{E}_{P}}\}_{P\in\mathcal{M}_S(\mathsf{V})}$ to $\mathcal{S}$ (Sec.~\ref{sec:primitiveFitting})\;
$E \gets $ $E_{\textrm{coverage}}(\mathcal{D}(\mathsf{V})) + c_1 E_{\mathrm{centrality}}(\mathsf{V}) + c_2 E_{\mathrm{count}}(\mathsf{V})$\;
\Return $E$
\end{algorithm}

To initialize the optimization, 
\hln{we begin with a predefined number of \hln{initial candidate points} $\mathsf{V}^{(0)}$ selected from the medial axis $\mathcal{M}$.}
The selection is performed by maximizing the shortest path distance on the $\mathcal{M}$ between each pair of points. 
Then, we use NM simplex algorithm to optimize objective~\eqref{eq:globalopt} evaluated by Alg.~\ref{alg:overview}.
\hln{
Since the NM simplex algorithm cannot guarantee any trial points are on $\mathcal{M}$,
we project the input set of points $\bar{\mathsf{V}}$ 
onto $\mathcal{M}$ at the beginning of each iteration as shown in Alg.~\ref{alg:overview}.
}
The iteration proceeds until 1) the objective~\eqref{eq:globalopt} no longer decreases
for a consecutive of five iterations;
2) the primitives fully cover the target mesh; 
or 3) the maximal number of iterations is reached.

To speed up the optimization, we incrementally optimize $\mathsf{V}^*$.
We configure the initial set $\mathsf{V}^{(0)}$ to be small, 
which can be as few as one vertex ($|\mathsf{V}^{(0)}|=1$). 
We then optimize Eq.~\eqref{eq:globalopt} for $t$ iterations, 
yielding the result $\tilde{\mathsf{V}}^{(t)}$. 
For every $t$ iteration,
we check if there are individual regions remaining uncovered.
If that is the case, 
we add a candidate vertex at the volume center for the first $n$ largest regions.
The newly added vertices $\mathsf{V}_{\mathrm{new}}^{(t)}$ 
combined with the existing candidate vertices
form the final candidate set $\mathsf{V}^{(t+1)} =
[\tilde{\mathsf{V}}^{(t)};\mathsf{V}_{\mathrm{new}}^{(t)}]$.
The iteration proceeds until the stopping criteria are met.
We only optimize $\mathsf{V}_{\mathrm{new}}^{(t)}$ while freezing the optimized $\tilde{\mathsf{V}}^{(t)}$ for performance consideration.
In our implementation, we set $c_1 = 1$ and $c_2 = 10^{-3}$ consistently for all the examples.
Fig.~\ref{fig:global-opt} shows the intermediate results and energy convergence curves for three examples.

\subsection{Feature-Preserving Refinement}\label{sec:refinement}
\input{sections/refinement}

%% file: sections/refinement.tex
Our optimization pipeline produces a medial skeletal diagram with maximal coverage of the target mesh. 
As we discretize the generalized enveloping primitives to a triangle mesh,
however,
despite the radii of primitive mesh vertices might align with the target mesh, 
the triangle faces of the primitive mesh could still significantly deviate from the target mesh due to the differences in tessellations of the primitive mesh and the target mesh.
To address this, we propose a \emph{robustness} refinement process that allows our representation to ``exactly'' match the target mesh in terms of geometry 
and ``almost exactly'' mirror its tessellation.
By ``almost exactly,'' we indicate that the tessellation of the target mesh is entirely encompassed within the refined primitive.
Moreover, our refinement method produces post-refinement primitives with  \emph{consistency}, which satisfy the definition of generalized enveloping primitive, retaining the advantages of these primitives as discussed in Sec.~\ref{sec:continuousEP}.
Guaranteeing consistency filters out many trivial refinement methods; e.g., simply dilating the primitive and boolean-intersecting with the target mesh may violate the consistency, as the resulting primitive may exhibit multiple radius values along one direction for a profoundly concave shape.

\begin{figure*}[t]
\centerline{\includegraphics[width=0.89\linewidth]{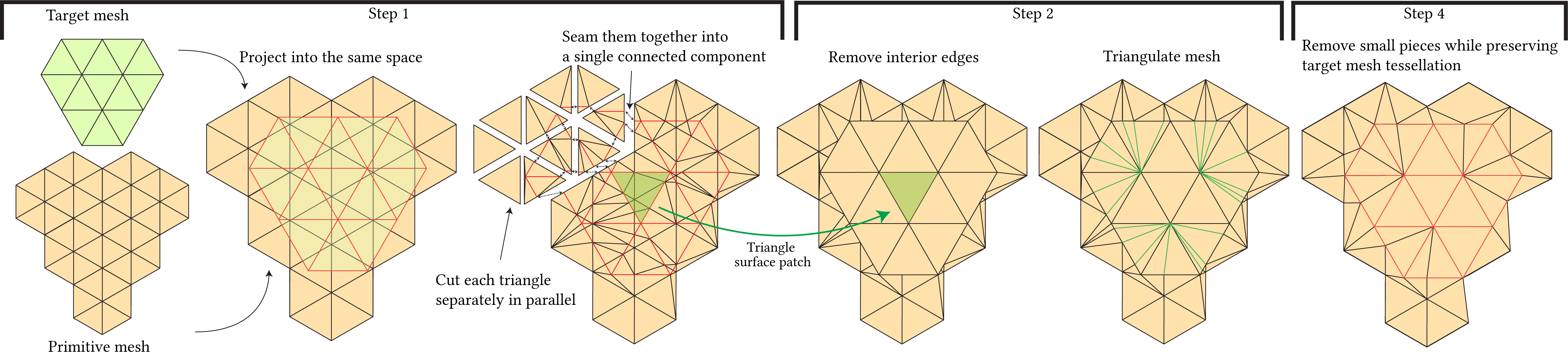}}
\vspace{-0.3cm}
\caption{\textbf{The procedure of our feature-preserving refinement.} 
The process consists of four steps. 
In the first step, we refine the primitive mesh such
that it contains the tessellation of the target mesh (step 1).
Next, we remove redundant edges in each triangle surface patch
and produce a triangulated primitive mesh (step 2).
Then, the vertex positions are updated based on their primitive type (step 3).
% This steps is not show here.
Finally, we clean up the small triangles and edges as much as possible
while preserving the target mesh tessellation (step 4).
Here for 2D illustration, we omit step 3 and refer the readers to Fig.~\ref{fig:coref-3d} for a 3D example.}
\vspace{-0.2cm}
\label{fig:coref-pipeline}
\end{figure*}

\begin{figure*}[t]
\centerline{\includegraphics[width=0.89\linewidth]{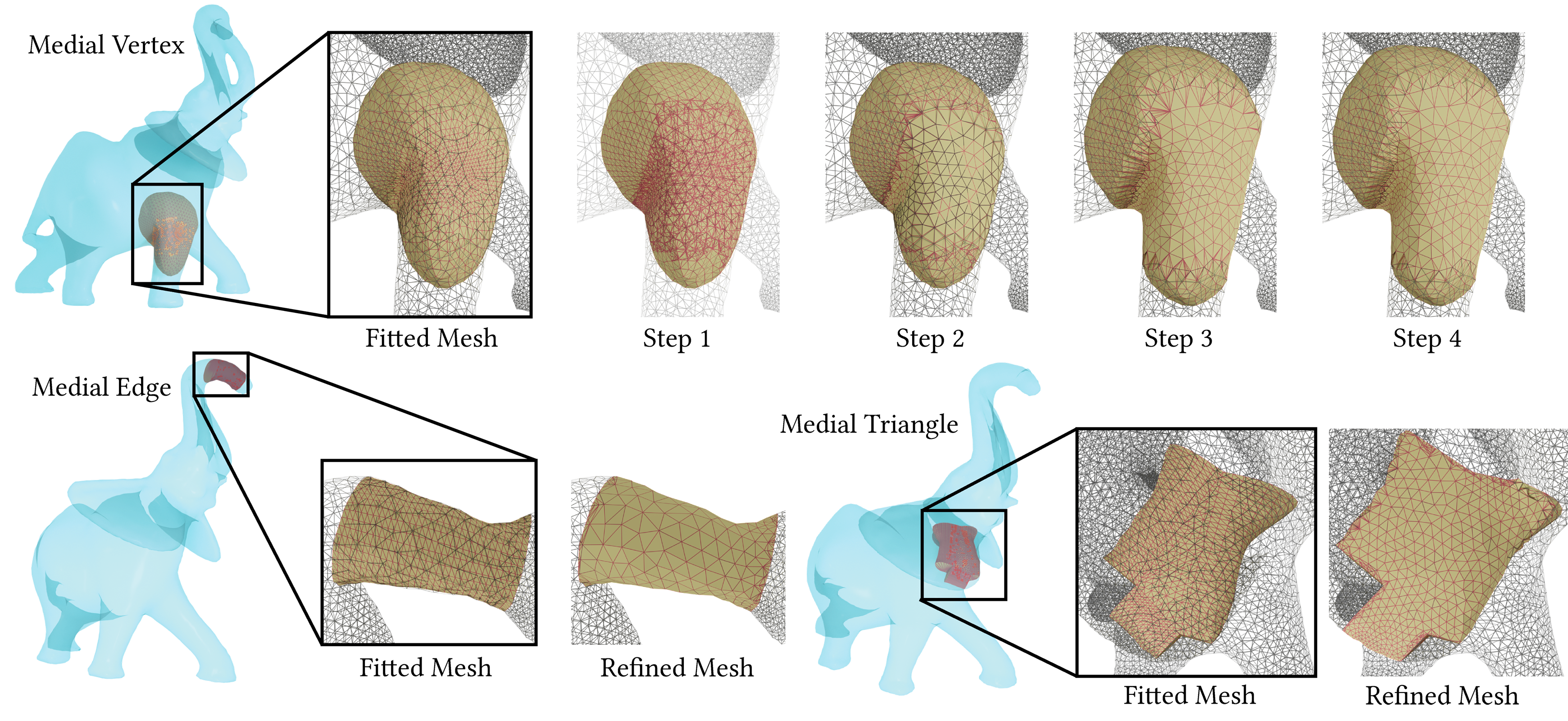}}
\vspace{-0.4cm}
\caption{\textbf{Results of our feature-preserving refinement on primitive meshes of all three types.}
In the first row, we show the step-by-step intermediate results obtained during the process of
refining a medial vertex primitive mesh.
The primitive mesh undergoes refinement aligning with the target mesh in terms of both geometry and tessellation.
The second row offers additional examples featuring the refinement process for the medial edge primitive mesh
and medial triangle primitive mesh.}
\vspace{-0.1cm}
\label{fig:coref-3d}
\end{figure*}

The core strategy is to refine each fitted generalized enveloping primitive mesh so that both its radius function and tessellation in the regions that are close to the target mesh are precisely matched with the target mesh.
To achieve both consistency and robustness, 
our refinement process consists of four steps,
as shown in Fig~\ref{fig:coref-pipeline}.
For each fitted generalized enveloping primitive $\mathcal{T}_{\mathcal{E}_{P}}$,
1) we initially identify the sub-surface of the target mesh that resides within a $\delta$ distance to $\mathcal{T}_{\mathcal{E}_{P}}$ and project its tessellation onto $\mathcal{T}_{\mathcal{E}_{P}}$.
This step ensures that the tessellation of the sub-surface, including vertices, edges, and triangles, is fully encompassed within $\mathcal{T}_{\mathcal{E}_{P}}$.
2) We then remove redundant edges that do not contribute to any geometric and tessellation features of the target mesh.
3) Next, we update the vertices positions of $\mathcal{{T}}_{\mathcal{E}_{P}}$ such that the geometry of the sub-surface exactly aligns with the target mesh.
Because the primitive mesh encompasses the tessellation of the target mesh,
updating the vertex position makes the primitive mesh match the target mesh exactly in the selected sub-surface.
4) We finally carry out a mesh clean-up by removing small edges and triangles without affecting the matched geometry and tessellation.
We implement the above four steps using exact rational arithmetic to ensure the robustness of our process.
\hl{
For more detailed information on the implementation, we direct readers to consult the supplementary materials.}
% For a detailed implementation, we refer readers to our Supplement and its references.
In Fig.~\ref{fig:coref-3d}, we show three example primitives undergoing the refinement process. 
\hln{
One benefit of our post-refinement is that it aligns the tessellation of each primitive with the corresponding local region of the input surface mesh.
% Due to our post-refinement, 
% the tessellation of each primitive aligns with the corresponding region of the input surface mesh.
Even when using primitives with a dense meshing resolution during the local primitive fitting stage, the resolution of the post-refined primitives is determined solely by the resolution of the input mesh. 
This allows for the use of densely meshed primitives in the local fitting stage to capture intricate geometric details of the input mesh while ensuring that the resolution of the final reconstructed mesh closely matches that of the input mesh, avoiding unnecessary complicated continuous parameters in MSD.
% Consequently, the resolution of the mesh of fitting primitives affects the resulting shape but not the number of vertices used.
% The primary factor influencing the resolution of each primitive is the resolution of the input mesh.
}

%% file: sections/experiments.tex
\section{Experiments}
\label{sec:exp}

\begin{table}[b]
	\captionof{table}{\hl{Average Timing for global optimization and refinement}.}
	% \vspace{-0.2cm}
	\small
	\begin{tabular}{c|l|c}
		\toprule
		% \multirow{2}{*}{Global Opt. \#iter.} &  \multirow{2}{*}{51}\\
		% \\\midrule
		% \multirow{2}{*}{\addstackgap{\Shortunderstack{Global Opt. Timing \\ (per iteration)}}} & \multirow{2}{*}{28.68s}\\
		% \\\midrule
		% \multirow{2}{*}{\addstackgap{\Shortunderstack{Refinement Timing \\ (per primitive)}}} & \multirow{2}{*}{10.78s}\\
		\parbox[t]{2mm}{\multirow{5}{*}{\rotatebox[origin=c]{90}{Per Iter.}}} 
		& Projection & 0.006s\\
		& MSD Construction & 4.12s\\
		& Primitive Fitting & 4.64s\\
		& Energy Evaluation & 2.71s\\
		\cline{2-3}
		& Total Global Opt. &  11.57s \\
		\midrule
		\midrule
		& Total Global Opt. \#iter. & 51\\
		\midrule
		& Refinement & 10.78s\\
		\bottomrule
	\end{tabular}
	\label{table:timimg2}
\end{table}

\begin{figure}[t]
	\centerline{\includegraphics[width=0.44\textwidth]{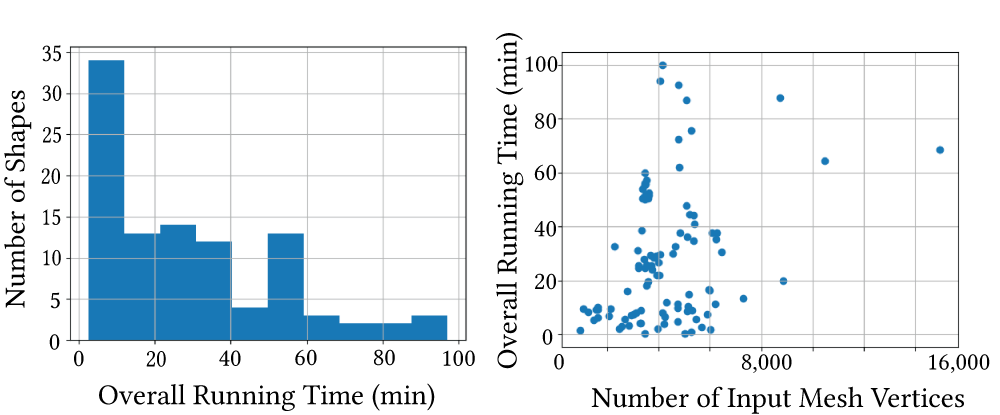}}
 \vspace{-0.3cm}
	\caption{\hl{\textbf{Statistics of the computational cost for constructing medial skeletal diagram from an input shape},
			including a histogram of overall running time across the $100$ shape benchmark and the complexity of input mesh v.s. running time.
			The computational time is principally influenced by the input mesh size (impacting feature-preserving refinement) 
			and the shape complexity (impacting global optimization).}}
	% \caption{\textbf{Statistics of the computational cost for constructing medial skeletal diagram from an input shape}, including the number of iterations required for global optimization and a histogram of overall running time across the $100$ shape benchmark.\bh{add timing: projection, constructing skeleton, fitting, evaluating energy}} \label{fig:timimg}
	\label{table:timimg}
\vspace{-0.4cm}
\end{figure}

Our experiments are run on a desktop with an AMD Ryzen 9 5950X 16-core CPU and 64GB RAM.
We use GMP, MPFR, and CGAL for low-level arithmetic and geometric kernels.
To demonstrate the effectiveness and robustness of our method,
we evaluate our algorithm on $100$ closed manifold triangle meshes.
They are sourced from V-HACD dataset~\citep{VHACD}, 
AIM@Shape Repository~\citep{cohen2003restricted}, 
and the McGill 3D shape benchmark~\citep{fang2008new}.
The size of all shapes is normalized to the $[0, 1]$ range.
\hl{
For all shapes in our benchmark, 
we apply the same set of parameters to the local fitting (mentioned in Sec.~\ref{sec:primitiveFitting}),
For the global optimization, we set the initial number of candidate vertices $|\mathsf{V}^0|=10$.
After the first $50$ iterations, we add either three new vertices or five new vertices every $30$ iterations ($n=3$ or $n=5$), 
depending on which one gives the better result.
The coefficient for each energy term is mentioned in Sec.~\ref{sec:globalOpt}.}
All the experimental results are conducted using a consistent parameter setting.
We choose to use the simplified medial axis obtained from~\cite{dou2022coverage} 
as the initial start of medial skeletal diagram construction 
to strike a balance between computational speed and skeleton accuracy.
Note that our approach is not strictly tied to this choice, 
but can readily accommodate other variants of medial axis representations as inputs.
Statistics of the computational cost required to construct our representation are shown in Fig.~\ref{table:timimg} and \hl{Table~\ref{table:timimg2}}.

% \begin{figure}[t]
%   \begin{minipage}{0.49\linewidth}
%     \small
%     \captionof{table}{Average Timing for Global Optimization (per iteration).}
%     \vspace{-2ex}
%     \begin{displaymath} 
%       \begin{tabular}{c|c}
%         \toprule
%         % \multirow{2}{*}{Global Opt. \#iter.} &  \multirow{2}{*}{51}\\
%         % \\\midrule
%         % \multirow{2}{*}{\addstackgap{\Shortunderstack{Global Opt. Timing \\ (per iteration)}}} & \multirow{2}{*}{28.68s}\\
%         % \\\midrule
%         % \multirow{2}{*}{\addstackgap{\Shortunderstack{Refinement Timing \\ (per primitive)}}} & \multirow{2}{*}{10.78s}\\
%         Projection & 0.006s\\
%         MSD Construction & 4.12s\\
%         Primitive Fitting & 4.64s\\
%         Energy Evaluation & 2.71s\\
%          \midrule
%         Total Global Opt. &  11.57s \\
%         \midrule
%         Total Global Opt. \#iter. & 51\\
%         \midrule
%          \midrule
%         Refinement & 10.78s\\
%         \bottomrule
%     \end{tabular}
%     \end{displaymath}
%   \end{minipage}
%   \begin{minipage}{0.50\linewidth}
%     \centerline{\includegraphics[width=1.0\textwidth]{./figures/TimingStat.png}}
%   \end{minipage} \hfill
%   \caption{\textbf{Statistics of the computational cost for constructing medial skeletal diagram from an input shape}, including the number of iterations required for global optimization and a histogram of overall running time across the $100$ shape benchmark.\bh{add timing: projection, constructing skeleton, fitting, evaluating energy}} \label{fig:timimg}
%   \label{table:timimg}
% \end{figure}

\paragraph{Baselines.}
% For comparison, 
We include six baselines: 
\hl{(1) MAT~\citep{amenta1998surface}, the raw medial axis transformation;}
(2) MATFP~\citep{wang2022computing}, 
a state-of-the-art technique for computing the MAT while preserving features;
(3) LS Skeleton~\citep{baerentzen2021skeletonization}, 
a curve skeletonization approach grounded in local separators;
(4) Coverage Axis (CA)~\citep{dou2022coverage}, 
a shape skeletonization method that simplifies from MAT;
(5) Point2Skeleton (P2S)~\citep{lin2021point2skeleton}, 
a deep learning-based method for constructing medial meshes from point clouds;
(6) CoACD (ACD)~\citep{wei2022approximate}, 
a state-of-the-art method for convex decomposition.

\paragraph{Metrics.}
Following~\citep{wang2022computing, dou2022coverage, lin2021point2skeleton}, 
we use the two-sided average vertex-to-surface distance error, 
denoted as $\bar{\epsilon}$,
to evaluate the surface reconstruction accuracy using medial meshes.
$\bar{\epsilon}^1$ is the one-sided average vertex-to-surface distance from the original surface to the reconstructed surface and $\bar{\epsilon}^2$ is the distance in the reverse direction.
We also report the number of medial elements, including the vertices $\#v$, edges $\#e$, and triangles $\#f$, to compare the \hlca{sparsity} of discrete elements in different representations.
To offer a more complete overview, we detail the number of continuous parameters ($\#$c) associated with different representations.
For baseline methods, we report $\#$c as the total degree of freedom of medial spheres.
\hln{It is computed by the multiplication of the number of medial vertices in the skeleton by four,
as each medial vertex is equipped with a radius (one scalar) and a 3D position (three scalars).
For our method, we report the total number of directional vectors for all primitives plus the number of medial vertices multiplied by three (each vertex has a 3D position).}
To ensure a fair comparison of reconstruction error, in the case of MAT, we increase the target mesh resolution without changing its geometry to obtain a skeleton that has a comparable magnitude of $\#$c to that of our method.
%We choose to increase $\#$c in MAT rather than using MATFP because the resolution of MATFP is determined by the predefined external sharp features of the target mesh. In the scenario where all edges of the target mesh are treated as sharp features, MATFP reduces to MAT~\citep{wang2022computing}.
For MATFP, LS, CA, and P2S,
following their original papers,
we reconstruct the mesh from the skeleton using medial spheres, cones, and slabs. 
In the case of ACD, where there is no reconstruction involved, we only report the number of decomposed domains (represented by $\#v$) and $\#e$ for every pair of adjacent domains.
The number of unique edges $\#e$ of MATFP and the number of faces $\#f$ of LS Skeleton are omitted from the table as these counts are zero across all shapes.
We consider an edge unique if it is not a triangle edge in MATFP.
Since LS only produces curve skeletons,
it does not contain any triangles.

\begin{table}[t]
    \caption{\textbf{Quantitative comparisons on the number of discrete and continuous parameters in the representation and shape reconstruction error on the benchmark of $\mathbf{100}$ shapes.}
    We count the number of vertices ($\#v$), the number of edges ($\#e$),
    the number of triangles ($\#f$), and the sum of them (\#d) for discrete elements.
    In addition, we provide the number of continuous parameters in row \#c.    
    For the reconstruction errors,
    $\bar{\epsilon}^1$ is the one-sided average vertex-to-surface distance from the original surface to the reconstructed surface,
    $\bar{\epsilon}^2$ is the distance in the reverse direction.
    $\bar{\epsilon}$ is the two-sided average vertex-to-surface distance error.
    We compare our method with six existing methods:
    MAT, MATFP (MFP), CoACD (ACD), LS Skeleton (LS), Point2Skeleton (P2S),
    and Coverage Axis (CA).
    Our medial skeletal diagram demonstrates superior \hlca{sparsity} on discrete elements and the highest reconstruction accuracy compared to other representations. }
    \centering
    \footnotesize
    %\scriptsize
    \vspace{-0.2cm}
    \setlength{\tabcolsep}{4pt}
    %@{}p{0.2cm}
\begin{tabular}{l|c|c|c|c|c|c|c|c}
    \toprule
    \multicolumn{2}{c|}{}& MAT & MFP & ACD & LS & P2S & CA & Ours\\
    \midrule
    \midrule
    \multirow{3}{*}{$\#v$} 
    & Min & 17,955 & 692 & 3 & 4 & 100 & 57 & 3\\
    & Max & 37,705 & 4,119 & 192 & 195 & 100 & 738 & 277\\
    \cmidrule{2-9}
    & Avg & 32,539 & 2,110 & 27 & 57 & 100 & 229 & 26\\
    \midrule
    \multirow{3}{*}{$\#e$} 
    & Min & 0 & -- & 2 & 3 & 0 & 0 & 0\\
    & Max & 144 & -- & 639 & 192 & 35 & 131 & 81\\
    \cmidrule{2-9}
    & Avg & 15 & -- & 57 & 57 & 3 & 18 & 9\\
    \midrule
    \multirow{3}{*}{$\#f$} 
    & Min & 31,042 & 1,445 & -- & -- & 136 & 0 & 0\\
    & Max & 82,297 & 10,232 & -- & -- & 346 & 1,437 & 457\\
    \cmidrule{2-9}
    & Avg & 63,279 & 4,835 & -- & -- & 233 & 385 & 24 \\
    \midrule
    $\#$d
    & Avg & 95,833 & 6,945 & 84 & 114 & 337 & 632 & \textbf{58}\\
    \midrule
    \midrule
    \multirow{3}{*}{$\#$c} 
   %  & Min & 692  & -- & 4 & 100 & 57 & 12\\
   %  & Max & 4119 & -- & 195 & 100 & 738 & 10548\\
   %  \cmidrule{2-8}
   % & Avg & 2110.09 & 4422.69 & \textbf{57.30} & 100 & 229.39 & 1166.08\\
    & Min & 71,820 & 2,768  & -- & 16 & 400 & 228 & 4,594\\
    & Max & 150,820 & 16,476 & -- & 780 & 400 & 2,952 & 497,770\\
    \cmidrule{2-9}
    & Avg & 130,154 & 8,440 & 4,423 & \textbf{229} & 400 & 918 & 94,168\\
    \midrule
    \midrule
    \multirow{3}{*}{$\bar{\epsilon}^1$} 
    & Min & 0.06 & 0.044 & -- & 0.59 & 0.76 & 0.17 & 0\\
    & Max & 3.09 & 3.18 & -- & 24.34 & 2.92 & 12.37 & 0.41\\
    \cmidrule{2-9}
    & Avg & 0.31 & 0.31 & -- & 4.42 & 1.31 & 0.54 & 0.029\\
    \midrule
    \multirow{3}{*}{$\bar{\epsilon}^2$} 
    & Min & 0.11 & 0.029 & -- & 0.39 & 1.14 & 0.14 & 0\\
    & Max & 0.48 & 0.68 & -- & 10.03 & 10.35 & 0.83 & 0.098\\
    \cmidrule{2-9}
    & Avg & 0.24 & 0.15 & -- & 2.54 & 2.22 & 0.34 & 0.008\\
    \midrule
    % \multirow{3}{*}{$\bar{\epsilon}$}
    $\bar{\epsilon}$
    & Avg &\makecell{0.40\\($+129\%$)} & 0.31 & -- & \makecell{4.43\\($+1324\%$)} & \makecell{2.22\\ ($+615\%)$} & \makecell{0.54\\ ($+74\%$)} & \makecell{\textbf{0.031}\\ ({$\mathbf{-90\%}$})}\\
    \bottomrule
\end{tabular}
    \label{tab:allResults}
    \vspace{-0.4cm}
\end{table}

\subsection{Comparisons and Discussions}
\label{sec:exp-cmp}
The quantitative comparisons between our method and the baselines,
conducted on the $100$ meshes, are presented in Table~\ref{tab:allResults}.
\hl{
% Our representation requires fewer continuous parameters compared to MATFP and ACD but more than other methods such as LS, P2S, and CA.
Our method uses the fewest skeletal elements and achieves the lowest two-sided mean distance error relative to all other methods examined.
Despite the increased number of continuous parameters in our representation,
as detailed in subsequent sections (Sec.~\ref{sec:exp-shapeopt} and~\ref{sec:mesh-compression}),
this does not introduce complications but rather aids in finding a better solution.
Our mean reconstruction error is 10\% of MATFP and 7.8\% of MAT,
the two methods yielding the least reconstruction error among all baselines.
While MATFP employs 91.0\% fewer continuous parameters with reconstruction errors that are 10 times greater compared to our method, 
increasing the number of MATFP's continuous parameters to roughly match those of our method--by densely sampling the input mesh--only reduces its reconstruction errors by 60\%.
which is still 6 times higher than that of our approach.
On the other hand, despite that MAT uses more continuous parameters,
it fails to seamlessly align with the target mesh given a finite number of medial primitives.
%it incurs higher error due to an inherent limitation: with a finite number of uniform spheres, MAT often fails to seamlessly align with the subtle contours of the target mesh, leading to a less accurate, and visually bumpy reconstruction.
}

\hln{
The simplicity of our representation's discrete parameters enhances the computational speed of shape reconstruction.
To highlight this advantage, we compared the shape reconstruction running time of our representation with those of MATFP.
We choose SDF with a $256$-resolution voxel grid and marching cubes to perform shape reconstruction in both methods,
because unioning primitives on SDF for MATFP is much faster than on meshes.
For MATFP, the SDF value of each voxel is determined by calculating the minimal distance to all medial vertices and subtracting the radius of the corresponding medial sphere.
For ours, we calculate the SDF value by querying the minimal value among all closest distances to all primitive meshes for each voxel.
We conduct this comparison using six shapes selected from the benchmarking dataset of 100 shapes, specifically choosing those with the highest numbers of primitives in our representation.
The results show that the average reconstruction speed of our method is $2.4\times$ faster MATFP (our method: $17.5$s vs. MATFP: $41.1$s).
Moreover, our method achieved $40\%$ fewer reconstruction errors (our method: $6.9 \times 10^{-4}$ vs. MATFP: $1.1 \times 10^{-3}$). 
% This highlights the efficiency and accuracy of our approach in comparison to the state-of-the-art method.
% This computational speed benefit stems from the simpler discrete parameters used in our representation, effectively balancing accuracy and efficiency. 
% To further demonstrate the advantages of our representation in reconstruction, 
% For both methods, the reconstruction is achieved using SDF 
% we conducted a series of controlled experiments 
% where shape reconstructions were performed under SDF with the same resolution for both MATFP and our method. 
% as it will penalize the advantage of our method.
% Given the same set of input shapes, o
}

To demonstrate the effectiveness of our feature-preserving refinement,
we also measure the two-sided mean distance error of the fitting-only results.
Among $100$ shapes, our fitting-only reconstruction reports $0.22$ error, 
$7\times$ larger than the error after the refinement post-processing.
Nonetheless, it is still smaller than the errors of all existing methods.
Furthermore, we also measure the two-sided Hausdorff distance between the input shape and the reconstructed shape.
Our representation reports $1.81$ on average over $100$ shapes, 
lower than the mean distance error of both LS and P2S.
Moreover, ours is also lower than the two-sided Hausdorff distance of both CA and MATFP, which are $3.64$ and $2.22$, respectively.

\begin{figure}[t]
\centerline{\includegraphics[width=0.9\linewidth]{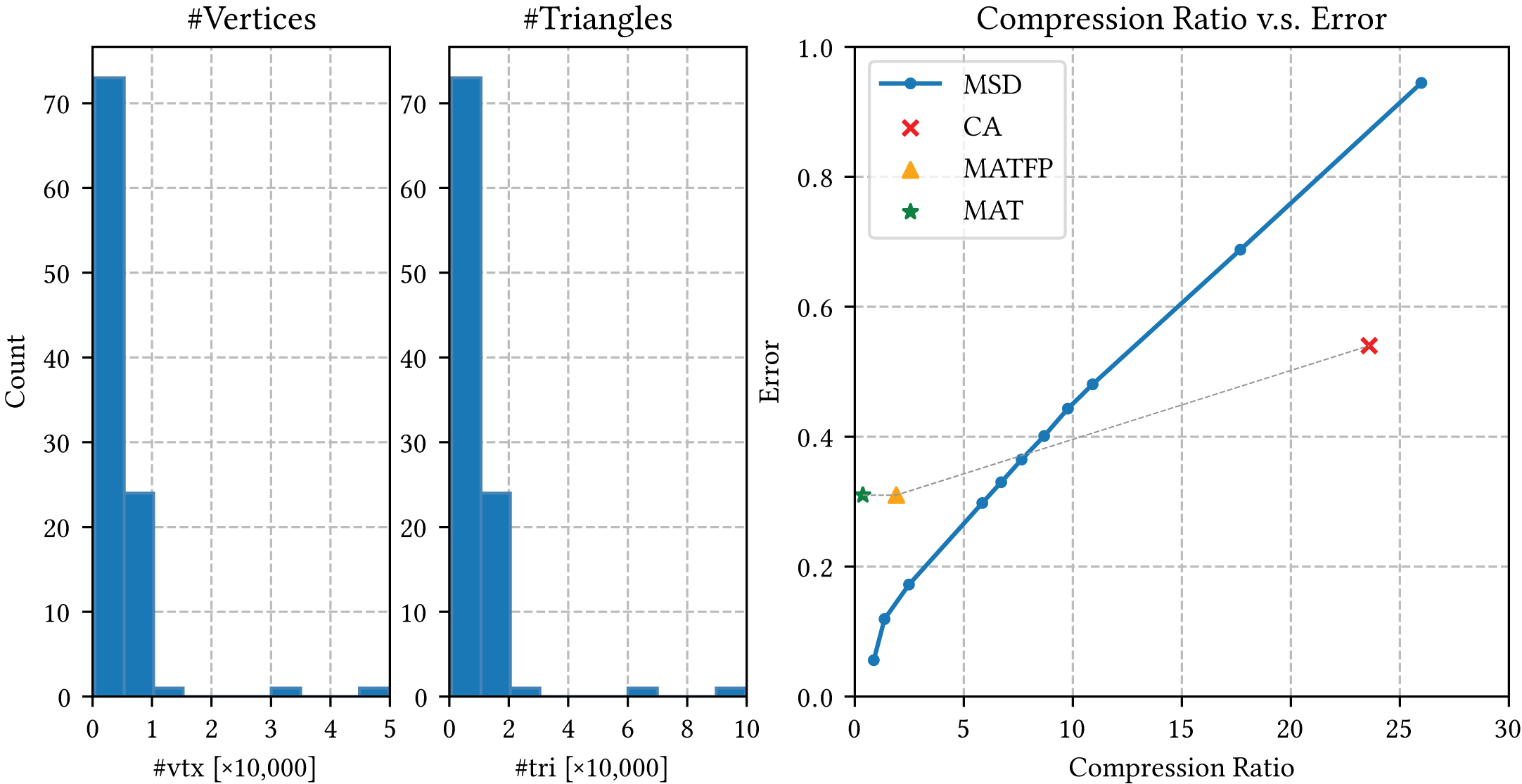}}
\vspace{-0.2cm}
\caption{
\hlca{\textbf{Statistics of the compression ratio v.s. reconstruction accuracy}.
The left panel shows histograms illustrating the distribution of the number of vertices (\#vtx) and the number of triangles (\#tri) across the triangle meshes of the 100 shapes used in the experiments.
The right panel presents a curve illustrating the relationship between the compression ratio and reconstruction accuracy.
Our method achieves superior reconstruction accuracy when the compression ratio is less than 8.}}
\label{fig:compression-accuracy}
\end{figure}

\hlca{To investigate the number of parameters used in each medial axis methods
compared with our MSD, 
we measure ``sparsity'' of the parameters by the compression ratio of the input mesh size ($\#v$$\times3+$$\#t$) to the number of representation parameters (\#c+\#d defined in Table 2) and ``completeness'' by the reconstruction error.
Our representation strikes a trade-off between ``sparsity'' and ``completeness''.
This trade-off can be visualized as a curve (Fig.~\ref{fig:compression-accuracy}):
At one extreme, our MSD achieves a low reconstruction error (0.031) at the sacrifice of compression ratio (0.9). 
To achieve such reconstruction accuracy, existing methods either require more DoFs or incur higher errors with the same DoFs as ours (such as Fig.~\ref{fig:teaser} (b)). 
Furthermore, for MSD, when achieving a reconstruction error similar to MATFP (0.31)
%by simplifying our primitives (as mentioned in Sec.~\ref{sec:mesh-compression}), 
our compression ratio is more efficient (6.6 vs. 1.9 for MATFP).
However, the advantage diminishes when the compression ratio exceeds 8, where CA achieves better accuracy. 
This occurs because, at high compression ratios (where both \#d and \#c are limited), the reconstruction accuracy depends more on the skeleton complexity. 
Since our MSD’s skeleton is relatively sparse, it results in a higher reconstruction error under these conditions.
}

% Nevertheless, the advantage falls off after the compression ratio exceeds 8,
% where CA has a better accuracy under the same compression ratio.
% This is due to the number of discrete elements and the number of continuous parameters 
% are comparable in our representation.
% Thus, shifting the parameters will not improve the performance.}

\begin{figure*}[t]
\centerline{\includegraphics[width=0.95\hsize]{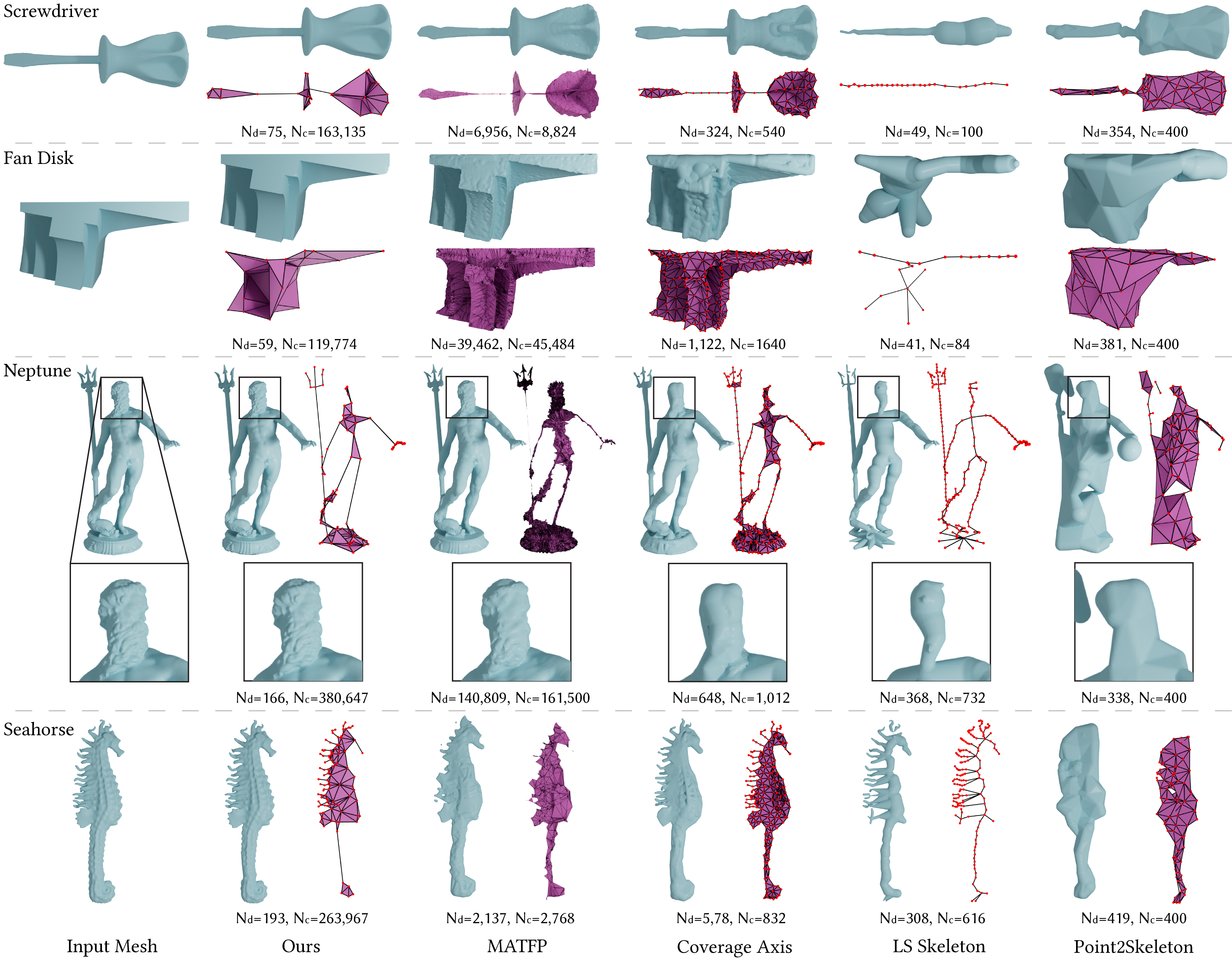}}
\vspace{-0.3cm}
\caption{
\textbf{Qualitative comparison of the skeletal representation and reconstructed mesh (part 1/2),} involving four baseline methods and our proposed method. We omit the visualization of medial vertices for MATFP due to their excessive quantity.
\hln{Our method achieves the highest qualitative reconstruction accuracy by using a relatively small number of discrete elements ($N_d$), albeit at the cost of using a larger set of continuous parameters ($N_c$).} 
% to achieve the highest reconstruction accuracy qualitatively.}
 % \mh{param for MATFP saying the result is disconnected}
 }
	\label{fig:cmp1}
\end{figure*}

\begin{figure*}[t]
	\centerline{\includegraphics[width=0.95\hsize]{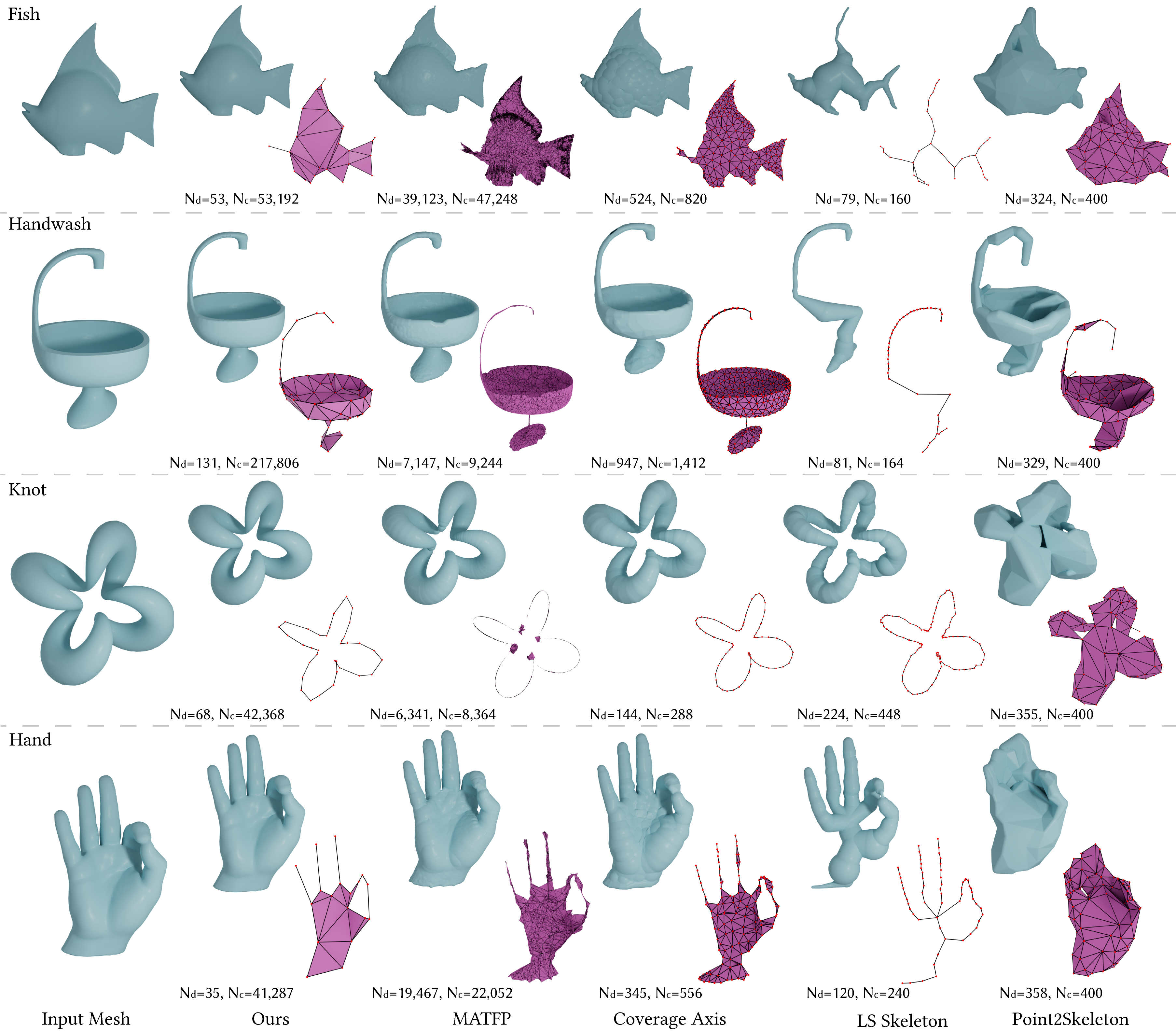}}
 \vspace{-0.2cm}
	\caption{
 \textbf{Qualitative comparison of the skeletal representation and reconstructed mesh (part 2/2),} involving four baseline methods and our proposed method. We omit the visualization of medial vertices for MATFP due to their excessive quantity. \hln{Our method achieves the highest qualitative reconstruction accuracy by using a relatively small number of discrete elements ($N_d$), albeit at the cost of using a larger set of continuous parameters ($N_c$).} }
	\label{fig:cmp2}
\end{figure*}

In addition to quantitative comparisons, we also show a set of qualitative results on various shapes 
from these $100$ meshes in Fig.~\ref{fig:cmp1} and~\ref{fig:cmp2}.
Compared with other baselines, our representation produces visually simpler skeletons and more accurate reconstruction.

\paragraph{Comparison with Medial Axis Transform}
We compare our method with MATFP~\citep{wang2022computing} in the context of the quality of the reconstructed mesh. 
MATFP is highly regarded for its remarkable precision and is tailored to preserve external features such as sharp edges and corners of the input mesh surface. 
This algorithm is essentially a superior variant of the medial axis transforms.
MATFP requires two input thresholds for controlling sharp feature detection. 
We use the default parameters provided within the official source code for all $100$ shapes.
Our method demonstrates a better performance with MATFP regarding reconstruction accuracy.
This is attributed to our refinement process which exactly matches the geometry and tessellation of the reconstructed shape from our method with the input mesh -- a feature that is notably missing in existing medial axis representations including MATFP.
What truly distinguishes our method is its ability to achieve these results using significantly fewer medial elements. 
This simplicity is largely due to our utilization of generalized enveloping primitives. 
Thus, while maintaining a high level of accuracy akin to MATFP, 
our method outperforms it in terms of the skeleton simplicity, 
with a reduction in the number of medial elements by two orders of magnitude.

\paragraph{Comparison with Simplified Skeleton-based Methods}
We consider two simplified skeleton-based methods: Coverage Axis~\citep{dou2022coverage} and Point2Skeleton~\citep{lin2021point2skeleton}. 
Both methodologies are based on a similar process that simplifies the raw medial axis to preserve the geometric and topological features of the input mesh surface.
Coverage Axis employs Mixed Integer Linear Programming (MILP) to minimize the count of medial vertices while ensuring comprehensive coverage of all the input mesh surface points. 
\hlca{We use scaling offset, whose value is set by binary search to ensure the number of vertices is about 400.}
This tactic results in superior reconstruction accuracy when compared to other baseline methods, owing to its explicit optimization for this particular objective.
On the other hand, Point2Skeleton adopts a data-driven strategy, leveraging deep neural networks to predict skeletal points and their interconnections. 
As it is a data-driven method, however, it does not generalize to arbitrary input shapes.
The output of a neural network necessitates a fixed number of vertices ($100$ in the official implementation), 
which can pose challenges when generalizing to unseen shapes that deviate from the training dataset.
Both simplified skeleton methods rely on spheres, cones, and slabs for surface reconstruction. 
The primary objective of these methods is to optimize point coverage, so 
inadvertently introduce errors in the surface reconstruction process. 
This stems from the nature of these primitives - their geometric shapes are fixed and lack the flexibility to adapt to arbitrary regions, which can result in an inaccurate approximation of the original mesh surface.
By contrast, our approach leverages the mesh representation of generalized primitives 
and benefits from the refinement technique. 
%This combination significantly augments the quality of the reconstruction process. 
%The refinement technique, in particular, enables our method to mirror the target mesh with an unprecedented level of accuracy, capturing both the points and the tessellation with exactitude.
Neptune and Seahorse in Fig.~\ref{fig:cmp1} are two evident examples.

\paragraph{Comparison with Curve Skeletons}
LS Skeleton~\citep{baerentzen2021skeletonization} is a state-of-the-art method for curve skeletonization. 
\hl{Notable for its simplicity and smoothness in representing shapes},
LS Skeleton shares a common drawback with other curve skeleton-based methods:
it struggles to accurately represent large flat regions present in the input shape. 
This is particularly problematic for CAD models that frequently feature such flat regions.
Screwdriver and Fan disk example in Fig.~\ref{fig:cmp1} provide evidence.
In this case, a noticeable advantage of our approach is its ability to represent an entire rectangular region with a few triangles while maintaining high reconstruction accuracy. 
By contrast, LS Skeleton falls short in this regard.

\hl{
Rather than explicitly reflecting detailed geometry features as curve skeletons/MAT, our skeleton captures topological structures while primitives effectively represent local geometry details of the shape, as determined through global optimization.}

\subsection{Applications}
\label{sec:app}
\hl{\hln{In this section, 
we attempt our representation through several applications,
demonstrating the simplicity of the discrete elements and the expressiveness of the generalized enveloping primitives.}
% Using our representation achieves superior results compared with existing skeleton-based methods.
}

\subsubsection{Shape Optimization}
\label{sec:exp-shapeopt}
In the previous sections, we have shown the remarkably \hlca{sparse} discrete structure of our medial skeletal diagram,
which is a direct result of our design intention - to shift complexity from discrete to continuous elements. 
In this application, we further demonstrate that the added complexity of these continuous elements does not impede the usability and applicability of our representation.
\hl{Conversely, it gives a better-converged compliance across all examples compared to the medial axis representation,
i.e., each domain being a sphere, a cone, or a slab.}

\begin{figure}[t]
	\centerline{\includegraphics[width=0.95\linewidth]{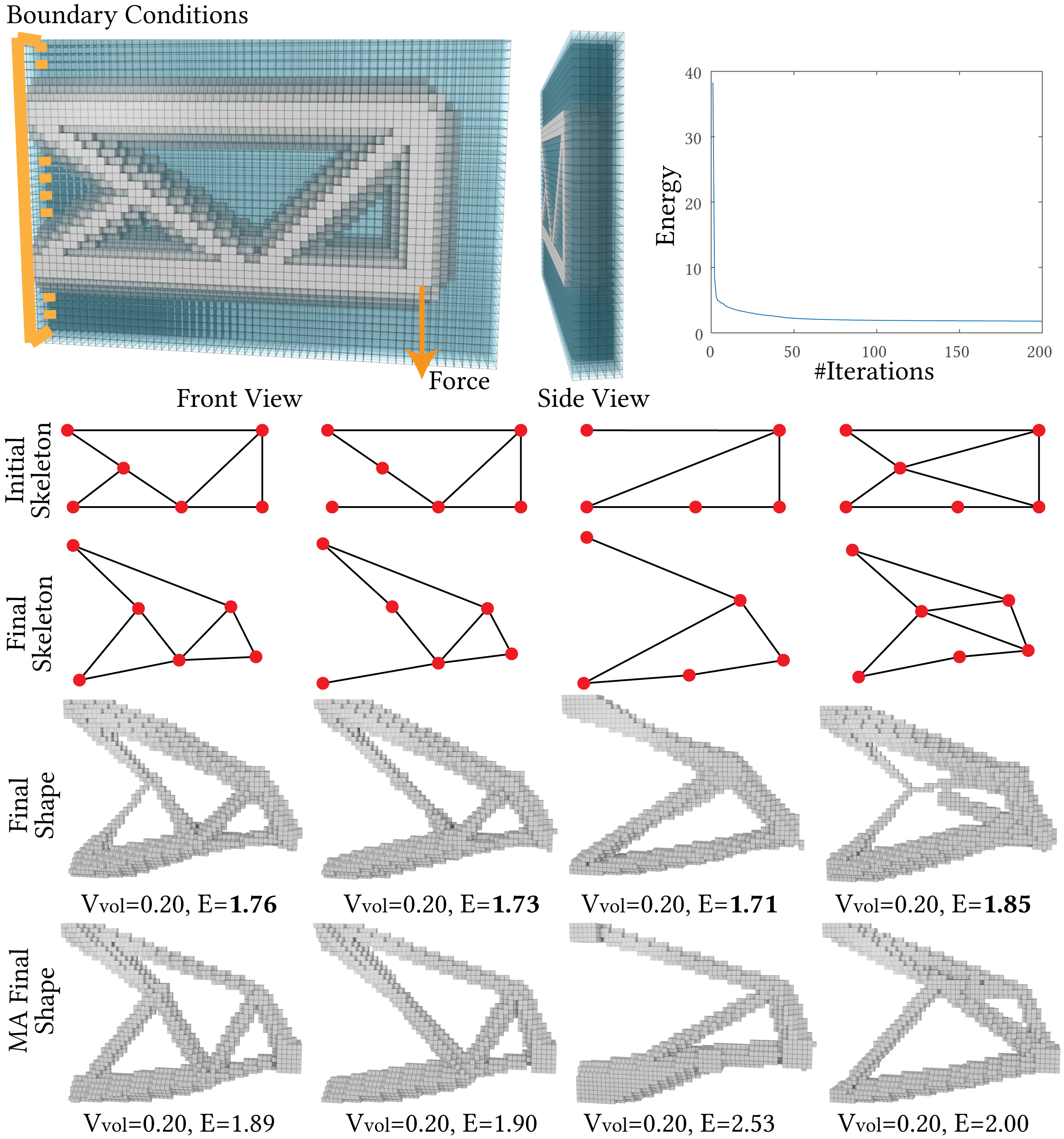}}
 % \vspace{-0.2cm}
	\caption{\textbf{Results of shape optimization.} 
    We run shape optimization using our medial skeletal diagram with a similar setup to topology optimization (top left), 
    but with four input skeletons as topology constraints. 
    The optimization converges fast (top right), 
    and each optimized shape shares the same topology as the corresponding input (bottom).
    \hl{Moreover, with our representation (Final Shape),
    we obtained a lower converged energy in all examples compared with the medial axis representation (MA Final Shape).} }
	\label{fig:to-case1}
 \vspace{-0.2cm}
\end{figure}

\begin{figure}[t]
	\centerline{\includegraphics[width=1.0\linewidth]{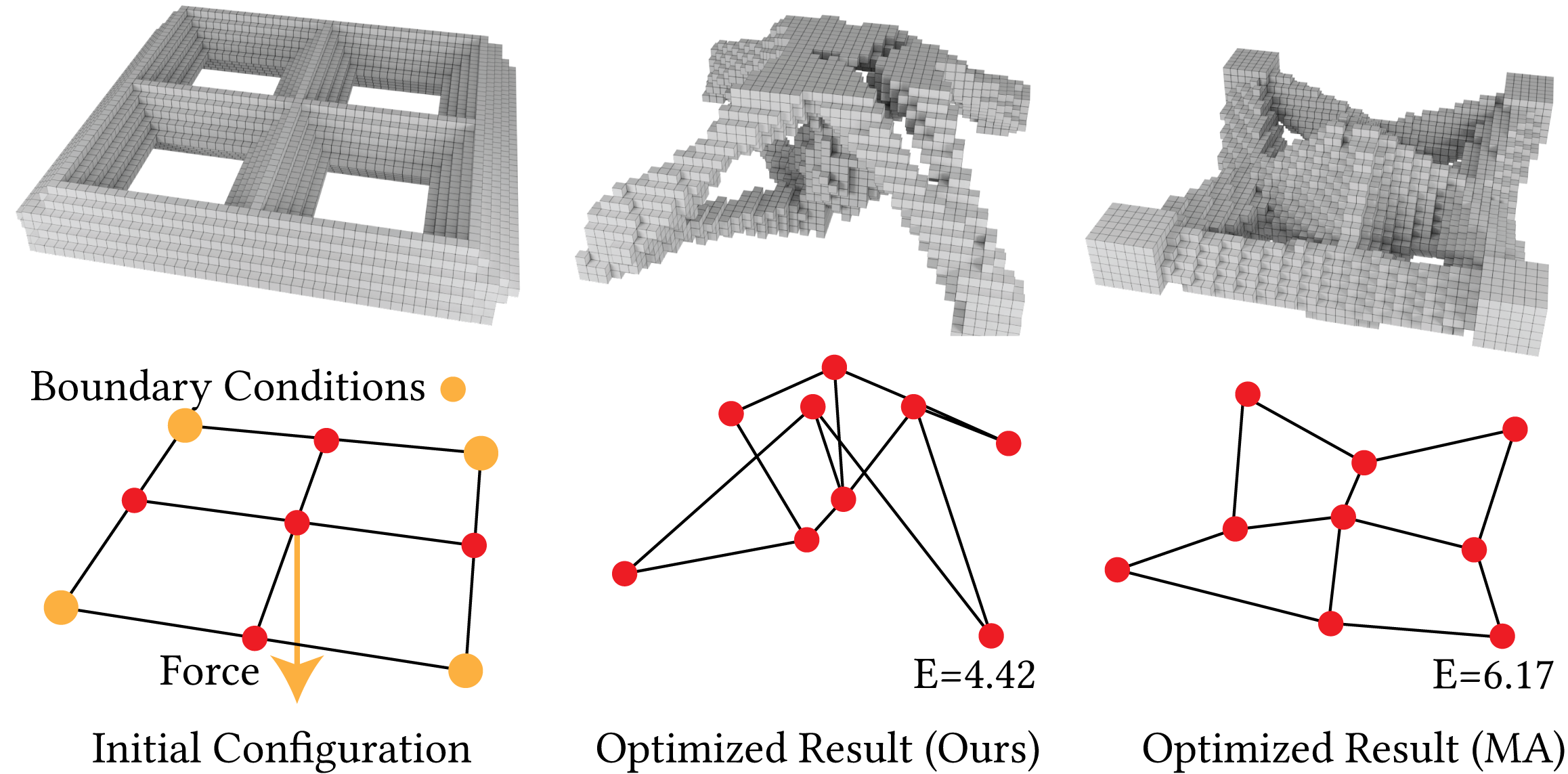}}
 % \vspace{-0.2cm}
	\caption{\textbf{The result of shape optimization with a complex topology as constraints.} 
    Our representation is capable of optimizing shapes with user-defined topology constraints, 
    a feature that is non-trivial to incorporate for voxel representation.
    Moreover, with our representation (middle),
    we obtained a lower converged energy in this complex example compared with the medial axis representation (right). 
    }
	\label{fig:to-case2}
 \vspace{-0.2cm}
\end{figure}

To substantiate this, we apply our representation to a topology-constrained shape optimization task.
This task takes an initial 3D skeletal graph, which specifies the topology of the shape, and optimizes the shape to achieve a predefined objective while maintaining the same topological structure.
We choose the objective to be compliance minimization, a well-known topology optimization task.
The goal of this task is to optimize a shape that can withstand input load 
with minimal compliance given the maximal allowed material usage.
Conventionally, such optimization is conducted within the space of a voxel grid, where each voxel carries a binary value to denote the presence of the material.
Here, we re-frame the optimization within the space of our representation.
We start by taking the initial 3D skeletal graph as the discrete structure of our medial skeletal diagram and constructing initial generalized enveloping primitives upon this discrete structure.
The optimization variables are the continuous parameters of our medial skeletal diagram, which include the positions of the vertices and the radii of each primitive.
Fig.~\ref{fig:to-case1} and~\ref{fig:to-case2} show the experimental setup.
% we can easily define 
% the topology of the shape through our representation.
% Combined with existing topology optimization techniques,
% we can obtain an optimized shape with a given topology.
% In addition, topology optimization is often 
% bottlenecked by the resolution $n$ of the space,
% because the number of DOFs is $O(n^3)$.
% As a result, it not only slows the performance, 
% but also increases the difficulty of optimization.
% On the contrary, the number of continuous parameters in our representation is much smaller.
% Moreover, the number of discrete parameters is also small in our representation 
% compared with existing representations.
% To demonstrate it, we combine our representation with the topology optimization.
% First, we optimize several shapes with a few user-defined skeletons to arrive at minimum compliance.

The optimization problem is formulated as follows:
\begin{align*}
	\argmin_{\hat{\mathbf{v}}, \hat{\mathbf{r}}}\ & E(\mathbf{u},\hat{\mathbf{v}}, \hat{\mathbf{r}})=\frac{1}{2} \mathbf{u}^\intercal \mathbf{K}(\mathbf{X}) \mathbf{u},\\
	\textrm{s.t.}\ & \mathbf{K}(\mathbf{X}) \mathbf{u} = \mathbf{f},\\
                      & \mathbf{X}_i = \theta(x_i;\hat{\mathbf{v}}, \hat{\mathbf{r}}), \forall x_i\in\Omega\\
	               & \sum \mathbf{X}_i \leq V_{\mathrm{vol}} \cdot |\Omega|,\\
	               & \mathbf{u}_{\mathrm{BC}} = 0,
\end{align*}
where $\hat{\mathbf{v}}$ is the concatenation of the medial vertex positions, which in turn determines the locations of generalized enveloping primitives,
$\hat{\mathbf{r}}$ is the concatenation of the radius values $\mathbf{r}$ of all primitives,
$\mathbf{X}_i$ is the density of voxel $x_i$ in the given domain $\Omega$,
$\mathbf{K}(\cdot)$ is the tangent stiffness matrix of a linear elastic material,
$\mathbf{u}$ is the displacement of the shape under external force $\mathbf{f}$,
$\mathbf{u}_{\mathrm{BC}}$ is the displacement of the grid points on the boundary condition,
and $V_{\mathrm{vol}}$ is the desired volume fraction of the final shape relative to the total voxel space.
We refer the readers to~\citet{chen2007shape} and~\citet{kumar2023honeytop90} for the definition of $\mathbf{K}(\cdot)$ and other standard aspects of the compliance minimization problem.
Note that $\theta(x_i;\hat{\mathbf{v}}, \hat{\mathbf{r}})$ bridges the continuous parameters of our medial skeletal diagram with the voxel grid space, defined as
\begin{align*}
    \theta(x_i;\hat{\mathbf{v}}, \hat{\mathbf{r}}) = H(f(\bigcup_{k} -\mathcal{E}_k(x_i;\mathcal{T}_{\mathcal{E}_{P}}(\hat{\mathbf{v}}, \hat{\mathbf{r}})))).
\end{align*}
Here, $\mathcal{E}_k(\cdot;\mathcal{T}_{\mathcal{E}_{P}}(\hat{\mathbf{v}}, \hat{\mathbf{r}}))$ is the implicit function (as defined in Sec.~\ref{sec:primitiveFitting}) of the $k$-th generalized enveloping primitive $\mathcal{T}_{\mathcal{E}_{P}}$ constructed using $\hat{\mathbf{v}}, \hat{\mathbf{r}}$. 
$f(\cdot)$ refers to Rvachev disjunction function~\citep{chen2007shape} converting boolean compositions into a real-value function, where the sign of the output value indicates the boolean value of the input compositions: $f(\phi_s\cup\phi_t)=\phi_s + \phi_t + \sqrt{\phi_s^2 + \phi_t^2}$.
$H(\cdot)$ is Heaviside step function.
In essence, $\theta(x_i;\hat{\mathbf{v}}, \hat{\mathbf{r}})$ outputs a value of one for voxels within the union of primitives, and zero otherwise.
We solve the optimization using gradient descent.
Appendix~\ref{app:to} provides details of the gradient calculation.
% To accommodate the primitives in our method, 
% for each primitive $i$, we define a implicit function 
% $\phi_i(x, p_i, d_i)= d^{\mathrm{max}}(x, p_i, r_i)-d(x, p_i, r_i)$,
% where $r^{\mathrm{max}}(\cdot)$ is the maximal radius 
% along the direction that $x$ is at 
% and $r(\cdot)$ is the distance to the skeleton along the same direction.
% In other words, if $x$ is inside the primitive 
% the function outputs a positive number, and \textit{vise versa}.
% According to our representation, the final structure is the union of all primitives.
% To do it in a differentiable manner, we adopt the idea of Rvachev function~\cite{chen2007shape}.
% Then, a union of $\phi_s,\phi_t$ can be written analytically as 
% $f(\phi_s\cup\phi_t)=\phi_s + \phi_t + \sqrt{\phi_s^2 + \phi_t^2}$.
% Then, the density $\theta_i$ of voxel $i$ 
% is defined as $\theta_i=H(f(\bigcup\limits_{k} \phi_k(x_i, p, r)))$,
% where $x_i$ is the center of voxel $i$ and
% $H(v)$ is an activation function 
% that outputs $\epsilon > 0$ when $v$ is negative and 1 when $v$ is positive.
% To make $H(\cdot)$ differentiable, 
% we approximate it using a piecewise cubic spline.
% With the definition above, 
% computing $\frac{\partial \theta}{\partial (p, r)}$
% becomes feasible.

We conduct two sets of experiments using the aforementioned formulation.
The first set is to optimize shapes under the same boundary conditions and external forces within a voxel space of $60\times40\times6$
while maintaining topology constraints using four distinct user-defined skeletal graphs built upon a shared set of vertices.
The input value $V_{\mathrm{vol}}$ is set to 0.2.
The optimization successfully finds the minimum in all four scenarios, 
with each converging to a valid solution that matches the topology of the corresponding input skeletal graph.
The optimized shapes, along with their compliance values, are shown in Fig.~\ref{fig:to-case1}.
\hl{Moreover, all minimizers from our representation are lower than those using the medial axis representation,
with an average $14\%$ lower compliance value.}
The second experiment optimizes a shape within a larger voxel space $60\times60\times40$ using a more complex skeletal structure as input.
The input value $V_{\mathrm{vol}}$ is set to $0.05$.
The optimizer also successfully yields a structure with minimal compliance ($E=4.42$), 
\hl{versus minimal compliance values $6.17$ from the medial axis representation}.
The final result is shown in Fig.~\ref{fig:to-case2}.
These experimental results demonstrate the capability of our representation 
to perform shape optimization with user-defined topology constraints,
a feature that is non-trivial to incorporate for voxel representation in a typical topology optimization setup.
\hl{Further, it also produces superior results compared to existing work with medial axis representation~\cite{bell2012geometry}}.
In our implementation, 
we use $|\hat{\mathbf{r}}_k|=8$ for a medial vertex and $|\hat{\mathbf{r}}_k|=12$ for a medial edge, 
which results in a total of approximately $250$ DOFs in both examples 
-- much smaller than the total number of voxels.

\subsubsection{Shape Generation}

\begin{figure}[t]
	\centerline{\includegraphics[width=1.0\linewidth]{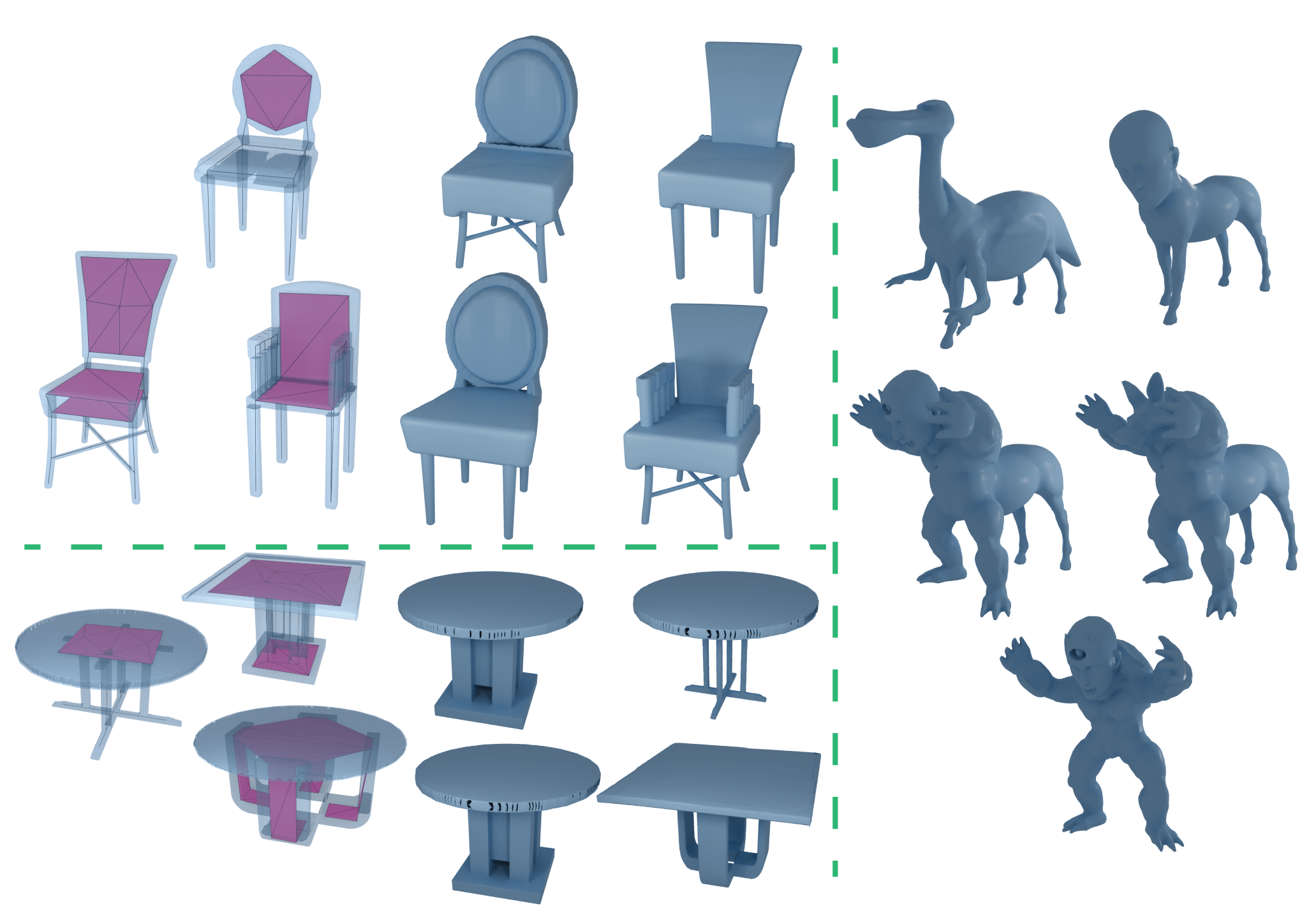}}
 \vspace{-3ex}
	\caption{\textbf{Results of shape generation.} We use a cut-and-paste method based on our medial skeletal diagram to generate new shapes. The chairs and tables are generated from shapes provided in~\citet{mo2019partnet}, with each row showing outputs generated using three mesh inputs (left). The bottom row shows five generated shapes derived from the $100$ shape benchmark.}
	\label{fig:shapeGen}
\end{figure}

We use the cut-and-paste method employed in existing shape modeling techniques~\cite{pandey2022face} to generate new shapes from a given set of input shapes.
We first construct the medial skeletal diagram for each shape in the input set. 
We then build a corpus of sub-skeletons by cutting subgraphs from these skeletons.
To generate new shapes, we assemble these sub-skeletons, adding edges between them to ensure the resulting skeleton is a single connected component. 
New shapes are then generated by pasting and uniting the fitted generalized enveloping primitives from the input shapes.
\hl{We use mesh boolean operations implemented from libigl to unite the primitives~\citep{libigl}.}
Using this process, we generate shapes from PartNet~\citep{mo2019partnet} and leverage the provided first-level part labels to perform the cutting on the skeleton.
Fig.~\ref{fig:shapeGen} illustrates the generated shapes for chairs and tables. 
This process of shape generation is remarkably straightforward and efficient, yet it ensures diverse changes in topology and geometry.
In Fig.~\ref{fig:shapeGen}, we also showcase several generated shapes derived from the $100$ shapes benchmark.
\hl{This streamlined process is made possible by the simplicity of the skeleton combined with the rich expressiveness of the primitives in our representation. In contrast, the medial axis representation demands a complex graph to accurately reconstruct the surface. Consequently, executing straightforward algorithms like the cut-and-paste method becomes challenging and necessitates considerable manual effort.}

\begin{figure}[t]
	\centerline{\includegraphics[width=0.98\linewidth]{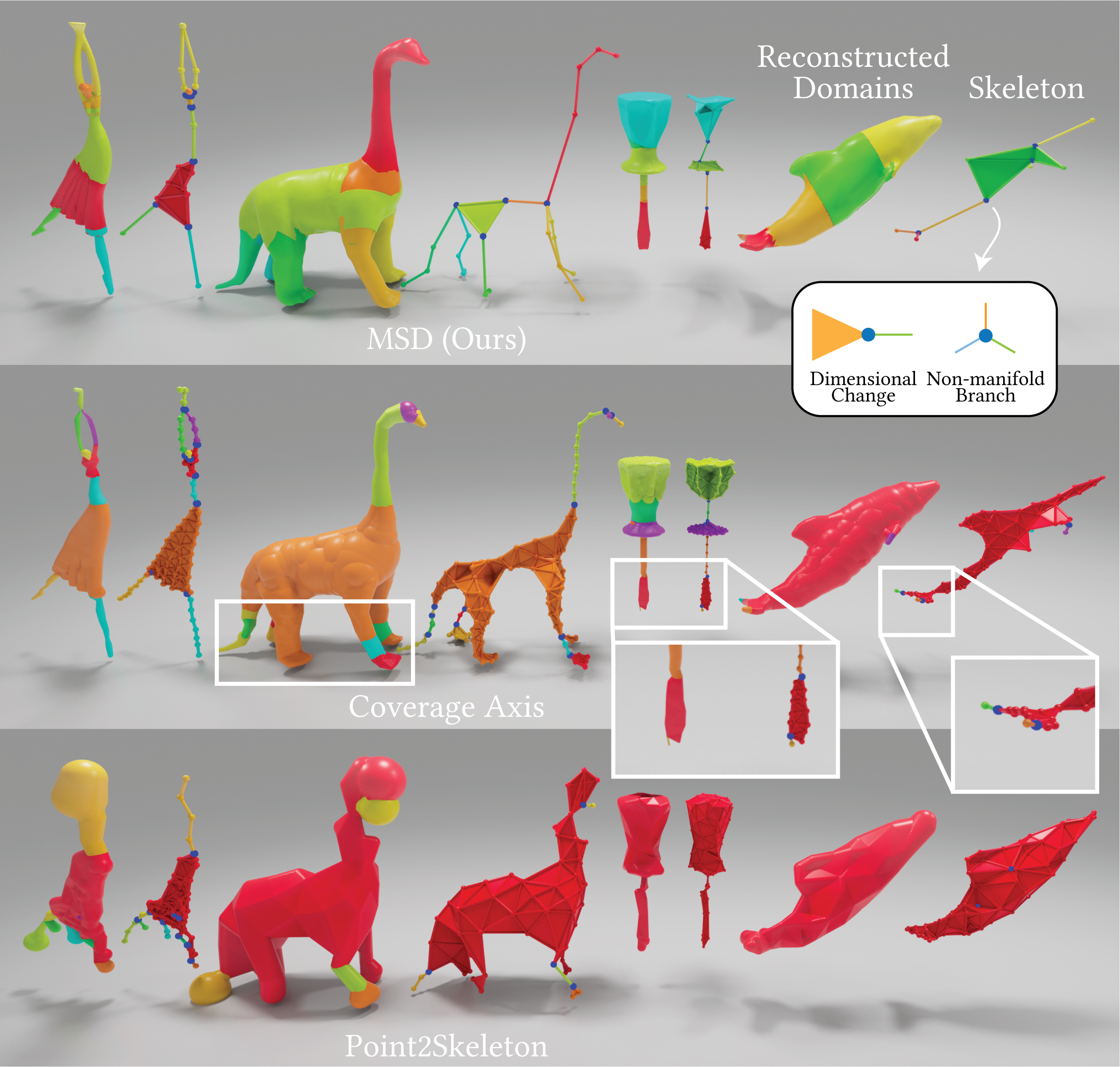}}
 \vspace{-0.2cm}
	\caption{\hl{\textbf{Domain Decomposition.} 
    We adopt the method from~\cite {lin2021point2skeleton} for domain decomposition. 
    Due to the simplicity of our skeleton and its precise shape reconstruction capabilities,
    our representation yields meaningful domain decomposition results. 
    Although the coverage axis typically provides high-quality decomposition in many instances,
    it struggles to produce accurate decomposition in certain highlighted areas (white rectangles).
    While Point2Skeleton excels in generating high-quality decomposition for specific shapes, like chairs, 
    as demonstrated in their paper~\cite{lin2021point2skeleton}, it falls short with more general shapes.}}
	\label{fig:domain-decomp}
\end{figure}
\begin{figure}[t]
	\centerline{\includegraphics[width=0.84\linewidth]{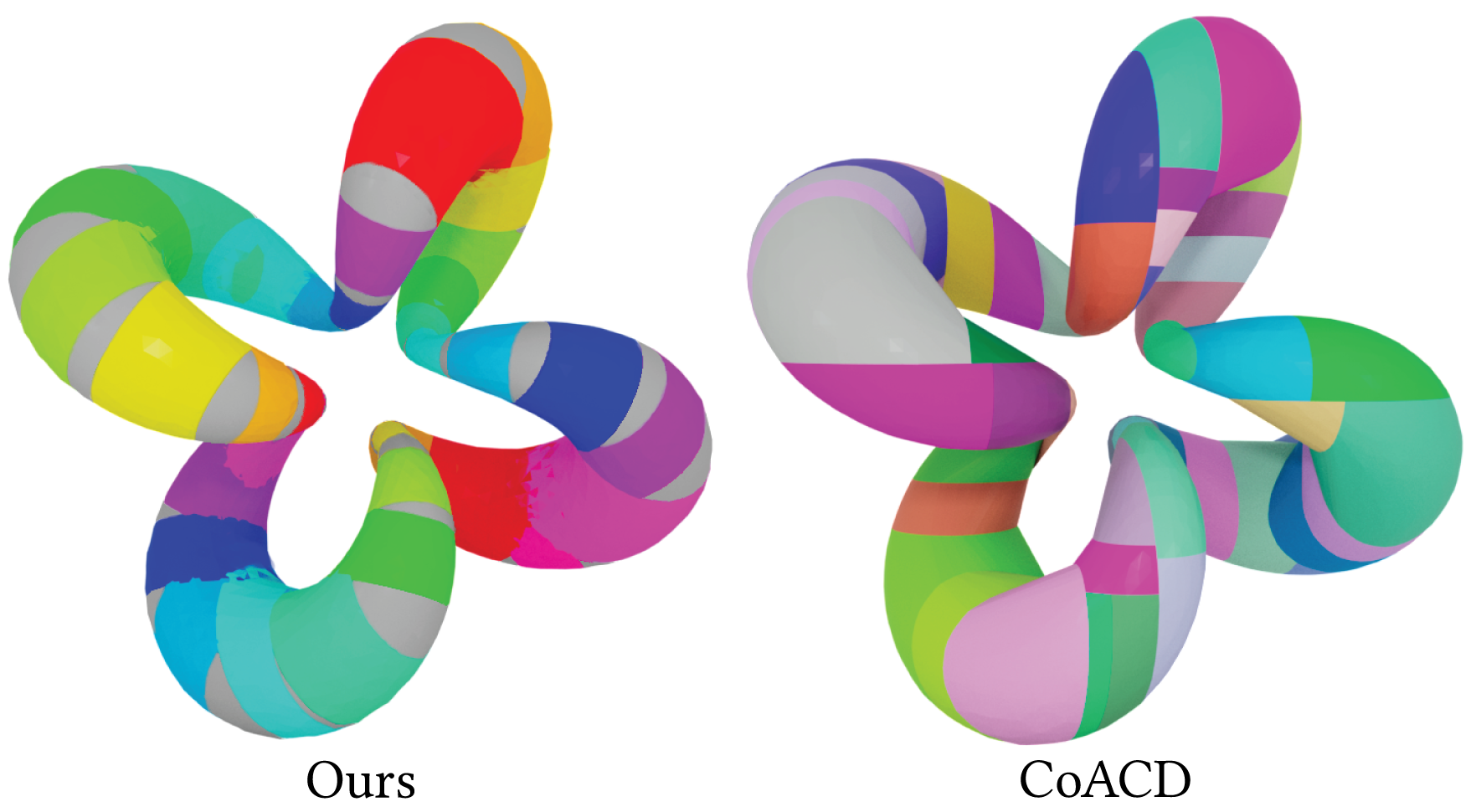}}
 \vspace{-0.3cm}
	\caption{\textbf{Comparison between medial skeletal diagram and CoACD on Knot example.} The number of generalized enveloping primitives in our representation is smaller than the domains decomposed by CoACD.}
 % The grey regions in our result are the overlaps between two adjacent primitives.}
	\label{fig:ddComp}
\end{figure}

\subsubsection{Mesh Decomposition}
\label{sec:mesh-decomposition}

\hl{Due to the simplicity of our skeleton and high accuracy in reconstructing the shape, 
our representation can also be used to perform domain decomposition of the shape.
Following the method from~\citep{lin2021point2skeleton},
we first identify the skeleton vertices that encounter a dimensional change on the neighboring elements
or that are non-manifold vertices and then decompose the skeleton at these vertices.
As shown in Fig.~\ref{fig:domain-decomp}, 
we can generate a meaningful and concise decomposition of the mesh,
while other methods fail to do so.}

Moreover, we compare the number of elements in our representation with the state-of-the-art domain decomposition method, CoACD~\citep{wei2022approximate}. 
Although these two methods do not share the same objective, a comparison can still be illuminating. 
Specifically, CoACD aims to decompose shapes into convex domains by cutting meshes with 3D planes,
whereas our approach works by performing a union of all fitted primitives to represent the shape.
Fig.~\ref{fig:ddComp} illustrates the comparison on the Knot example.
While our method, unlike CoACD, does not ensure the convexity of each primitive, 
it outperforms in terms of simplicity, 
demonstrating a reduced number of elements compared to CoACD. 
\hln{The overlapping region in our representation can be fixed by mesh boolean operations.}
% Furthermore, our method excels in capturing the topology of 3D shapes,
% a feature not adequately addressed by domain decomposition methods.
% \bh{is this sentence ok}.

\subsubsection{Mesh Alignment}
\label{sec:mesh-alignment}

\begin{figure}[t]
	\centerline{\includegraphics[width=1.0\linewidth]{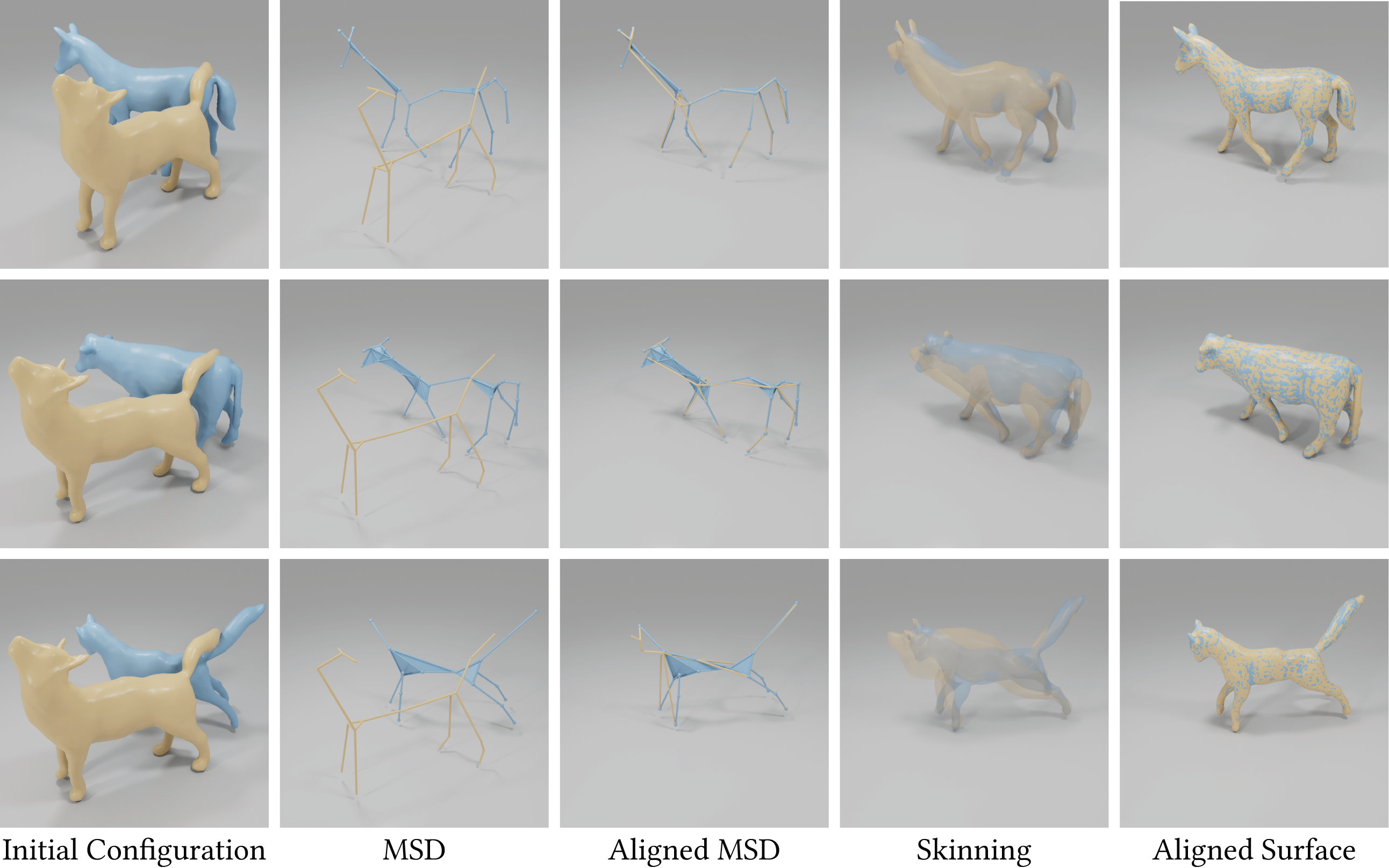}}
 \vspace{-0.2cm}
	\caption{\hl{\textbf{Non-rigid Alignment.} Using our representation, one can initially align the medial skeletal diagrams to simplify subsequent surface non-rigid alignment.
    After the skeletons are matched, the surfaces deform in line with their corresponding skeletons, providing an initial estimate for the final non-rigid alignment.
    This preliminary alignment considerably eases the non-rigid alignment process for the surfaces, reducing the need for manual intervention.
    }}
	\label{fig:alignment}
 \vspace{-0.2cm}
 % \vspace{-0.6cm}
\end{figure}

\begin{figure}[t]
	\includegraphics[width=1.0\linewidth]{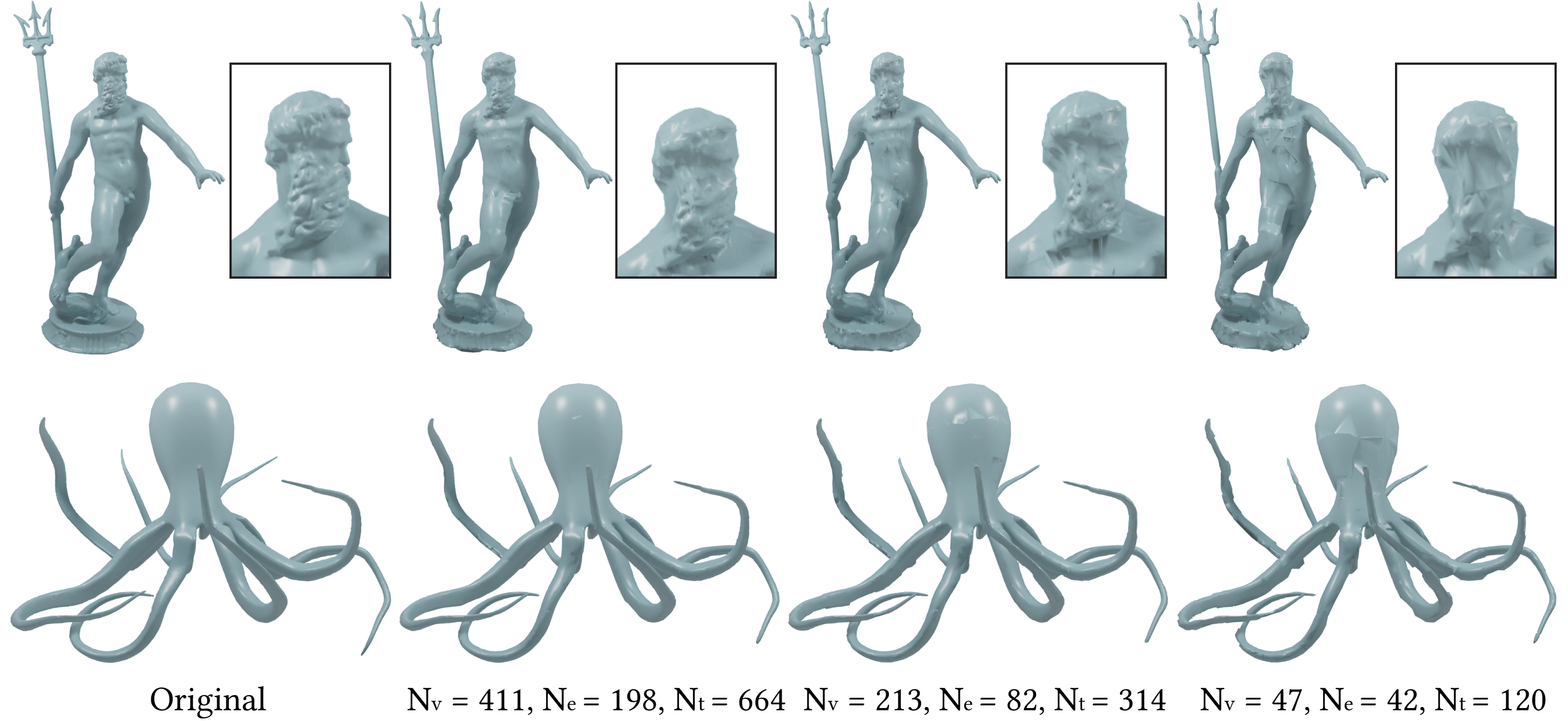}
 \vspace{-0.3cm}
	\caption{\hl{\textbf{Mesh Compression.} 
    Our MSD can be easily used for mesh compression.
    For each primitive type, we define a canonical mesh with three different resolutions.
    Primitives at high resolution can be directly downsampled to lower resolutions using the predefined canonical mesh, thereby achieving mesh compression.
    }}
 \vspace{-0.1cm}
	\label{fig:mesh-compression}
\end{figure}

\hl{
Using our representation, we can ease non-rigid alignment between different shapes.
To showcase its efficacy, we select four shapes from different species within the Animal3D dataset~\cite{xu2023animal3d}. 
We then non-rigidly align the source mesh (highlighted in yellow) with the other three meshes (depicted in blue), as illustrated in Fig.~\ref{fig:alignment}.
Instead of directly aligning the source mesh to the target,
we first compute our MSD for each shape and then non-rigidly align their skeletons. 
Due to the simplicity of these skeletons, 
alignment is more straightforward than with surface meshes.
Once the skeletons are aligned,
the surface mesh is deformed according to the skeleton's deformation.
In our examples, we treat the skeleton as the kinematic object,
with the surface mesh representing the surface of a tetrahedral mesh anchored to the skeleton.
This mesh is then deformed using a physically-based simulation.
The deformed mesh serves as an initial guess for the non-rigid alignment of the surface meshes,
reducing the manual effort needed to select landmark pairs between the source and target meshes.
}

\begin{figure}[t]
	\centerline{\includegraphics[width=0.9\linewidth]{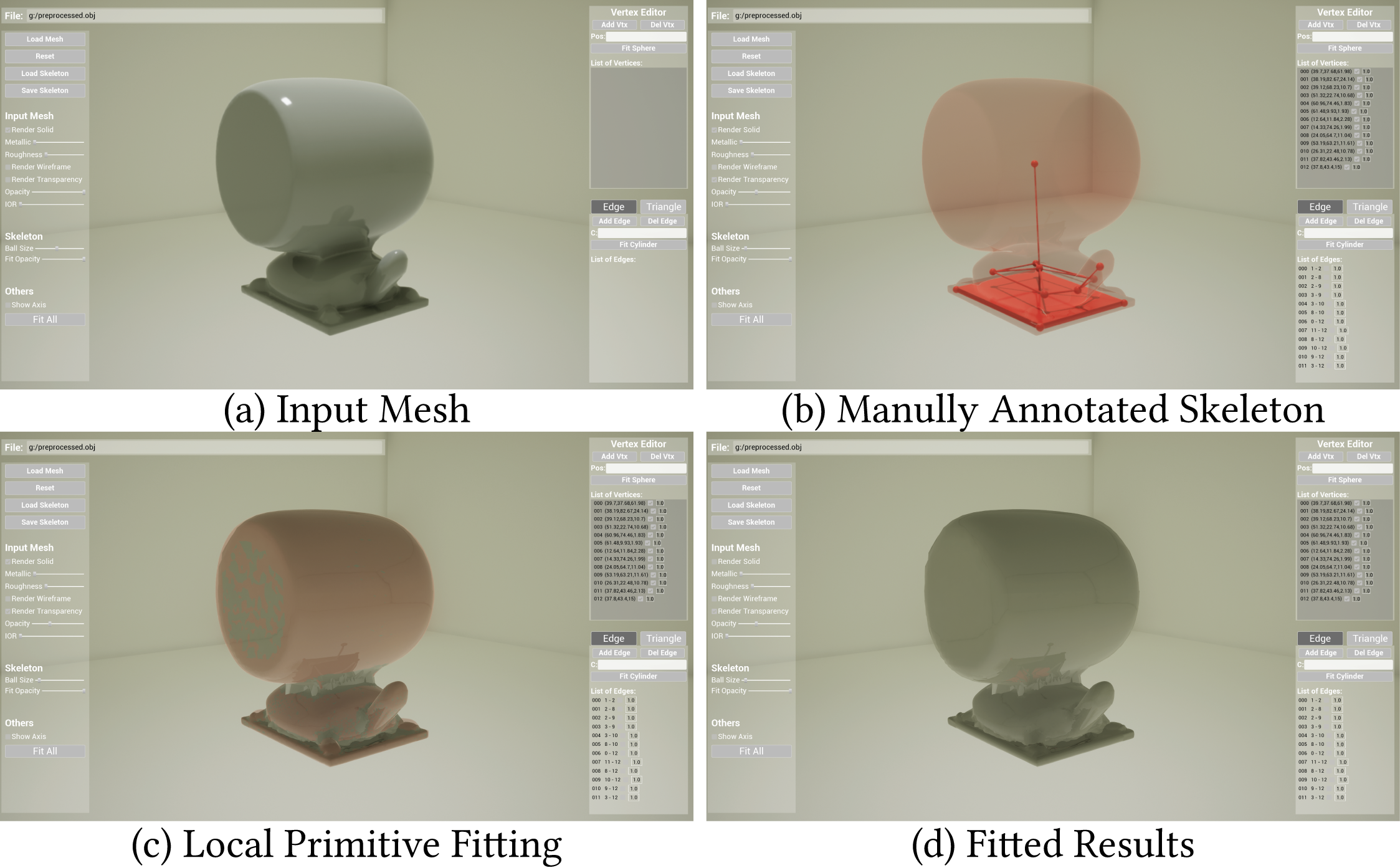}}
 \vspace{-0.2cm}
	\caption{\textbf{Constructing medial skeletal diagram using our GUI.} Users can manually annotate all the skeletal elements and use our proposed local primitive fitting to obtain each primitive.}
	\label{fig:ui-manual}
\end{figure}

\begin{figure}[t]
	\centerline{\includegraphics[width=0.9\linewidth]{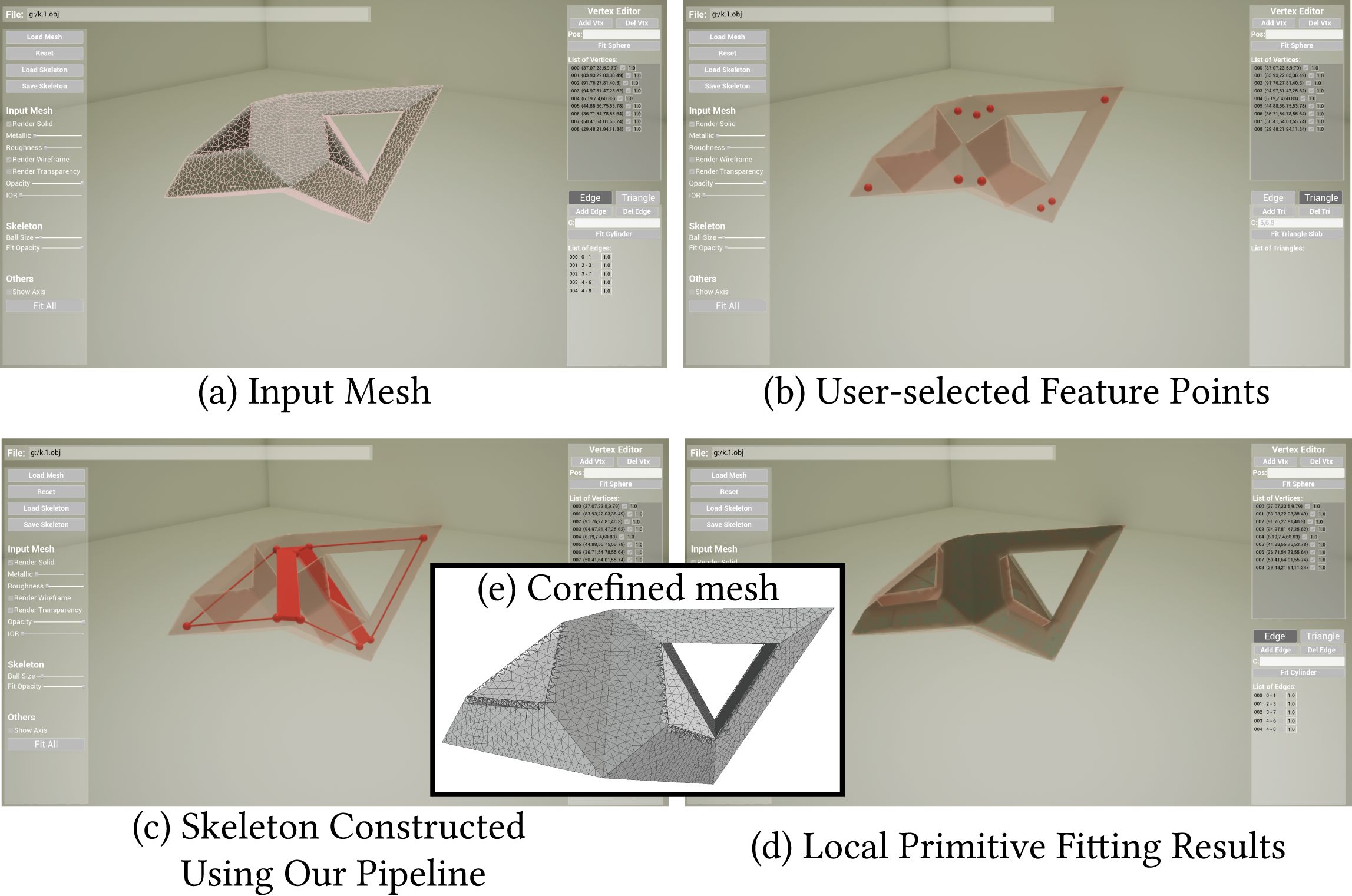}}
 \vspace{-0.2cm}
	\caption{\textbf{Constructing medial skeletal diagram with user-selected feature points using our GUI.} Users can designate only the skeletal vertices where our medial skeletal diagram construction pipeline automatically produces the corresponding skeleton and the reconstructed mesh.}
	\label{fig:ui-points}
\end{figure}

\subsubsection{Mesh Compression} 
\label{sec:mesh-compression}

We also propose a simple approach to control the number of continuous parameters associated with each primitive, facilitating mesh compression. 
For each example shown in Fig.~\ref{fig:mesh-compression}, 
we first compute its MSD. 
Subsequently, mesh compression is achieved by 
adjusting the mesh resolution for each primitive. 
\hlca{This can be achieved by isotropic remeshing~\citep{botsch2010polygon} or
Garland-Heckbert simplification~\citep{garland1997surface}.}
For each primitive type, we define three canonical meshes with varying resolutions. 
Specifically, for a sphere mesh, the resolutions are $411$, $213$, and $47$; 
for a cylinder mesh, they are $198$, $82$, and $42$; 
and for a prism mesh, they are $664$, $314$, and $120$.
Each calculated generalized enveloping primitive is mapped to its corresponding canonical mesh.
As shown in Fig.~\ref{fig:mesh-compression},
\hln{reducing the number of continuous parameters still preserves the overall original shape.
}

\subsubsection{User Interface}
\label{sec:ui}

The \hlca{sparsity of the skeleton in our medial skeletal diagram} offers users an intuitive 
and convenient method for shape modeling.
To demonstrate it, we implement a graphical user interface (GUI).
Within this GUI, users can create skeletal elements and apply our proposed local primitive fitting to represent a given input shape.
As shown in Fig.~\ref{fig:ui-manual}, 
the user was able to manually construct a skeleton with
$13$ vertices, $12$ edges, and $4$ triangles to represent the Drum shape.
Moreover, users can manually identify and mark necessary skeletal feature points of the shape, such as topological features, and place points for an input mesh using our GUI.
These selected points are then used as inputs for 
our construction pipeline which automatically produces a high-quality skeleton and the reconstructed mesh. An example of a double torus is illustrated in Fig.~\ref{fig:ui-points}.

\subsection{Additional Results}
We also show a gallery of additional results from the benchmark of $100$ shapes in Fig.~\ref{fig:more-ret1} and~\ref{fig:more-ret2}. 
These results underscore the versatility of our method and its ability to handle a wide variety of shapes.

%% file: sections/limitation.tex
\section{Limitations and Future Work}\label{sec:limitations}
% topology
% exact geodesic VD
% curves, surfaces -> edges, triangles
% numerical stability, thin regions, meshing quality
% timing
\noindent\emph{\hlca{Topology preservation.}} 
Our current approach does not guarantee topological preservation, 
which is also a limitation of the state-of-the-art work~\cite{wang2022computing, wang2022restricted}. 
A key reason for this limitation lies in the sparsity of our medial skeletal diagramm, which contradicts the sufficient condition for maintaining homotopy, which requires that sampled points be sufficiently dense to satisfy the $\epsilon$-sampling condition, 
as detailed in~\cite{amenta1998surface, yan2009isotropic}. 
This leads to situations where the topology of the reconstructed 3D shape might be compromised.
Additionally, while we ensure that 
% submersion for 
each individual primitive is homotopic to its corresponding skeleton element, 
% Therefore, t
the final reconstructed shape -- which is a union of all the primitives -- is not guaranteed to be homotopic to the skeleton.
Future research may delve deeper into these topological considerations, 
investigating potential modifications to the current method that could ensure topological preservation, even with the sparse skeleton of our medial skeletal diagram.

\noindent\emph{\hlca{RVD computation.}} 
% In addition to the topological preservation issue, t
The computational approach for calculating the shortest path during the construction of the medial skeletal diagram could be optimized. 
While the currently used Dijkstra's algorithm provides a functional solution, it is not ideal for this particular application as mentioned in~\cite{wang2020robustly, xin2022surfacevoronoi}. 
A theoretically more suitable approach would be to employ geodesic Voronoi diagram (GVD).
However, this presents a challenge due to the non-manifoldness of the medial axis, the space where GVD is computed. 
Future work could thus seek methods for replacing RVD with GVD in non-manifold domains. 
Possible avenues for this could include using a non-manifold Laplacian coupled with heat diffusion~\cite{sharp2020laplacian} or the development or adaptation of an exact geodesic distance calculator~\cite{surazhsky2005fast}.

\noindent\hlca{\emph{Centrality and uniqueness.} 
Unlike the uniform, linearly interpolated primitives typically used in conventional medial axes, 
which generally align with the geometric center of a shape, 
our generalized enveloping primitives are non-convex. 
As a result, defining centrality for our primitives is challenging. 
To ensure our approach still captures the essence of centrality, 
we focus on maintaining local centering for each primitive during optimization. 
This is achieved through: 1) all vertices of the medial skeletal diagram are constrained to lie on the raw medial axis; 2) the smoothness energy used during local primitive fitting; and 3) $E_{\mathrm{centrality}}$ in global optimization.
Despite these efforts, the MSD may still exhibit a loss of ``global centrality,''
where the overall skeleton structure may not be centered within the global shape. 
This could result in an increased number of discrete elements in the MSD, 
as more primitives might be necessary to fully cover the input mesh.
Additionally, this could potentially degrade performance in applications such as shape alignment. 
Future work could explore integrating global information, such as skinning weights, into the MSD construction process.}
Moreover, while our objective aims to achieve one optimal skeleton, 
it does not guarantee the uniqueness of the global minimal solution. 
Future research could focus on incorporating shape properties as constraints to enhance the centrality and uniqueness of the optimized skeleton.

\noindent\hlca{\emph{Overlaps of primitives.} 
From the application results, one can find that there are overlapping regions between neighboring primitives. 
Since the final shape is the union of these primitives,
this overlapping does not pose significant problems in applications such as shape optimization and shape generation. 
However, we acknowledge that this overlap may violate intersection-free constraints in certain tasks, 
such as mesh decomposition and physical simulation. 
Additionally, the lack of one-to-one mapping could introduce ambiguity in shape alignment. 
These issues could be addressed by post-processing methods, 
including collision handling, Boolean operations, or multi-material meshing~\citep{MultimaterialMeshing24}, to generate non-overlapping primitives.}

\noindent\emph{\hlca{Nonlinear skeleton.}} Our current medial skeletal diagram consists solely of piecewise linear elements. 
Although this configuration has demonstrated solid performance, 
the inclusion of higher-order surfaces like polynomial elements~\cite{marschner2021sum} 
could potentially enhance the representation further. 
Higher-order surfaces could provide more precise coverage of local regions \hlca{with sparser skeletons}, 
contributing to both the accuracy and efficiency of our representation.

\noindent\emph{\hlca{Computational cost.}} Currently, the time efficiency of computing MSD is less than ideal. 
Future work that incorporates data-driven methods and graph generation may enhance computational speed, using our current method to supply training data pairs.

\begin{figure*}[t]
	\centerline{\includegraphics[width=0.95\linewidth]{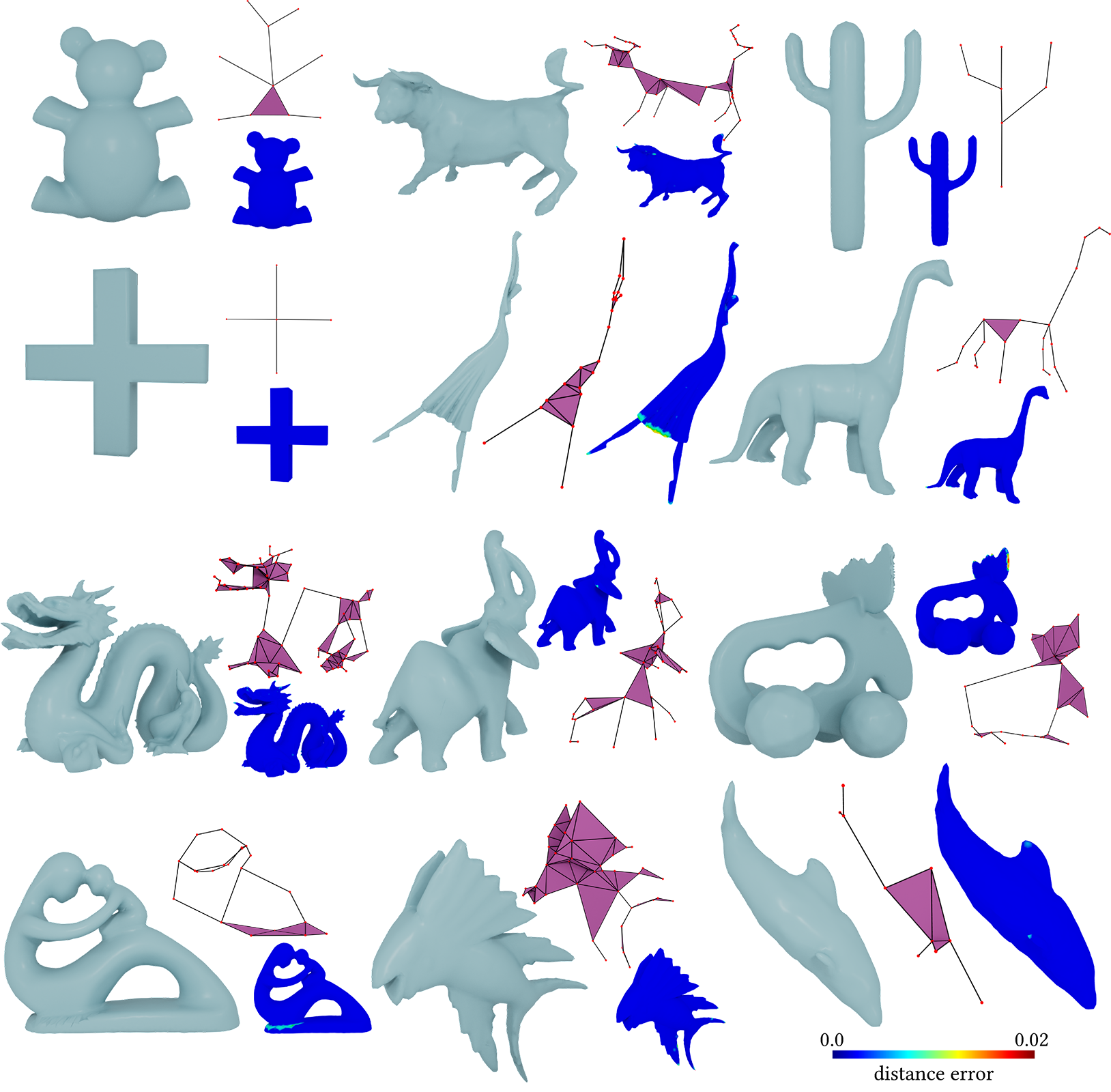}}
	\caption{\textbf{A gallery of more results of our medial skeletal diagram (part 1/2).} For each sample, we show the reconstructed shape, the discrete skeleton, and the reconstruction error map on the input mesh.}
	\label{fig:more-ret1}
\end{figure*}

\begin{figure*}[t]
	\centerline{\includegraphics[width=0.94\linewidth]{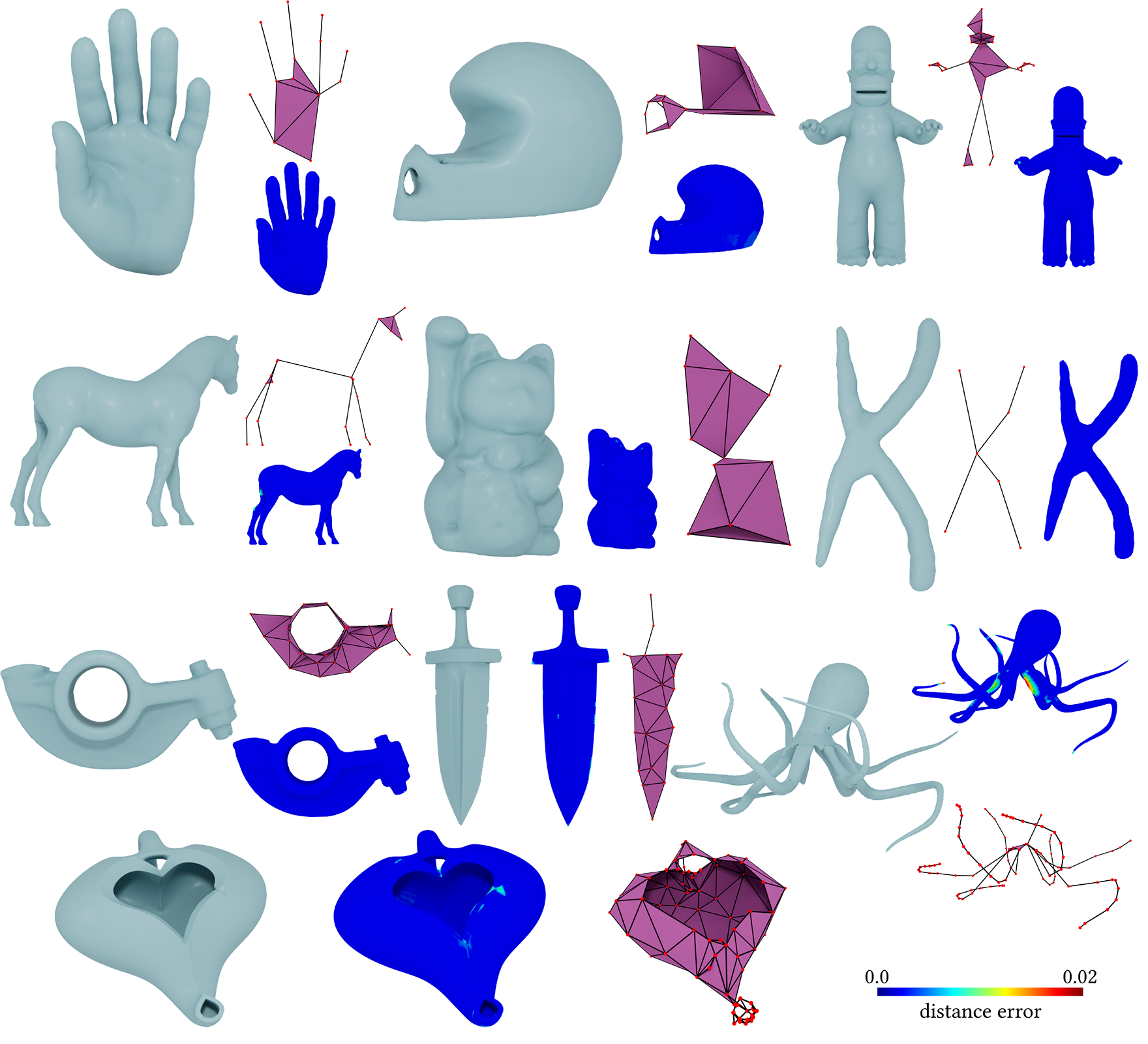}}
	\caption{\textbf{A gallery of more results of our medial skeletal diagram (part 2/2).} For each sample, we show the reconstructed shape, the discrete skeleton, and the reconstruction error map on the input mesh.}
	\label{fig:more-ret2}
\end{figure*}

%% file: sections/conclusion.tex
\section{Conclusion}
In this paper, we present the medial skeletal diagram, a skeletal representation that strives for both \hlca{sparsity in the skeletal structure and completeness in the reconstruction}. 
Our approach significantly reduces the number of discrete elements while preserving completeness by augmenting the complexity of the continuous elements.
Our medial skeletal diagram extends the medial axis 
by introducing generalized enveloping primitives which effectively cover complex local shape regions.
We also present a computational pipeline to construct the medial skeletal diagram from an input shape by employing a local-global optimization paradigm to maximize shape coverage and minimize the count of discrete elements.
A post-refinement process is conducted to guarantee an accurate match of our representation with the target mesh. 
We demonstrated the effectiveness of our method on a benchmark of $100$ shapes and highlighted its diverse applications \hln{in shape optimization, shape generation, mesh decomposition, mesh alignment, mesh compression, and user-interactive design.}

%% file: sections/appendix.tex
\section*{Appendix}

\section{Shape Optimization Using Gradient Descent}
\label{app:to}
\begin{figure}[t]
	\centerline{\includegraphics[width=0.5\linewidth]{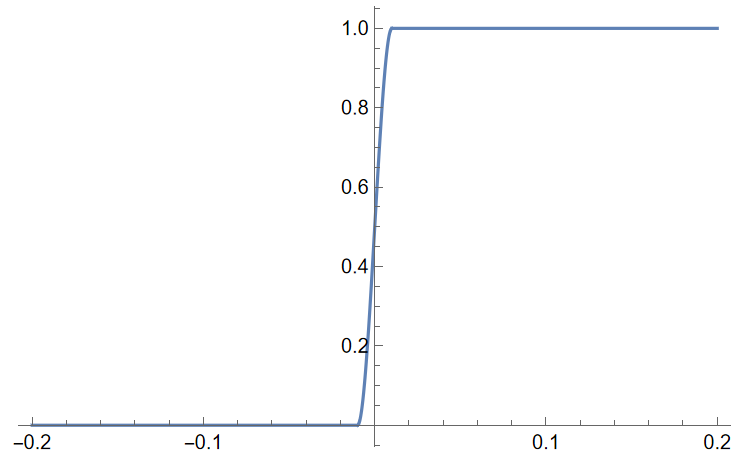}}
	\caption{\textbf{The modified function $H(\cdot)$} }
	\label{fig:H}
\end{figure}

% \mh{align notations}
We make the Heaviside step function differentiable by modifying $H(\cdot)$ as
\begin{equation}
	H(x) =\begin{cases}
		0 \ &x < \gamma,\\
		-\frac{1}{4 \gamma^3} x^3 + \frac{3}{4 \gamma} x + 0.5 \ &-\gamma \leq x \leq \gamma,\\
		1 \ &x > \gamma
	\end{cases},
\end{equation}
where $\gamma$ is a small number.
We choose $\gamma=0.01$ as the cell size of each voxel.
The function is $C^1$ continuous in $\mathbb{R}$.
Fig.~\ref{fig:H} shows the plot of the function $H(\cdot)$.

To perform gradient descent, 
we first convert the inequality constraint into a quadratic energy term,
\begin{equation}
	E_v(\hat{\mathbf{v}}, \hat{\mathbf{r}})=
	\begin{cases}
		\frac{w}{2}(\sum_{x_i} \frac{\theta(x_i;\hat{\mathbf{v}}, \hat{\mathbf{r}})}{|\Omega|} - V_{\mathrm{vol}})^2 &\mathrm{if } \sum_{x_i} \frac{\theta(x_i;\hat{\mathbf{v}}, \hat{\mathbf{r}})}{|\Omega|} > V_{\mathrm{vol}},\\
		0 &\mathrm{o.w.}
	\end{cases}
\end{equation}
The energy is non-zero only if the current volume exceeds the user-defined one.
Next, we differentiate the static equilibrium constraints
\begin{equation}
\frac{\partial \mathbf{K}(\theta)}{\partial \theta} \mathbf{u} +
\mathbf{K}(\theta) \frac{\partial \mathbf{u}}{\partial \theta} = 0.
\end{equation}
The gradient of the objective w.r.t. $(\hat{\mathbf{v}}, \hat{\mathbf{r}})$ is
\begin{align}
\frac{d E}{d (\hat{\mathbf{v}}, \hat{\mathbf{r}})}&=\frac{\partial E}{\partial \mathbf{u}} \frac{\partial \mathbf{u}}{\partial (\hat{\mathbf{v}}, \hat{\mathbf{r}})} +  \frac{\partial E}{\partial (\hat{\mathbf{v}}, \hat{\mathbf{r}})} \\
&=\left(-\frac{1}{2}\mathbf{u}^T \frac{\partial \mathbf{K}(\theta)}{\partial \theta} \mathbf{u} + 
wS^T (S \frac{\theta}{|\Omega|} -  V_{\mathrm{vol}})\right) \frac{\partial \theta}{\partial (\hat{\mathbf{v}}, \hat{\mathbf{r}})},
\end{align}
where $S$ is the summation linear operator.
% Once we have the gradient, we can apply the gradient descent method.
We apply gradient descent to optimize the objective of shape optimization.
At each gradient descent iteration,
we also perform a back-tracing line search to guarantee a decrease in the total energy.
At each line search iteration,
we first compute $\mathbf{K}$ from a new $(\hat{\mathbf{v}}, \hat{\mathbf{r}})$ and 
then solve the static equilibrium to obtain $\mathbf{u}$.
The energy is computed as $\frac{1}{2}\mathbf{u}^\intercal\mathbf{K}\mathbf{u} + E_v$.

\section{Well-definedness of Directional Vectors in Generalized Enveloping Primitive}\label{app:welldefine}
% TODO change references, explain the well-definedness
\begin{definition}%[~\cite{munkres2018analysis}]
\cite[$\epsilon$-neighborhood]{munkres2018analysis}\label{def:eNeighbor}
Let ${P}$ be a smooth submanifold in $\mathbb{R}^3$ and $\epsilon:{P}\mapsto\mathbb{R}^{+}$ be a smooth positive function. The \emph{$\epsilon$-neighborhood} of ${P}$, denoted by ${P}^{\epsilon}$, is defined as,
\begin{align}
{P}^{\epsilon}:=\big\{y\in\mathbb{R}^3 \big| ||y - x||_2 < \epsilon(x) \ \text{for some} \ x\in {P}\big\}.
\end{align}
\end{definition}
\noindent Furthermore, the following theorem asserts the existence of a surjective map between ${P}^{\epsilon}$ and ${P}$:
\begin{theorem}%[$\epsilon$-neighborhood theorem~\cite{munkres2018analysis}]
\cite[Proposition of Tubular Neighborhood Theorem]
{lee2012smooth}\label{thm:eNeighbor}
\hlca{For a smooth submanifold ${P}$ in $\mathbb{R}^3$, there exists a smooth positive function $\epsilon:{P}\mapsto\mathbb{R}^{+}$ such that the map $\pi_{\epsilon}:{P}^{\epsilon}\mapsto{P}$, which maps each point $y\in{P}^{\epsilon}$ to the closest point $\pi_{\epsilon}(y)$ in ${P}$, is a submersion. Moreover, if ${P}$ is compact, then $\epsilon$ can be taken to be a constant.}
\end{theorem}
\noindent 
% This proposition is also referred to as the Tubular Neighborhood Theorem~\cite{lee2012smooth}.
% (1) each $y\in{P}^{\epsilon}$ has a unique closest point $\pi_{\epsilon}(y)$ in ${P}$; (2)
\hlca{Note that $\pi_{\epsilon}$ is well-defined due to the fact that ${P}$ is convex, so for each $y\in{P}^{\epsilon}$, there is a unique closest point in ${P}$.}
This theorem ensures that for a medial mesh element ${P}$, there exists an appropriate constant ${\epsilon}$ to construct a smooth closed submanifold ${P}^{\epsilon}$ and establish a surjective map $\pi_{\epsilon}$ between ${P}^{\epsilon}$ and ${P}$. 
For more details of choosing ${\epsilon}$, we refer the readers to the proof of the theorem in ~\cite{lee2012smooth}.
% \footnote{\mh{the way to choose ${\epsilon}$ is in the proof of the theorem: https://math.stackexchange.com/questions/3653887/the-epsilon-neighborhood-theorem}}
If we consider the unit normal vector for each point $y$ on the boundary $\partial{P}^{\epsilon}$ of ${P}^{\epsilon}$, we obtain a submersion $g: N{\partial P}^{\epsilon}\mapsto{P}$,
\begin{align*}
    g=\pi_{\epsilon}\circ f^{-1},
\end{align*}
where $N{\partial P}^{\epsilon}= \{(y, n)|y\in\partial{P}^{\epsilon}, n\in N_p\partial{P}^{\epsilon}\}$ denotes the unit \emph{normal bundle} of ${\partial P}^{\epsilon}$, and $f:\partial{P}^{\epsilon}\mapsto N{\partial P}^{\epsilon}$, which is constructed from Gauss map of $\partial{P}^{\epsilon}$, maps between $\partial{P}^{\epsilon}$ and $N{\partial P}^{\epsilon}$.
Here $f$ is bijective since ${\partial P}^{\epsilon}$ is convex.
% both ${P}^{\epsilon}$ and ${\partial P}^{\epsilon}$ are smooth and closed.
The map $g$ serves as a crucial connector between the medial mesh element ${P}$ and the normal vectors of its $\epsilon$-neighborhood boundary $\partial{P}^{\epsilon}$. 
Essentially, $g$ associates each point in ${P}$ with a set of unit directional vectors -- the normal vectors of $\partial{P}^{\epsilon}$, while the surjection of $\pi_{\epsilon}$ concurrently guarantees each such normal vector corresponds to a specific point in ${P}$.
This provides a theoretical guarantee that our primitives are mathematically well-defined in the continuous setting, i.e., every point in $P$ correlates with at least one directional vector through the surjection of $g$.
% For any point $p\in{P}$, we denote its associated directional vectors as
% \begin{align}
%     d_p({P}) := \big\{||n||_2 = 1\big|g(y, n)=p, \forall (y, n)\in N{\partial P}^{\epsilon}\big\},
% \end{align}
% and denote the set of all directional vectors of ${P}$ as
% \begin{align*}
%     d({P}) = \big\{d_p({P})| \forall p\in{P}\big\}.
% \end{align*}
% This relationship enables a comprehensive understanding of the geometric association between the medial mesh element and its surrounding $\epsilon$-neighborhood, which is essential for constructing accurate generalized enveloping primitives that can seamlessly and efficiently cover local shape features.

\hlca{
Due to the convexity of ${P}^{\epsilon}$, the  directional vectors $d(P)$ are not affected by the choice of ${\epsilon}$ in the continuous case.
In our implementation for discrete primitives, as shown in Fig.~\ref{fig:discreteEPs}, we choose $\epsilon=0.01$ for input shapes normalized to the range [0,1].
}

%% file: sections/supp-refinement.tex
\section{Implementation Details of Feature-preserving Refinement}
We have developed a robust implementation of our feature-preserving refinement, 
which benefits from exact rational arithmetic, thereby ensuring robustness. 
In the following sections, we will explain each step in detail.

\begin{figure}[t]
\centerline{\includegraphics[width=0.95\linewidth]{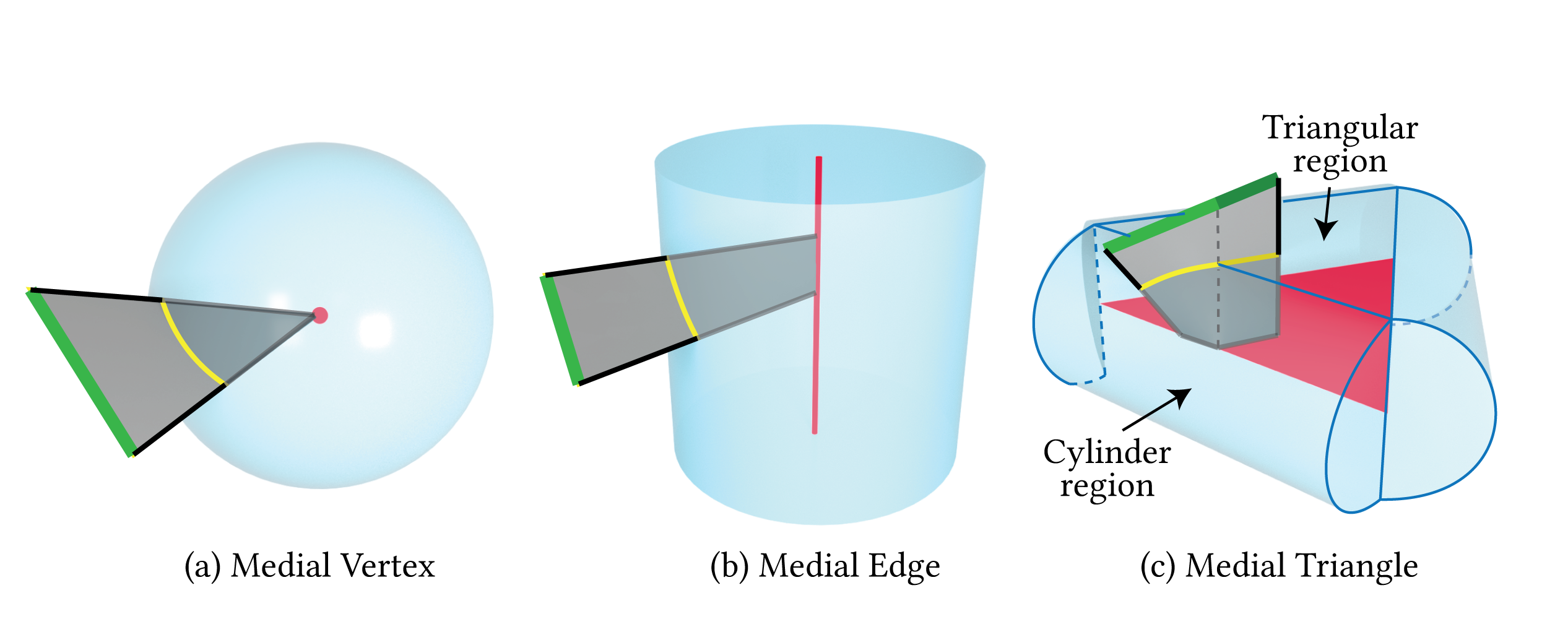}}
\caption{\textbf{Projection trajectory of one target edge for each type of primitive.}
The projection trajectory is shown in gray,
the medial element is shown in red,
and the projected curve is shown in yellow.
For a medial vertex, the projection trajectory of a target edge is a triangle.
For a medial edge, the projection trajectory of a target edge is nontrivial,
because the target edge and its projection are not coplanar.
For a medial triangle, the projection trajectory is a mixture of the
previous cases.
In our implementation, we omit the sphere portion of the primitive in (b) and (c) for convenience.}
\label{fig:coref-cutting}
\end{figure}

\begin{figure*}[t]
\centerline{\includegraphics[width=0.95\linewidth]{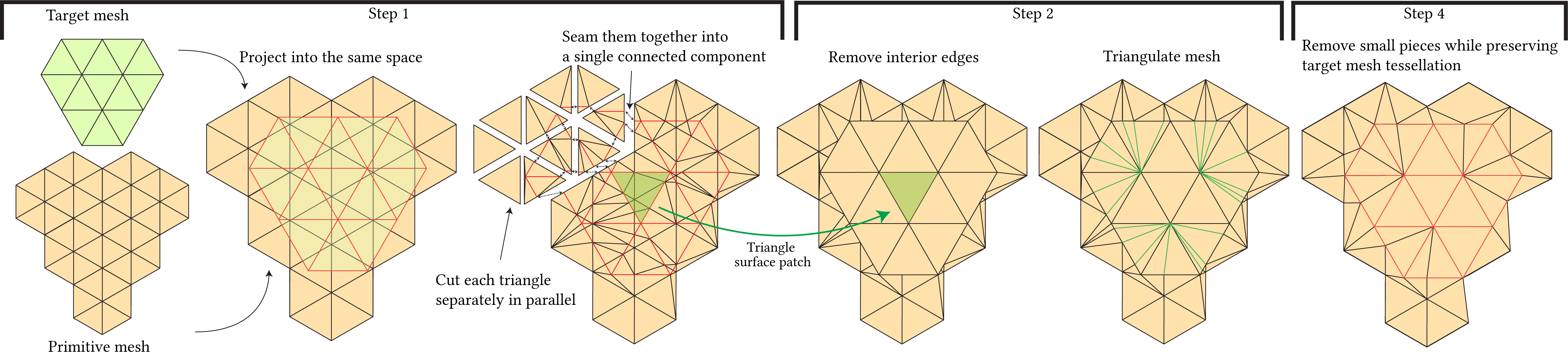}}
\caption{\textbf{The procedure of our feature-preserving refinement.} 
The process consists of four steps. 
In the first step, we refine the primitive mesh such
that it contains the tessellation of the target mesh (step 1).
Next, we remove redundant edges in each triangle surface patch
and produce a triangulated primitive mesh (step 2).
Then, the vertex positions are updated based on their primitive type (step 3).
% This steps is not show here.
Finally, we clean up the small triangles and edges as much as possible
while preserving the target mesh tessellation (step 4).}
\label{fig:coref-pipeline}
\end{figure*}

\subsection{Tessellation Projection and Corefinement}
The sub-surface of the target mesh $\mathcal{S}$ for projection is constructed from those edges of $\mathcal{S}$ which
1) reside within the distance of $\delta$ from $\mathcal{{T}}_{\mathcal{E}_{P}}$ or penetrate with the primitive mesh;
and 2) have an edge projection trajectory that does not intersect with $\mathcal{S}$.
The edge projection trajectory refers to the area swept during the projection of the edge of $\mathcal{S}$ onto its corresponding medial element (medial vertex, edge, or triangle) of $\mathcal{{T}}_{\mathcal{E}_{P}}$, where the endpoints of the edges are mapped to their nearest points on the medial element.
The projected edge on $\mathcal{{T}}_{\mathcal{E}_{P}}$ is determined as the intersection between the trajectory and $\mathcal{{T}}_{\mathcal{E}_{P}}$.
We denote the edges of the sub-surface as $\{\hat{e}\}$, the projection trajectory of each edge $\hat{e}$ as $S_{\hat{e}}$, and each projected curve on $\mathcal{{T}}_{\mathcal{E}_{P}}$ as $\gamma_{\hat{e}}^{\mathrm{proj}}$.
%which is essentially a discretized curve, 

The difficulty lies in how to use exact rational numbers and arithmetic operations to obtain a refined $\mathcal{{T}}_{\mathcal{E}_{P}}$ whose edge set contains $\gamma_{\hat{e}}^{\mathrm{proj}}$.
To achieve this, we define an operator $C(\mathcal{T}_1, \mathcal{T}_2)$, 
which takes two discretized surfaces $\mathcal{T}_1$ and $\mathcal{T}_2$ represented by exact rational numbers as inputs and performs \emph{corefinement} on $\mathcal{T}_2$ using exact arithmetic 
so that their intersection polylines are a subset of edges in $\mathcal{T}_2$.
Thus, the output of $C(\mathcal{T}_1, \mathcal{T}_2)$ is corefined $\mathcal{T}_2$
that contains these intersection polylines.
This operator $C(\cdot, \cdot)$ can be found in modern geometric processing libraries such as CGAL, 
which supports exact computation and exact rational numbers.
We refer the readers to the documentation for more details\footnote{\url{https://doc.cgal.org/latest/Polygon_mesh_processing/index.html\#coref_def_subsec}}.
We use this operator to compute the corefined primitive mesh $\hat{\mathcal{T}}_{\mathcal{E}_{P}}$ as 
$\hat{\mathcal{T}}_{\mathcal{E}_{P}} = C(S_{\hat{e}}, \mathcal{{T}}_{\mathcal{E}_{P}})$, where
the projected curve $\gamma_{\hat{e}}^{\mathrm{proj}}$ is 
a subset of edges of $\hat{\mathcal{T}}_{\mathcal{E}_{P}}$.
% The computation is determined by the corresponding medial element of $\mathcal{{T}}_{\mathcal{E}_{P}}$, as illustrated in Fig.~\ref{fig:coref-cutting}:
The construction of $S_{\hat{e}}$ with exact rational numbers
is determined by the corresponding medial element of $\mathcal{{T}}_{\mathcal{E}_{P}}$, as illustrated in Fig.~\ref{fig:coref-cutting}:
1) For a medial vertex, $S_{\hat{e}}$ is simply the triangle formed by the two endpoints of $\hat{e}$ and the medial vertex itself. 
2) For a medial edge, in cases where $\hat{e}$ is not coplanar with the medial edge in the 3D Euclidean space, $S_{\hat{e}}$ becomes a curved surface in 3D that cannot be parameterized easily.
%and therefore cannot directly represented using exact rational numbers.
We address this problem by performing $C(\cdot, \cdot)$ using the cylindrical coordinates instead of Cartesian coordinates, 
where both $S_{\hat{e}}$ and $\hat{e}$ are line segments and can be used to perform $C(\cdot,\cdot)$.
The output is then transformed back to the Cartesian coordinate system.
Further details about this process can be found in Sec.~\ref{sec:a1}.
3) For a medial triangle, $S_{\hat{e}}$ is divided into various pieces based on 
where the projected endpoints of $\hat{e}$ are located on the medial triangle.
For the portion falling into a cylinder region (Fig.~\ref{fig:coref-cutting}), 
$S_{\hat{e}}$ is constructed and handled similarly to the medial edge case.
For the part falling into a triangle plane region, $S_{\hat{e}}$ is a square plane region formed by $\hat{e}$ and the projected line segment on the triangle plane, which can be represented directly using exact rational numbers.
We process these distinct portions individually and then seam them back together into one $\gamma_{\hat{e}}^{\mathrm{proj}}$. 
Using exact rational numbers and exact arithmetic throughout ensures flawless seaming without numerical error.
%, which can be performed by simply comparing the numbers directly.

To improve computational performance, 
instead of directly taking the whole mesh of $\mathcal{{T}}_{\mathcal{E}_{P}}$ as the input to $C(\cdot, \cdot)$, 
we separately input each triangle of $ \mathcal{{T}}_{\mathcal{E}_{P}}$ and compute the corefinement in parallel.
We then seam together all resulting corefined triangles to obtain $\hat{\mathcal{T}}_{\mathcal{E}_{P}}$,
as shown in Fig.~\ref{fig:coref-pipeline}~(step 1).
Once more, the use of exact rational numbers and arithmetic proves instrumental, allowing us to exactly compare the numbers without any numerical error.
The seamless merging result aligns perfectly with the one obtained from directly operating $C(\cdot, \cdot)$ on the entire $\mathcal{{T}}_{\mathcal{E}_{P}}$, but with significantly enhanced computational speed.

\subsection{Redundant Edge Removal}
The corefinement step introduces redundant edges that do not contribute to any geometric feature of the target shape $\mathcal{S}$.
As shown in Fig.~\ref{fig:coref-pipeline} (middle left), the geometry of the green triangle (denoted as \emph{triangle surface patch}) can be directly deduced from the red edges (projected from $\mathcal{S}$) by using their corresponding radius values with piecewise linear interpolation.
All the black edges and the vertices lying in the interior of the triangle surface patch inside it do not contribute to any geometric feature of $\mathcal{S}$, so we can safely remove them.

To identify a triangle surface patch, we use a BFS process, which starts from an arbitrary triangle that contains an edge derived from the target mesh edge and expands to its neighboring triangles that do not contain an edge from the target mesh edge.
When the BFS terminates, if there are three visited target mesh edges in total and they form a target mesh triangle, 
all the visited triangles are identified as a single triangle surface patch.
We continue this process until all the triangle surface patches have been identified.
We then simplify each triangle surface patch by removing all interior vertices and 
the edges except those corresponding to three target edges.
% In the case that a triangle surface patch is surrounded by other three triangle surface patches,
% we can further simplify it by removing the vertices and edges corresponding to three target edges can be removed, 
However, if a triangle surface patch is surrounded by three triangle surface patches,
those vertices and edges corresponding to three target edges can be removed, 
as shown in Fig.~\ref{fig:coref-pipeline} (green).
Accordingly, the resulting mesh may contain non-triangular polygons.
We then perform a triangulation of the mesh by adding necessary edges to 
ensure the resulting modified primitive mesh is a triangle mesh.

\subsection{Vertex Update}
Once the tessellation of the target mesh is included in the primitive mesh, 
we update the vertex position of the primitive mesh to match the target geometry.
We leverage the correspondence between the vertices of triangle surface patches and 
vertices of $\mathcal{S}$ established in step 1 and 
update the vertex positions of triangle surface patches in the modified primitive mesh 
by moving them to the corresponding vertex positions of $\mathcal{S}$.
This step ensures the radius of updated $\mathcal{{T}}_{\mathcal{E}_{P}}$ are matched with the target mesh $\mathcal{S}$.

\begin{figure}[t]
\centerline{\includegraphics[width=1.0\linewidth]{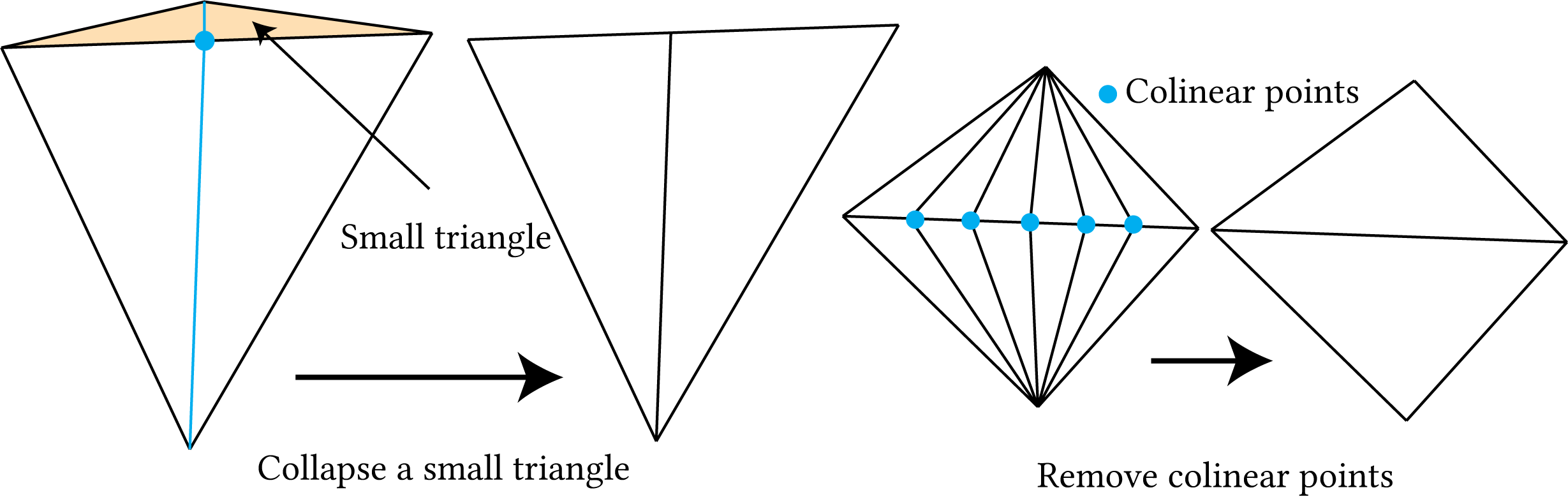}}
\caption{\textbf{Two of our mesh cleanup operations.}
For a small triangle (orange), we first add a vertex to the shared edge of the two triangles, then construct two new edges (blue) by connecting the newly added vertex with the two vertices opposite to the shared edge, and collapse the top edge to remove the traingle.
% by first constructing
% two new edges to the bottom edge of the selected triangles from two opposite triangles.
% Then, the edge on the top is collapsed and the selected triangle is removed.
For the vertices that are colinear, we collapse as many as possible
without destroying target mesh features or violating consistency.}
\label{fig:coref-cleanup}
\end{figure}

\subsection{Mesh Cleanup}
\label{sec:mesh-cleanup}
We finally perform a cleanup process on the tessellation of the primitive mesh that takes into account the geometry updates from the previous step.
We try to remove all minute edges and triangles not corresponding to the shape features and tessellation of $\mathcal{S}$,
with lengths and areas below a certain threshold, 
as well as redundant colinear vertices, 
but forego those removals that would either violate link condition 
or induce triangle flipping.

These operators shown in Fig.~\ref{fig:coref-cleanup}
are implemented based on the edge collapsing operation.
To keep our consistency requirement, 
% the cleanup should not remove any tessellation obtained from the previous refinement steps.
% To achieve this, 
we perform a validation check before performing a standard edge collapse.
Specifically, for an edge to be removed, we check if 
1) the link condition is satisfied~\cite{dey1999topology};
2) the collapsing does not cause triangle flipping; and
3) the collapsing does not destroy the tessellation obtained from the previous refinement steps.
The edge can be collapsed only if all the above conditions are satisfied
% , then the edge can be collapsed.
% If only the link condition is not satisfied,
% we propose a simple fix to the problem.
% We attempt to remesh the neighboring region of the edge to be collapsed.
% If such remeshing is successful, then we can perform the edge collapsing.
% Otherwise, we skip this edge.
% To remesh the local region, we check if such remeshing removes a vertex or a edge
% that corresponds to the target mesh feature.
% If so, we cannot remesh the region.
% After the sanity check, we remove all neighboring triangles of the input edge
% that invalidate the condition.
% After removal, if the hole is quadrilateral, 
% then we fill the hole and it will guarantee to satisfy the link condition~\cite{botsch2010polygon}.
% Otherwise, we roll back to the beginning of the edge collapsing, and process the next edge.

The overall mesh cleanup process is conducted as follows, 
in an order of dealing with small edges, small triangles, and colinear vertices:
We first remove all small edges.
We then traverse all small triangles.
% which guarantees that any small triangle does not contain a small edge.
For each encountered small triangle, 
we select one of the triangle vertexes such that
it can be projected onto the edge opposite it.
Then, we add this projected vertex to the vertex set and construct a new edge $e$ from the selected vertex to its projection.
This minimizes the effect of the edge collapsing as this newly added edge $e$ is the shortest.
After that, we connect the newly added vertex to its opposite vertex in the neighboring triangle.
This guarantees that the current mesh is still a triangle mesh.
Next, we try to collapse $e$ adhering to our aforementioned validation checklist with an extra validation ensuring that the collapsing operation does not introduce any new small triangles.
% In addition, we also check if the collapsing operator introduces anny  
% are introduced during this operation.
This guarantees that this triangle  
% repeating small triangle 
elimination can terminate without inducing an endless loop.
Finally, once all small edges and triangles have been removed, 
we proceed to collapse all colinear vertices.
The collinearity verification is carried out using exact rational numbers derived from the previous refinement steps.
% Since all operations in the previous steps are performed using exact rational numbers,
% the detection of colinear vertices is simple.
% Once they are detected, we perform the same edge collapsing.
Edge collapsing is employed to remove the detected colinear vertices.
In our implementation, we use a removal threshold of 
$5\times10^{-3}$ with the target mesh $\mathcal{S}$ normalized to $[0, 1]$.

\subsection{Consistency of the Refinement process}
% check the robustness and the consistency at the last
Now we validate the consistency requirement of the refinement process:
In tessellation projection (step 1), we consider only sub-surfaces whose edge projection trajectories do not intersect with the target mesh.
This restriction ensures that each newly added vertex after tessellation corefinement associates with only one corresponding radius value, i.e., the distance traveled by this vertex along the projection trajectory.
Vertex update (step 3) only adjusts the radius value of the primitive mesh vertices, thereby preserving consistency.
Redundant edge removal (step 2) only removes newly added vertices and edges from step 1 and thus does not breach consistency.
In mesh cleanup (step 4), our operations only alter the tessellation of the primitive mesh
and thereby do not invalidate consistency.
Consequently, our process maintains consistency throughout.

\subsection{Cylindrical Coordinate Conversion}
\label{sec:a1}
We elaborate on the use of the cylindrical coordinate system, which allows for tessellation projection and corefinement of a medial edge primitive using exact rational numbers and computation. 
For a primitive of medial edge $\mathbf{e}=(\mathbf{v}_1, \mathbf{v}_2)$, the projected curve $\gamma_{\hat{e}}^{\mathrm{proj}}$ of a target edge $\hat{e}=(\hat{v}_1, \hat{v}_2)$ cannot be computed using exact rational number computation since the projection trajectory $S_{\hat{e}}$ is not a flat surface.
To address this issue, instead of performing the projection in the Cartesian coordinate system, we consider a cylindrical coordinate system with the medial edge as the longitudinal axis and $\mathbf{v}_1$ as the origin.
% reparametrize $S_{\hat{e}}$ using a 
Here $\hat{v}_1$, $\hat{v}_2$ have the coordinates $(\rho_0, \varphi_0, z_0)$ and $(\rho_1, \varphi_1, z_1)$, respectively.
The projection trajectory $S_{\hat{e}}$ can be then reparameterized as $S_{\hat{e}}(s, t) = ((1-t)\rho_0 + t\rho_1, (1-s)\varphi_0 + s\varphi_1, (1-t)z_0 + tz_1), s\in[0,1], t\in[0,1]$.
Note that $S_{\hat{e}}$ is a linear surface in the cylindrical coordinate system.
We compute the projected curve $\gamma_{\hat{e}}^{\mathrm{proj}}$ by using the tessellation corefinement operator
$C(S_{\hat{e}}, \mathcal{{T}}_{\mathcal{E}_{P}})$, where the vertex positions of both inputs are represented using the cylindrical coordinates. 
The output corefined mesh $\hat{\mathcal{T}}_{\mathcal{E}_{P}}$ is then converted back to the Cartesian coordinate system for the use of the following steps.
%%%%%
% The coordinate Each interior point of $\hat{e}$
% In this cylindrical coordinate system, $S_{\hat{e}}$ becomes a flat surface.
% while both $\gamma_{\hat{e}}^{\mathrm{proj}}$ and $\hat{e}$ become straight lines.
% cannot be obtained in the Cartesian coordinate system since the projection trajectory $S_{\hat{e}}$ is not a flat surface in Euclidean space.

Although cylindrical coordinate reparametrization enables tessellation corefinement with exact-number precision, the conversion between the cylindrical coordinates and the Cartesian coordinates involves irrational number computation due to the use of trigonometric functions and the square root function to calculate the coordinate value of azimuth and axis distance.
To address this issue, we implement two techniques to ensure that the two-way conversion between coordinates does not sacrifice any accuracy.
The total vertices involved in this two-way conversion can be divided into two sets:
\begin{enumerate}[leftmargin=*]
    \item [(1)] The first set $V_s^{1}$ consists of the two projected endpoints of $\hat{e}$ on $\mathcal{{T}}_{\mathcal{E}_{P}}$ and all the vertices of $\mathcal{{T}}_{\mathcal{E}_{P}}$. These vertices are both transformed from the Cartesian coordinate system to the cylindrical coordinate system (taken as input by corefinement operator $C(\cdot, \cdot)$) and the other way round (they remain within the corefined mesh $\hat{\mathcal{T}}_{\mathcal{E}_{P}}$ which is converted back to Cartesian coordinate system for the following refinement steps).
    \item [(2)] The second set $V_s^{2}$ consists of all the newly added vertices during the tessellation corefinement created by the intersection between $\hat{e}$ and the edges of $\mathcal{{T}}_{\mathcal{E}_{P}}$. For these vertices, we only need to handle the conversion from cylindrical coordinates to Cartesian coordinates.
\end{enumerate}

For the first vertex set $V_s^{1}$, 
we can establish a bijective mapping for these vertices between 
the two coordinate systems 
so that the forward and backward conversion does not introduce any numerical error.
We achieve this by using multiple-precision floating-point numbers 
to represent both coordinates with a chosen precision 
such that the closest pair of vertices in $V_s^{1}$ 
in either coordinate system is distinguishable.
%,i.e., they have a non-zero distance in both coordinate systems.
The use of multiple-precision floating-point numbers also 
provides a rational form of vertices for the input of $C(\cdot, \cdot)$.

For every $v_s^{2}$ from the second vertex set $V_s^{2}$, since $v_s^{2}$ lies on the projected target edge $\hat{e}$.
% there are two possible cases of this $v_s^{2}$ to be addressed in the following refinement process: 
% \begin{enumerate}[leftmargin=*]
%     \item [1)] If $v_s^{2}$ is identified as an interior vertex of the BFS visited triangle surface patches in the following refinement step 2, it will be removed from the final refined mesh.
    % \item [2)] Otherwise, 
$v_s^{2}$ will be placed onto $\hat{e}$ of the target mesh in step 3.
% \end{enumerate}
% Note that in case 1), the condition for judging if $v_s^{2}$ will be removed only depends on the tessellation of $\hat{\mathcal{T}}_{\mathcal{E}_{P}}$ and does not depend on the position of $v_s^{2}$.
% So for case 1) we do not need to convert the cylindrical coordinates $v_s^{2}$ back to Cartesian coordinates with exact-number precision, but instead, only maintain its tessellation in $\hat{\mathcal{T}}_{\mathcal{E}_{P}}$.
We need to obtain the corresponding point of $v_s^{2}$ (in Cartesian coordinates of exact rational numbers) on the target edge $\hat{e}$, denoted as $\hat{v}_s^{2}$, where it will be moved to in step 3.
For clarity, we use $v_{s, \mathrm{cld}}^{2}$ and $v_{s, \mathrm{cts}}^{2}$ to represent the cylindrical coordinates and the Cartesian coordinates of $v_s^{2}$, and similarly for $\hat{v}_s^{2}$.
To obtain $\hat{v}_{s, \mathrm{cts}}^{2}$ with exact-number precision, we leverage two facts: 
\begin{enumerate}[leftmargin=*]
    \item [(i)] For any point, the coordinate conversion in-between the Cartesian system and the cylindrical system preserves exact-number precision at the axial coordinate.
More specifically, given the axial coordinates of the two endpoints of the medial edge as $0$ and $1$, the axial coordinate of 
an arbitrary point $\tilde{e}$ is the barycentric weight of its projection on the medial edge in the Cartesian coordinate system, where only exact rational numbers and exact-number computation are involved; 
    \item [(ii)] In the cylindrical coordinate system, $\hat{v}_{s, \mathrm{cld}}^{2}$ and $v_{s, \mathrm{cld}}^{2}$ share the same axial coordinate value, and correspondingly in the Cartesian coordinate system, the direction from $\hat{v}_{s, \mathrm{cts}}^{2}$ to $v_{s, \mathrm{cts}}^{2}$ is orthogonal to the medial edge (i.e., the projection direction of $\hat{v}_s^{2}$ on $\hat{e}$).
\end{enumerate}
$\hat{v}_{s, \mathrm{cts}}^{2}$ is obtained by the following procedure:
In the cylindrical coordinate system, we first find the edge that contains $v_{s, \mathrm{cld}}^{2}$ in ${\mathcal{T}}_{\mathcal{E}_{P}}$ and calculate the barycentric weight of $v_{s, \mathrm{cld}}^{2}$ on the edge.
We then use the same barycentric weight to compute the Cartesian coordinates of ${v}_{s, \mathrm{cts}}^{2}$ by interpolating between the endpoints of the same edge in the Cartesian coordinate system.
$\hat{v}_{s, \mathrm{cts}}^{2}$ is computed as the intersection of the target edge $\hat{e}$ and the orthogonal plane of the medial edge that contains $v_{s, \mathrm{cts}}^{2}$.

Here we elaborate on how the aforementioned procedure provides an exact-number precision when converting $\hat{v}_{s, \mathrm{cld}}^{2}$ to $\hat{v}_{s, \mathrm{cts}}^{2}$. 
The first two steps utilize facts (i) and (ii) and make sure the computed $v_{s, \mathrm{cts}}^{2}$ has a corresponding axial value in the cylindrical coordinate system that exactly matches both $v_{s, \mathrm{cld}}$ and $\hat{v}_{s, \mathrm{cld}}$.
The last step is constraining $\hat{v}_{s, \mathrm{cts}}^{2}$ to share the same corresponding axial value with $v_{s, \mathrm{cld}}$ in the cylindrical coordinate system and ensuring it lies on the target edge $\hat{e}$.
% Note that the match may not be capable of the azimuth value and radial value since exact-number precision is only preserved for the axial coordinate.
In this way, $\hat{v}_{s, \mathrm{cts}}^{2}$ has a corresponding axial value that exactly matches $\hat{v}_{s, \mathrm{cld}}$.

Together considering both vertex sets in 1) and 2), the two-way conversion between the cylindrical coordinates and the Cartesian coordinates introduces no numerical error to the refinement process, satisfying the robustness constraint.

\subsection{Handling Periodic Domain in the Cylindrical Coordinate System}
The periodic nature of the cylindrical coordinate system, i.e., $(\rho, \varphi, z)$ and 
$(\rho, \varphi + 2n\pi, z),n\in\mathbb{Z}$ represent the same vertex, poses challenges in tessellation corefinement for the medial edge primitive in step 1, as 
% This poses difficulties in corefining a target edge with primitive triangles,
% because 
it may cause false positive/negative intersection tests in the cylindrical coordinate system.
For instance, supposing the azimuth axis has range $\phi \in [-\pi, \pi]$, 
a short edge $e_1$ with endpoints $(-0.9\pi,1,1)$ and $(0.9\pi,1,1)$ can be treated as a long edge spanning from $-0.9\pi$ to $0.9\pi$ when performing intersection test with other edges.
% a short edge whose end points map to $(-0.9\pi,0.5,1)$ and $(0.9\pi, 0.5,1)$
% becomes a long edge that spans from $-0.9\pi$ to $0.9\pi$ in the cylindrical coordinate system.
% Assume $\phi \in [-\pi, \pi]$.
% The root cause 
The essential cause
of this problem is due to its crossing the periodic boundary $\pm \pi$.
To resolve this issue, we simultaneously use two different cylindrical coordinate systems, where
one is obtained by rotating the other by $\pi$. % along the medial edge.
Under the assumption that the tessellation of both the primitive mesh and the target mesh is reasonably dense,
% By constructing such two coordinate systems,
for any pair of the target edge $\hat{e}$ and primitive mesh triangle $\hat{\Delta}$, there always exists at least one coordinate system where neither $\hat{e}$ or $\hat{\Delta}$ cross the periodic boundary.
% successfully projected into at least one coordinate system without crossing the periodic boundary.
The corefinement operator $\mathcal{C}(\cdot,\cdot)$ can thus output faithful results within one of these two coordinate systems.
% This guarantees that $\mathcal{C}(\cdot,\cdot)$ is correct in at least in one coordinate system.

In step 1 of the refinement process, 
seaming the triangles together into one CC requires exact comparisons between vertices using exact rational numbers.
In the case of using two cylindrical coordinate systems, 
When comparing two identical vertices belonging to two different coordinate systems,
we adjust the azimuth coordinate value for one of the vertices by $\pi$.
To use exact rational numbers, we project $\pi$ to a reasonably accurate rational number $\pi^R$.
Therefore, if the coordinate is $(\rho, \varphi, z)$ in the first system, 
then the counterpart in the second system is $(\rho, \varphi + \pi^R, z)$ if $\varphi < 0$ and
$(\rho, \varphi + \pi^R, z)$ otherwise.

% In addition to ensuring the correctness of the corefinement, 
% the conversion between the two coordinate systems is also necessary.
% This is because we need to compare 
% if a pair of vertices are identical, which is frequently performed in step 1.
% Since we only rotate one coordinate system by $\pi$ to arrive at the other, 
% the conversion between two systems is simple.
% Therefore, if the coordinate is $(\theta,t,r)$ in the first system, 
% then the counterpart in the second system is $(\theta + \pi^R, t,r)$ if $\theta < 0$ or
% $(\theta - \pi^R, t,r)$ otherwise.

\subsection{Refinement of Medial Triangle Primitive Mesh}
% For each triangle, we first compute intersected cutting meshes for target edges.
% The cutting mesh is a triangle for medial vertex, a 2D line segment for medial edge, 
% or a rectangle mesh for the edge in the triangular region for medial triangle.
% Although the corefinement of each primitive mesh triangle with the target mesh is performed individually,
% the newly added vertices along the triangle borders will be the same as
% the ones on the borders of the neighboring triangles (Fig.~\ref{fig:coref-pipeline}). 
% This is because all the corefinement operations are performed using exact rational numbers.
% This makes it easy to seam all the corefined triangles back into a connected manifold primitive mesh.
% Note that for the case of medial edge primitive, since all operations are performed exactly
% in the 2D space, there are no mismatches.
For the case of medial triangle primitive, 
we define three regions on the primitive mesh
as shown in Fig.~\ref{fig:coref-cutting}.
For each target edge, it is first broke into several pieces with each one falling into
one of the regions.
As known, we need to seam all corefined triangles together in step 1.
The vertices on the edges of each neighboring triangles must be comparable.
Because the corefinement is performed in different coordinate systems in different regions,
an additional handling is needed for vertices on the borders between different regions
to ensure that they can be compared.
For the vertices of the corefined triangles in the triangular region,
the exact 3D Cartesian coordinate can be directly computed.
On the other hand, for the vertices of the corefined triangles 
in the cylinder region,
we give a method to compute the exact 3D Cartesian space position of each vertex
on the region border.
Denote such vertex by $v_c$.
%and its position in the Cartesian coordinate and cylindrical coordinate
%by $v_{c,\mathrm{cts}}$ and $v_{c,\mathrm{cld}}$, respectively.
Our method first computes the exact barycentric weights of $v_c$ 
with respect to the containing primitive triangle.
Since we only focus on the vertices on the region border,
$v_c$ must lie on the edge of a primitive triangle in $\mathcal{T}_{\mathcal{E}_P}$.
Therefore, the barycentric weights can be computed using only the end vertices of the resident edge.
This primitive triangle edge is special because it is also on the region border.
Therefore, the azimuth $\phi$ of any point on the edge is the same (either 0 or $\pi$).
Benefiting from it, this computation can be done by merely considering the axial coordinates $\rho$.
As mentioned in Appendix~\ref{sec:a1}, axial coordinates are exact.
Thus, the barycentric weights can be computed exactly by the ratio of the axial coordinates.
Again, since the azimuth $\phi$ of any point on the edge is the same (either 0 or $\pi$)
and axial coordinates are linearly related to the 3D positions,
the barycentric weights can be used to compute the 3D position in the Cartesian space
by linear interpolation.
Accordingly, we can obtain the exact 3D position of $v_c$.